\documentclass[]{aastex7}

\usepackage{blindtext}
\usepackage{longtable}
\usepackage{graphicx}
\usepackage{amssymb}
\usepackage{amsmath}
\usepackage{afterpage}
\usepackage{rotating}
\usepackage{pifont}

\def\lya{Ly$\alpha$}

\makeatletter
\DeclareRobustCommand{\NHI}{%
  $N_{\mbox{\scriptsize H\,\check@mathfonts\fontsize\sf@size\z@\selectfont I}}$}
\makeatother

\makeatletter
\DeclareRobustCommand{\OmHI}{%
  $\Omega_{\mbox{\scriptsize H\,\check@mathfonts\fontsize\sf@size\z@\selectfont I}}$}
\makeatother

\makeatletter
\DeclareRobustCommand{\OmHIa}{%
  $\Omega^{\mathcal{A}}_{\mbox{\scriptsize H\,\check@mathfonts\fontsize\sf@size\z@\selectfont I}}$}
\makeatother

\makeatletter
\DeclareRobustCommand{\OmHIb}{%
  $\Omega^{\mathcal{B}}_{\mbox{\scriptsize H\,\check@mathfonts\fontsize\sf@size\z@\selectfont I}}$}
\makeatother

\makeatletter
\DeclareRobustCommand{\OmHIc}{%
  $\Omega^{\mathcal{C}}_{\mbox{\scriptsize H\,\check@mathfonts\fontsize\sf@size\z@\selectfont I}}$}
\makeatother

\makeatletter
\DeclareRobustCommand{\OmHId}{%
  $\Omega^{\mathcal{D}}_{\mbox{\scriptsize H\,\check@mathfonts\fontsize\sf@size\z@\selectfont I}}$}
\makeatother

\makeatletter
\DeclareRobustCommand{\HI}{%
  \mbox{H\,\check@mathfonts\fontsize\sf@size\z@\selectfont I}%
}
\makeatother

\makeatletter
\DeclareRobustCommand{\hi}{%
  \mbox{H\,\check@mathfonts\fontsize\sf@size\z@\selectfont I}%
}
\makeatother

\makeatletter
\DeclareRobustCommand{\MgII}{%
  \mbox{Mg\,\check@mathfonts\fontsize\sf@size\z@\selectfont II}%
}
\makeatother

\makeatletter
\DeclareRobustCommand{\HII}{%
  \mbox{H\,\check@mathfonts\fontsize\sf@size\z@\selectfont II}%
}
\makeatother

\newcommand{\cii}{[C{\sc ii}] 158$\mu$m}

\received{21 July 2026}
\revised{14 September 2026}
\accepted{16 September 2026}

\shorttitle{The KAGG survey: KCWI Analysis of Gas around Galaxies at $z\approx 3-5$}

\shortauthors{Oyarz\'un et al.}

\begin{document}

\title{The KAGG survey: KCWI Analysis of Gas around Galaxies at $z\approx 3-5$}

\author[0000-0003-0028-4130]{Grecco A. Oyarz\'un}
\affiliation{Homer L. Dodge Department of Physics and Astronomy, The University of Oklahoma, 440 W. Brooks St., Norman, OK 73019, USA}
\affiliation{Department of Physics and Astronomy, Johns Hopkins University, Baltimore, MD 21218, USA}
\email{grecco.oyarzun@ou.edu}

\author[0000-0002-9946-4731]{Marc Rafelski}
\affiliation{Space Telescope Science Institute, 3700 San Martin Drive, Baltimore, MD 21218, USA}
\affiliation{Department of Physics and Astronomy, Johns Hopkins University, Baltimore, MD 21218, USA}
\email{mrafelski@stsci.edu}

\author[0000-0002-7738-6875]{J. Xavier Prochaska}
\affiliation{Department of Astronomy \& Astrophysics, UCO/Lick Observatory, University of California, 1156 High Street, Santa Cruz, CA 95064, USA}
\affiliation{Kavli Institute for the Physics and Mathematics of the Universe (Kavli IPMU), 5-1-5 Kashiwanoha, Kashiwa, 277-8583, Japan}
\affiliation{Division of Science, National Astronomical Observatory of Japan, 2-21-1 Osawa, Mitaka, Tokyo 181-8588, Japan}
\email{xavier@ucolick.org}

\author[0000-0002-9757-7206]{Nissim Kanekar} 
\affiliation{National Centre for Radio Astrophysics, Tata Institute of Fundamental Research, Pune University, Pune 411007, India}
\email{nkanekar@ncra.tifr.res.in}

\author[0000-0002-9838-8191]{M. Neeleman}
\affiliation{National Radio Astronomy Observatory, 520 Edgemont Road, Charlottesville, VA, 22903, USA}
\email{mneeleman@nrao.edu}

\author[0000-0002-6505-9981]{M.E. Wisz}
\affiliation{Maria Mitchell Observatory, 4 Vestal St. Nantucket, MA 02554, USA}
\affiliation{Department of Physics, University of California, Merced, CA 95343, USA}
\email{wimarie01@gmail.com}

\author[0000-0001-6676-3842]{Michele Fumagalli}
\affiliation{Dipartimento di Fisica G. Occhialini, Universit\`a degli Studi di Milano-Bicocca, Piazza della Scienza 3, I-20126 Milano, Italy}
\affiliation{INAF - Osservatorio Astronomico di Trieste, via G. B. Tiepolo 11, I-34143 Trieste, Italy}
\email{michele.fumagalli@unimib.it}

\author[0000-0003-2973-0472]{Regina A. Jorgenson}
\affiliation{Department of Physics \& Astronomy, Cal Poly Humboldt, 1 Harpst Street, Arcata, CA 95521}
\affiliation{Maria Mitchell Observatory, 4 Vestal St. Nantucket, MA 02554, USA}
\email{regina.jorgenson@gmail.com}



\begin{abstract}
We present the KCWI Analysis of Gas around Galaxies (KAGG) survey of Ly$\alpha$ emission around Damped \lya\ Absorbers (DLAs) at $z\approx3-5$ with the Keck Cosmic Web Imager (KCWI) integral-field spectrograph. The KAGG survey is characterized by the sample of 36 DLAs spanning a wide metallicity range, the wide search area for associated Ly$\alpha$ emission (impact parameter $b\lesssim120$~kpc), and the high emission sensitivity, reaching Ly$\alpha$ luminosities $\leq 10^{42}$~erg~s$^{-1}$. We identify 7 DLA Ly$\alpha$ emitters that are likely DLA galaxy candidates due to the good match between the Ly$\alpha$ emission and DLA redshifts. Based on a composite sample of 11 DLA \lya\ emitters from KAGG and MAGG (MUSE Analysis of Gas around Galaxies), we find that the impact parameters of DLA \lya\ emitters decrease with DLA metallicity, from $b>50~$kpc at [M/H]~$\approx-2$ to $b<50~$kpc at [M/H]~$\approx-1$. Including \lya\ emitters at larger velocity offsets and at larger impact parameters from the DLAs, we construct samples of 35 ``Group'' \lya\ emitters and 17 ``Large-scale'' \lya\ emitters. We find that their impact parameters and incidence rates also depend on DLA metallicity. Based on simulations of our survey and comparisons with models of galaxy formation, we conclude that higher DLA metallicities are associated with higher halo masses. The low Ly$\alpha$ emission detection rate at low impact parameters suggests that most DLA galaxies either have low stellar masses (M$\lesssim 10^{8}$~M$_{\odot}$) or have their \lya\ emission suppressed due to \hi\ resonant scattering and/or dust absorption.
\end{abstract}


\section{Introduction} 
\label{1}

Damped Ly$\alpha$ absorbers (DLAs) are clouds of atomic hydrogen gas (\HI) of particularly high column density (\NHI~$>2\times 10^{20}$~cm$^{-2}$) identified in the absorption spectra of background quasars (QSOs). Since their detection over 40 years ago (\citealt{wolfe1986,wolfe1995,wolfe2005,lanzetta1991}), DLAs have garnered great interest from the astrophysical community for several reasons. DLAs dominate the \HI\ mass density of the Universe, and thus their incidence rate constrains the cosmic \HI\ mass budget (\citealt{prochaska-wolfe2009,noterdaeme2012,oyarzun2025}). In addition, DLAs enable studies of the physical conditions of \HI\ gas across cosmic history, yielding measurements of dust depletion, H$_2$ fraction, gas temperature, and metal enrichment up to $z\sim 5.5$ (e.g. \citealt{prochaska-wolfe1997,rafelski2012,rafelski2014,prochaska2013,neeleman2013,neeleman2015,balashev2017,klimenko2020,kanekar2003,kanekar2014}). And lastly, by accessing high \NHI\ gas that is beyond the reach of \HI\ 21~cm emission surveys today ($z<1.5$; \citealt{chowdhury2020,chowdhury2022}), DLAs enable characterization of the \hi\ gas reservoirs of galaxies up to $z\sim8$ (e.g. \citealt{fynbo2010,fynbo2018,fynbo2023,neeleman2017,neeleman2019,neeleman2020,neeleman2025,mackenzie2019,prochaska2019,kanekar2018,kanekar2020,kaur2021,kaur2024,kaur2025,bordoloi2022,heintz2024,oyarzun2024}). This endeavor, the identification and characterization of the galaxies associated with DLAs, is the main focus of this paper. 

The search for high-redshift DLA galaxies started before the turn of the century (e.g. \citealt{moller-warren1993}). For the first two decades, it was widely believed that the high \NHI\ of DLAs had to originate in the interstellar medium (ISM) of galaxies along the line of sight, and therefore the majority of searches were conducted with slit-based spectrographs that are sensitive to \lya\ emission at small projected physical separations from the DLA (i.e., small impact parameters $b$). Few of these campaigns succeeded at detecting DLA galaxies at $z>2$ (e.g. \citealt{fynbo2013,krogager2016,joshi2021}), with a collective detection rate of $\sim 10\%$. Complementary approaches (e.g. narow-band imaging and early integral-field units) similarly yielded little success in detecting \lya\ emission near DLAs at cosmic noon and beyond (\citealt{kulkarini2000,kulkarni2006,warren2001,chun2006,fumagalli2010,peroux2011}).

Among the reasons brought forward to explain the low detection rates of DLA galaxies was the brightness of the QSO in the rest-frame UV and optical (e.g. \citealt{fynbo2010}). This motivated studies to search for DLA galaxies in the sub-mm, a wavelength regime in which the QSO brightness is much lower. \citet{neeleman2017} utilized the Atacama Large Millimeter array (ALMA) to identify two DLA galaxies via their \cii\ emission at $z\sim 4$. Soon after, \citet{neeleman2018} and \citet{fynbo2018} identified two additional DLA galaxies with ALMA, this time through CO emission at $z= 2-2.5$. These studies have since been expanded upon, with over 10 DLA galaxies detected by their \cii\ and/or CO emission to date (\citealt{neeleman2018,neeleman2025,kanekar2020,kaur2022b,kaur2022a,kaur2025}). These sub-mm DLA galaxies stand out for their high star-formation rates, bright dust continuum emission, and large molecular gas masses (\citealt{kaur2021}). 

Interestingly, the detection rate of sub-mm DLA galaxies strongly depends on DLA metallicity: while $\gtrsim 50\%$ of DLAs with [M/H]~$>-0.8$ that have been surveyed with ALMA show an emission counterpart in the sub-mm, this number plummets to $\lesssim 10\%$ for DLAs with [M/H]~$<-0.8$. This result appears to be consistent with a DLA metallicity-luminosity relation, i.e., the hypothesis that metal-rich DLAs are associated with more massive, brighter galaxies \citep[e.g.][]{moller2004,moller2013,ledoux2006,fynbo2008,christensen2014}. Also, the impact parameters of sub-mm DLA galaxies are surprisingly large: CO emitters at $z\sim 2$ are typically found $b\sim 10-20$~kpc away from DLAs \citep{kanekar2020,kaur2025}, whereas \cii\ emitters at $z\sim 4$ are found even further away, $b\sim 20-60$~kpc from DLAs \citep{neeleman2017,neeleman2025}. This finding suggests that metal-rich ([M/H]$~\gtrsim -0.5$) DLAs may often probe \HI\ gas beyond the ISM of galaxies, likely in the circumgalactic medium (CGM). 

These insights can, at least in part, explain the low detection rate of DLA galaxies through \lya\ emission. First, the high \hi\ and dust contents of massive galaxies are bound to scatter and absorb \lya\ photons, severely lowering the escape of \lya\ radiation from the host galaxies of metal-enriched DLAs \citep[e.g.][]{oyarzun2016,oyarzun2017,oyarzun2024}. The \lya\ emission study that stands out for its detection rate (7 \lya\ emitters in 11 DLAs; \citealt{krogager2017}) revealed low \lya\ luminosities ($L_{\mbox{\small\lya}} = 10^{41}-10^{42}$~erg~s$^{-1}$) around DLAs of intermediate metallicity ([M/H]~$\sim-1$), in agreement with this interpretation. Second, any DLA galaxies located at $b\gtrsim 30$~kpc --- as is the case for half of the sub-mm DLA galaxies detected to date \citep[e.g.][]{neeleman2025}--- are beyond the impact parameter coverage of slit spectrographs or infrared integral-field units (IFUs; \citealt{christensen2007,peroux2011,peroux2012}). For this reason, sensitive optical IFUs have become important in the search for \lya\ DLA galaxies out to $b \gtrsim 50$~kpc \citep{fumagalli2017,mackenzie2019,nielsen2022,lofthouse2023,oyarzun2024}.

\citet{fumagalli2017} found early success in IFU-based searches for \lya\ emission with VLT/MUSE, identifying three Ly$\alpha$ emitting sources at $z\sim 3.25$ associated with a DLA of intermediate metallicity ([M/H]~$=-1.1$). All three sources are $b=18-35$~kpc away from the DLA, highlighting the need for IFUs to identify these galaxies. In the following years, a systematic search for DLA galaxies in \lya\ emission with VLT/MUSE was conducted: the MUSE Analysis of Gas around Galaxies (MAGG) survey (\citealt{mackenzie2019,dutta2020,lofthouse2020,lofthouse2023,fossati2021,galbiati2023,galbiati2024}). The MAGG survey detected over 20 DLA \lya\ emitters out to $b\approx 300$~kpc around low-metallicity ([M/H]~$\lesssim-1.5$) DLAs (\citealt{mackenzie2019,lofthouse2023}). Similarly, \citet{oyarzun2024} conducted a search with the Keck Cosmic Web Imager Integral Field Spectrograph (KCWI; \citealt{kcwi}) for \lya\ emission from DLA galaxies around 14 DLAs at $z\sim 2$. Five galaxies around intermediate- or high-metallicity DLAs ($-1.5<$~[M/H]~$<-0.3$) were detected in total, three of which were already known and all of which are found over a wide range of impact parameters ($b=0-70$~kpc). 

In this paper, we introduce a new survey for DLA galaxies in \lya\ emission: the KCWI Analysis of Gas around Galaxies (KAGG) survey. KAGG is designed to exploit the wide wavelength coverage of KCWI to search for \lya\ emitters around DLAs at $z\sim 2-5$. The KAGG survey is unique because of the unprecedented number of DLAs (36 absorbers in 17 QSO fields), the absence of any DLA metallicity pre-selection in the sample (which is key to further our understanding of the DLA galaxy population), and the sensitivity of our observations (reaching \lya\ detection luminosities below $10^{42}$~erg~s$^{-1}$ out to impact parameters of at least $b = 75$~kpc). 

This paper is structured as follows. We introduce the survey in Section~\ref{2} and outline our methodology in Section~\ref{3}. The results and discussion are presented in Section~\ref{4} and \ref{5}, respectively. We summarize the results of the paper in Section~\ref{6}. We assume a Lambda Cold Dark Matter cosmology, with $\Omega_m=0.3$, $\Omega_\Lambda = 0.7$, and $H_0=70$~km~s$^{-1}$~Mpc$^{-1}$ throughout the paper. All magnitudes are in the AB system \citep{oke1983}.

\section{The KAGG survey} 
\label{2}

The KAGG survey exploits the capabilities of KCWI, the integral field spectrograph at the Keck-II Telescope, to search for \lya\ emission around DLAs at high redshift. KCWI has a wide field of view ($\approx 150$~kpc at $z > 2$), and thus is well-matched to search for redshifted \lya\ emission from DLA fields. More importantly, KCWI has two spectral arms, each of which yield high sensitivities at good spectral resolutions over wide \lya\ redshift ranges. This uniquely wide spectral coverage enabled us to efficiently observe a large sample by targeting QSO sightlines with multiple DLAs.  

 \begin{figure*}[h!]
\vspace{-0.4em}
\centering 
\includegraphics[width=7in]{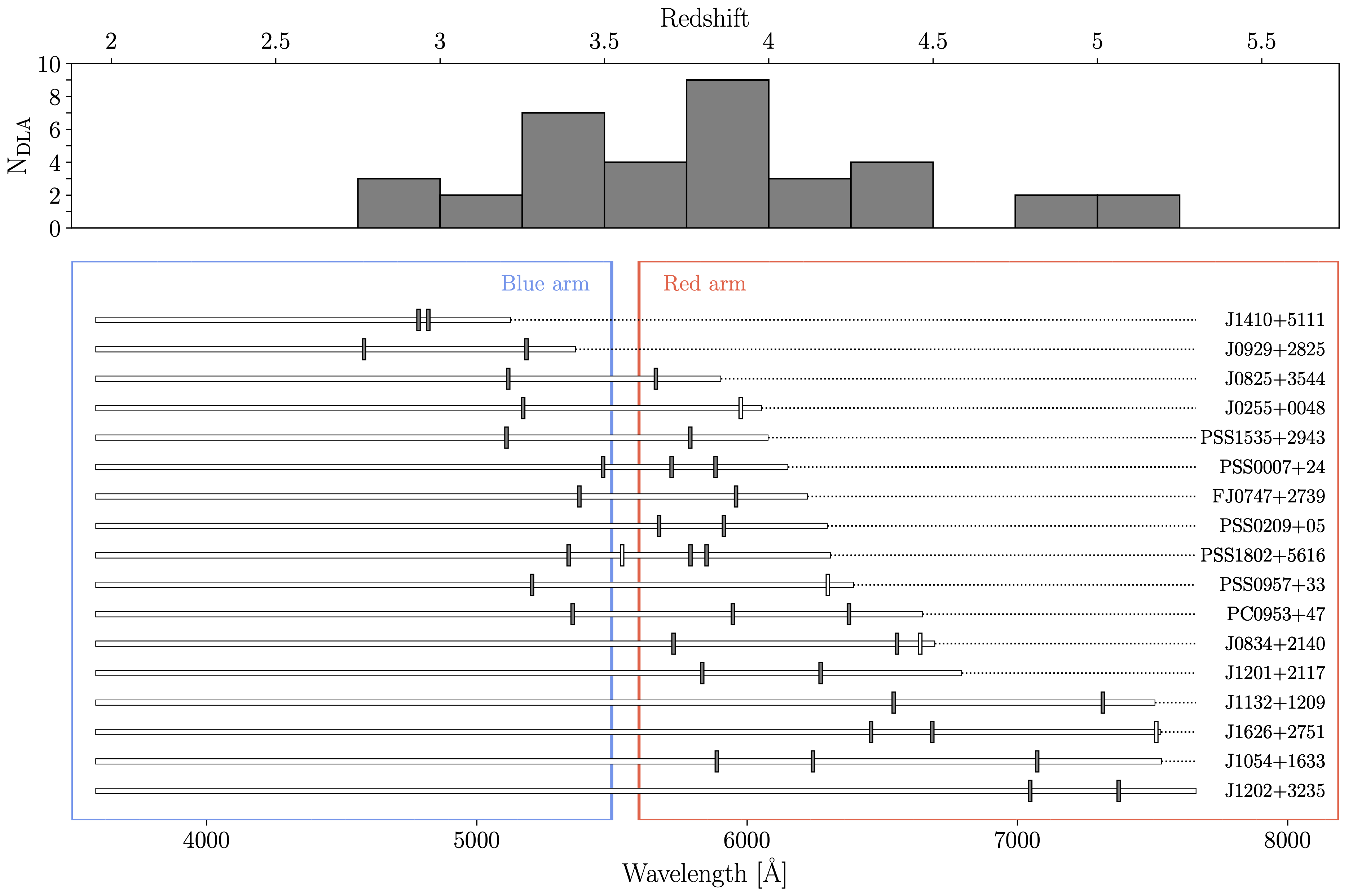}
\vspace{-0.6em}
\caption{Top: Redshift distribution of the DLAs comprising the KAGG survey. Bottom: Schematic view of the DLAs in KAGG. Each horizontal line represents a QSO sightline, where the line end corresponds with the redshifted wavelength of the \lya\ emission line of the QSO. Each DLA is plotted as a vertical gray bar at the corresponding absorption wavelength of the \lya\ line (x-axis) and at the corresponding QSO sightline (y-axis). The five DLAs plotted as vertical white bars are excluded from this paper because the \lya\ emission line fell between the two detectors or because they are proximate to the QSO. The DLA fields were observed with either the blue or the red arm of KCWI, as respectively denoted by the blue and red boxes. }
\label{fig_qsos}
\end{figure*}

 \begin{figure*}[h!]
\vspace{0em}
\centering 
\includegraphics[width=7in]{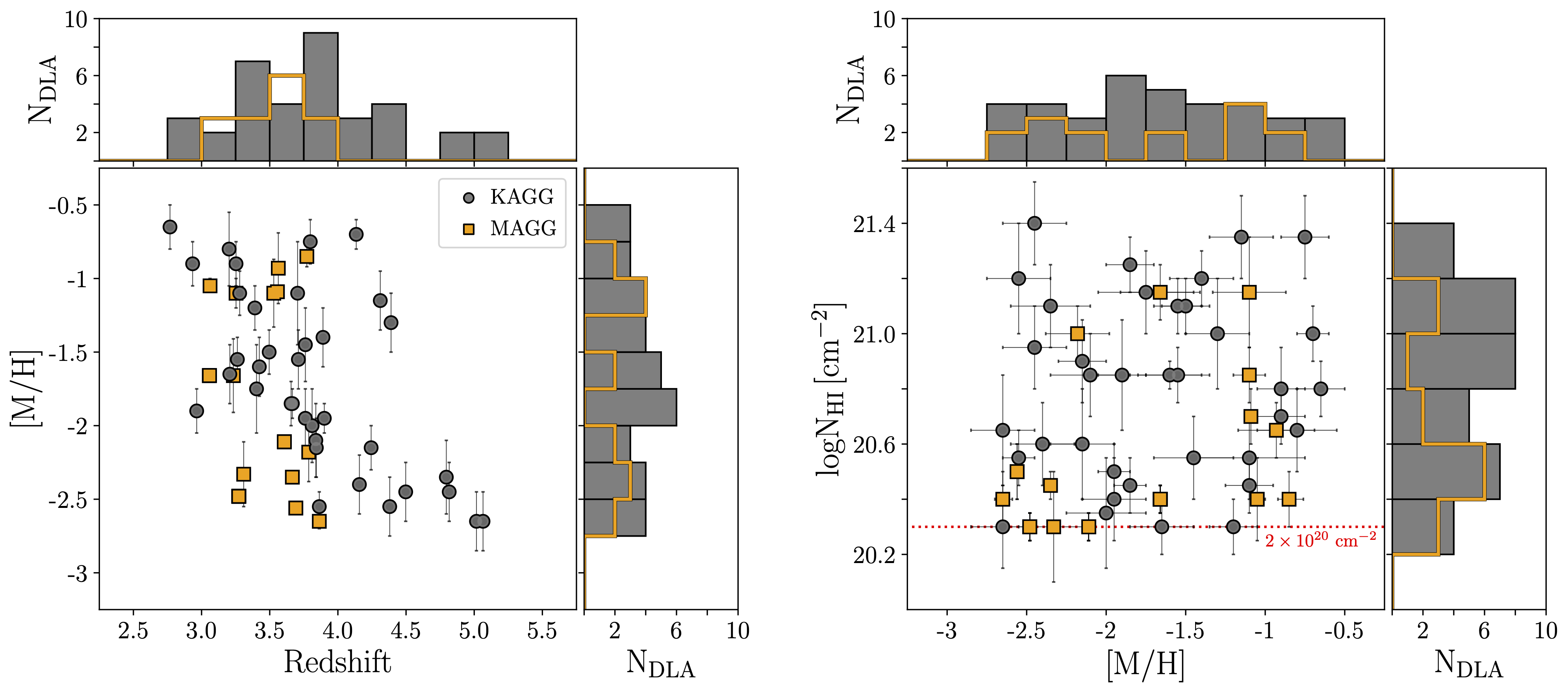}
\vspace{-0.7em}
\caption{Redshifts, metallicities and \NHI\ of the DLAs in the KAGG and MAGG surveys. KAGG DLAs are shown as gray circles and MAGG DLAs are shown as yellow squares. The \NHI\ limit that separates between DLAs and subDLAs is plotted as a red line. The smaller panels show the distribution of the samples in metallicity and \NHI.}
\label{fig_sample}
\vspace{-1em}
\end{figure*}

\subsection{Sample}
\label{2.1}

We selected our DLA sample from the DLA catalogs of \citet{rafelski2012,rafelski2014}. These studies characterized hundreds of DLAs at $z\sim 2-5$ with high resolution ($R>7000$) spectroscopy, which enabled them to obtain reliable \NHI\ and metallicity estimates (see \citealt{neeleman2016,oyarzun2025,wisz2026} for the importance of spectral resolution for DLA characterization). Our criteria for constructing the KAGG DLA sample from the \citet{rafelski2012,rafelski2014} parent sample were the presence of at least two DLAs along the QSO sightline (to maximize survey efficiency) and observability with the Keck telescopes. We observed a total of 36 DLAs along 17 of these QSO sightlines within the allocated observing time for the KAGG survey. This observed sample will be referred to as the KAGG survey sample throughout this paper. 

The DLA redshifts, \NHI, and metallicities were determined by \citet{rafelski2012,rafelski2014}. The redshifts were measured from low-ion metal transitions, the \NHI\ values were estimated through Voigt profile fitting, and the metallicities were calculated from the column densities of different element transitions (O, S, Si, Zn). The values of these quantities for the KAGG survey sample are presented in Table~\ref{table_sample}, Figure \ref{fig_qsos}, and Figure \ref{fig_sample}.

Our DLAs lie in the redshift range $z\approx 2-5$, with most at $z\approx 3.5-4$ (see Figure~\ref{fig_qsos}). This figure also shows the five DLAs that are not included in the final sample of the KAGG survey: (i)~a DLA towards PSS1802+5616 that could not be observed because its redshifted \lya\ wavelength falls in the wavelength gap between the two KCWI spectral arms and (ii)~four ``proximate'' DLAs toward J0255+0048, PSS0957+33, J0834+2140, and J1626+2751 that were observed but lie within 5000~km~s$^{-1}$ of the QSO redshift. These proximate DLAs are not included in the KAGG sample to prevent biases from the environments of luminous QSOs
\citep[e.g.][]{prochaska2008,ellison2010,ellison2011}.

Our QSO sightlines were selected merely based on the presence of multiple DLAs. Thus, the KAGG DLA sample was not subject to any pre-selection based on DLA metallicity. However, low \NHI\ DLAs (\NHI$~\approx 10^{20.3}~\rm cm^{-2}$) are underrepresented, just as in the parent sample from \citet{rafelski2012,rafelski2014}. As originally pointed out by \citet{jorgenson2013}, this is because spectroscopic high-resolution follow-up studies tend to avoid borderline DLAs (i.e., systems with \NHI$~\approx 10^{20.3}~\rm cm^{-2}$). This bias is clearly visible in the KAGG \NHI\ distribution, plotted in Figure~\ref{fig_sample}. Unlike samples that are randomly drawn from the DLA population, most of our DLAs have \NHI$~\approx 10^{20.5}-10^{21}~\rm cm^{-2}$ instead of \NHI$~\approx 10^{20.3}-10^{20.5}~\rm cm^{-2}$. 

The KAGG DLA sample contains several intermediate metallicity DLAs ([M/H]~$\approx -1$), some of which have yielded \cii\ emission detections at $z \approx 4$ \citep[e.g.][]{neeleman2017,neeleman2025}. Five \cii\ emitters have been detected with ALMA in the fields of the DLAs at $z_{abs} \approx 3.8$ toward J1201+2117, $z_{abs} \approx 4.13$ toward J1054+1633, $z_{abs} \approx 4.31$ toward J1626+2751, and $z_{abs} \approx 4.39$ toward J0834+2140 \citep{neeleman2017,neeleman2025}.

\begin{table*}[h!]
\centering
\tabletypesize{normalsize}
\caption{The KAGG survey sample. The columns are (1)~QSO name, (2)~right ascension, (3)~declination, (4)~QSO redshift, (5)~DLA redshift, (6)~DLA \NHI, and (7)~DLA metallicity.
\label{table_sample}}
\setlength{\tabcolsep}{0.2in}
\begin{tabular}{ccccccc}
\hline
\hline
QSO & RA & Declination & $z_{em}$ & $z_{abs}$ & log\NHI~[cm$^{-2}$] & [M/H] \\
\hline
& & & & & & \\
PSS0007+24 & 00:07:38.62 & +24:17:24.8 & 4.05 & 3.4959 &  $ 21.10\pm 0.10$ & $-1.50 \pm 0.15$ \\
... & ... & ... & ... & 3.7045 &  $ 20.55\pm 0.15$ & $ -1.25\pm 0.35$ \\
... & ... & ... & ... & 3.8382 &  $ 20.85\pm 0.15$ & $ -2.10\pm 0.25$ \\
PSS0209+05 & 02:09:44.62 & +05:17:13.7 & 4.17 & 3.6662 &  $ 20.45\pm 0.10$ & $-1.85 \pm 0.10$ \\
... & ... & ... & ... & 3.8636 &  $ 20.55\pm 0.10$ & $ -2.55\pm 0.10$ \\
J0255+0048 & 02:55:18.58 & +00:48:47.5  & 3.97 & 3.2529 &  $ 20.70\pm 0.10$ & $-0.90 \pm 0.15$ \\
FJ0747+2739 &  07:47:11.15 & +27:39:03.3  & 4.11 & 3.4237 &  $ 20.85\pm 0.05$ & $ -1.60\pm 0.20$ \\
... & ... & ... & ... &  3.9004 &  $ 20.50\pm 0.10$ & $ -1.95\pm 0.10$ \\
J0825+3544 & 08:25:40.13 & +35:44:14.2  & 3.85 & 3.2073 &  $ 20.30\pm 0.10$ & $-1.70 \pm 0.20$ \\
... & ... & ... & ... & 3.6567 &  $ 21.25\pm 0.10$ & $-1.85 \pm 0.15$ \\
J0834+2140 & 08:34:29.45 & +21:40:24.7 & 4.50 & 3.7102 &  $ 20.85\pm 0.10$ & $-1.55 \pm 0.20$ \\
... & ... & ... & ... & 4.3900 &  $ 21.00\pm 0.20$ & $ -1.30\pm 0.20$ \\
J0929+2825 & 09:29:14.50 & +28:25:29.1 & 3.40 & 2.7684 &  $ 20.80\pm 0.10$ & $ -0.65\pm 0.15$ \\
... & ... & ... & ... & 3.2627 &  $ 21.10\pm 0.10$ & $-1.55 \pm 0.15$ \\
PC0953+47 &  09:56:25.17 & +47:34:42.5 & 3.46 & 3.4033 &  $ 21.15\pm 0.15$ & $ -1.75 \pm 0.30$ \\
... & ... & ... & ... & 3.8907 &  $ 21.20\pm 0.10$ & $ -1.45\pm 0.20$ \\
... & ... & ... & ... & 4.2439 &  $ 20.90\pm 0.15$ & $ -2.15 \pm 0.15 $ \\
PSS0957+33 & 09:57:44.47 & +33:08:20.8  &  4.25 & 3.2796 &  $ 20.45 \pm 0.10$ & $ -1.10\pm 0.15$ \\
J1054+1633 & 10:54:45.43 & +16:33:37.4 & 5.19 & 3.8420 &  $ 20.60\pm 0.20$ & $-2.15 \pm 0.20$ \\
... & ... & ... & ... & 4.1346 &  $ 21.10\pm 0.10$ & $ -0.70\pm 0.10$ \\
... & ... & ... & ... & 4.8166 &  $20.95 \pm 0.15$ & $ -2.45\pm 0.20$ \\
J1132+1209 & 11:32:46.50 & +12:09:01.7  & 5.17 & 4.3802 &  $ 21.20\pm 0.20$ & $ -2.55\pm 0.20$ \\
... & ... & ... & ... & 5.0165 &  $ 20.65\pm 0.20$ & $ -2.65\pm 0.20$ \\
J1201+2117 &  12:01:10.31 & +21:17:58.6  & 4.58 & 3.7975 &  $21.35 \pm 0.15$ & $-0.75 \pm 0.15$ \\
... & ... & ... & ... & 4.1578 &  $ 20.60\pm 0.15$ & $ -2.40\pm 0.20$ \\
J1202+3235 & 12:02:07.78 & +32:35:38.8 & 5.29 & 4.7955 &  $ 21.10\pm 0.15$ & $ -2.35\pm 0.25$ \\
... & ... & ... & ... & 5.0647 & $ 20.30\pm 0.15$ & $ -2.65\pm 0.20$ \\
J1410+5111 &  14:10:30.61 & +51:11:13.6 & 3.21 & 2.9344 & $20.80 \pm 0.15$ & $-0.90 \pm 0.15$ \\
... & ... & ... & ... & 2.9642 &  $ 20.85 \pm 0.20 $ & $ -1.95 \pm 0.15$ \\
PSS1535+2943 & 15:35:53.80 & +29:43:13.7 & 3.99 & 3.2020 &  $20.65 \pm 0.15$ & $-1.05 \pm 0.25$ \\
... & ... & ... & ... & 3.7612 &  $ 20.40\pm 0.15$ & $ -1.95\pm 0.20$ \\
J1626+2751 & 16:26:26.50 & +27:51:32.5  & 5.18 & 4.3110 & $ 21.35\pm 0.15$ & $-1.35 \pm 0.20$ \\
... & ... & ... & ... & 4.4975 &  $ 21.40\pm 0.15$ & $ -2.45\pm 0.20$ \\
PSS1802+5616 &  18:02:48.84 & +56:16:50.5  & 4.18 & 3.3912 &  $ 20.30\pm 0.10$ & $ -1.35\pm 0.15$ \\
... & ... & ... & ... & 3.7617 &  $ 20.55 \pm 0.15$ & $ -1.45 \pm 0.25$ \\
... & ... & ... & ... & 3.8109 &  $ 20.35 \pm 0.20$ & $ -2.00 \pm 0.25$ \\
& & & & & & \\
\hline
\hline
\end{tabular}
\vspace{0.3cm}
\end{table*}

To date, 15 of the 17 QSO sightlines in our survey have not been searched for \lya\ emission at the redshifts of the intervening DLAs. The two remaning sightlines, toward PSS0209+05 and J0255+0048, were part of the MAGG survey and its precursors \citep{fumagalli2017,mackenzie2019,lofthouse2023}. For PSS0209+05, a single \lya\ emitter was detected at $z\approx 3.67$ and no \lya\ emitters were detected at $z \approx 3.86$ \citep{lofthouse2023}. For J0255+0048, one \lya\ emitter was detected at $z\approx 3.26$ \citep{fumagalli2017,mackenzie2019}. For reference and comparison with KAGG, the complete MAGG DLA sample is shown in the two panels of Figure~\ref{fig_sample}. We note that we obtain slightly different estimates of \NHI\ and DLA metallicity for the DLA along J0255+0048: we measure $\log{\mbox{\NHI} [\rm cm^{-2}]} = 20.7\pm 0.15$ and [M/H]~$=-0.9 \pm 0.15$, whereas MAGG report $\log{\mbox{\NHI} [\rm cm^{-2}]} = 20.85\pm 0.1$ and [M/H]~$=-1.1 \pm 0.1$ \citep{lofthouse2023}.

\subsection{The Keck/KCWI observations}
\label{2.2}

The Keck/KCWI observations were carried out during 6 nights between 2023 December and 2025 May. KCWI was deployed in its large slicer configuration, which yields a field of view of $33\arcsec \times 20\farcs 4$ and a slice width of $1\farcs 35$. At the average redshift of our sample, this configuration corresponds to a field of view of $\approx 230\times 140$~kpc, with a spaxel size of $\approx 10$~kpc.

We chose different instrument configurations for the blue and red arms. For the blue arm, we used the BM grating, which yields a spectral resolution of $R\sim 2000$ and a wavelength range of $\Delta \lambda \sim 800$~\AA\ within the operable wavelength range of the grating ($3500-5500$~\AA). For the red arm, we opted for the RM1 grating, which yields a spectral resolution of $R\sim 1400$ and a wavelength range of $\Delta \lambda \sim 1300$~\AA\ within the operable wavelength range of the grating ($5500-8600$~\AA).

Our observational strategy was different for each spectral arm. For the blue arm, we aimed for 10--15~min exposures, which helped minimize readout overheads in the blue detector. For the red arm, we aimed for 4-6~min exposures to reduce cosmic ray contamination. We dithered between each set of exposures by $2^{\prime\prime}$ following a 5-point pattern. For each sightline, the total exposure time was in the range $\approx 2-5$~hr, with longer exposure times for sightlines featuring DLAs at $z\gtrsim 4$. The observational strategy resulted in $\approx 90\%$ completeness with a line signal-to-noise ratio~$\approx 4$ for line fluxes of~$4 \times 10^{-18}$~erg~s$^{-1}$~cm$^{-2}$ per $1\farcs 35$ spaxel (see Section~\ref{3.3}). 

The KAGG survey data were reduced with \texttt{PypeIt}\footnote{\href{https://zenodo.org/records/19208465}{https://zenodo.org/records/19208465}}, a Python package designed for the reduction of spectroscopic data (\citealt{pypeit}). \texttt{PypeIt} was run in its standard form to perform dark correction, bias subtraction, and trace pattern identification. For wavelength calibration, we used observations of FeAr lamps that often required manual line identification with the \texttt{pypeit\_identify} routine. Our approach to flat-fielding depended on the spectral arm. For the blue arm data, we corrected for pixel-by-pixel variations using internal flats and for variations in the spectral illumination pattern with twilight flats. For the red arm data, we only corrected for pixel-by-pixel variations using internal flats. The absence of bright objects in our target fields allowed us to use the science exposures for sky subtraction. One-dimensional spectra for standard stars were extracted with \texttt{PypeIt} from individual, reduced exposures. These one-dimensional standard star spectra were then used to generate a flux sensitivity function for flux calibration. 

We generated independent flux-calibrated data cubes for each individual exposure in the blue and red arms. All cubes were constructed in a previously defined (ra, dec, wavelength) grid, which was necessary to account for dithering between exposures. These individual cubes were then combined to produce the final blue and red arm cubes for every target. For this final step, we implemented 3$\sigma$ clipping of the cubes in the wavelength dimension. We then revised the astrometric solutions of the data cubes using the QSO coordinates.

Finally, we estimated the accuracy of our flux calibration by comparing the spatially-integrated QSO fluxes from our KCWI cubes with the QSO fluxes from the SDSS. Of the 17 QSOs of our sample, 15  have a flux-calibrated SDSS spectrum; the exceptions are PSS0007+24 and PSS1802+5616. We quantified the differences in the absolute fluxing between the two KCWI and SDSS datasets by computing the metric
\begin{gather}
f_{\rm SDSS/KCWI} = \rm median (f_\lambda^{SDSS} / f_\lambda^{KCWI}) 
\end{gather}
for every spectrum, where $f_\lambda^{\rm SDSS}$ and $f_\lambda^{\rm KCWI}$ are the flux-calibrated QSO spectra from SDSS and KCWI, respectively. To maximize the reliability of the estimate, the median was computed only with pixels in the top two S/N quartiles. 

We find no wavelength dependence in the value of this metric. The values fluctuate in the range $f_{\rm SDSS/KCWI} \approx 1- 2.5$, depending on the QSO. Because some of these differences could arise from intrinsic QSO variability, we refrain from scaling every cube by the measured $f_{\rm SDSS/KCWI}$ value. Instead, we used the median across the whole sample ($f_{\rm SDSS/KCWI} = 1.9$) to rescale the flux densities for all cubes. The standard deviation ($\sigma_{f_{\rm SDSS/KCWI}} = 0.4$) was added in quadrature to the measurement error in the flux densities for all cubes. All errors on the line flux that are reported in this paper take into account this uncertainty on the flux density scaling.

\section{Identifying \lya\ emitters} 
\label{3}
In this section, we describe the initial processing of the data cubes to remove contaminating features (primarily due to the QSO itself), the line search and identification algorithms, simulations to estimate the  line signal-to-noise and line flux upper limits, and the identification of line interlopers from objects at other redshifts.

\subsection{QSO and background subtraction} 
\label{3.1}

The steps described in this subsection aim to prepare the data cubes for the execution of the line search algorithm. With this in mind, we start by processing the spaxels featuring sources of contamination, primarily the QSO. Although the QSO emission is entirely blocked at the DLA wavelengths, removal of the QSO is necessary to search for lines at the edges of the DLA absorption trough and to accurately quantify the significance of any emission features by understanding the noise properties across the whole spectrum. The following steps are distinct depending on whether the spectral pixel falls near the DLA's central wavelength (hereafter, the DLA spectral region) or far from the DLA's central wavelength (hereafter, the field spectral region). We distinguish between the two regions by adopting a conservative cutoff in velocity of $|\Delta v |= |v_{\rm pixel} - v_{\rm \small DLA}| < 16,000$~km~s$^{-1}$ for the DLA spectral region.

The DLA spectral region represents the wavelength range of interest in our search for \lya\ emission around the DLA redshift. We start by co-adding the spectra from all the spaxels featuring significant QSO emission. This co-addition yields a high signal-to-noise, co-added QSO spectrum that we fit with a Chebyshev polynomial in the wavelength dimension \citep[see, e.g., ][]{oyarzun2024}. The resulting QSO continuum model is then scaled and subtracted from all QSO continuum spaxels. Then, we proceed to subtract the emission from the sky background. To this end, we obtain a  median spectrum for each cube, fit this with a Chebyshev polynomial in the wavelength dimension, and then subtract this function from all spaxels in the DLA spectral region.

The field spectral region represents the extended wavelength range away from the DLA redshift, and it is necessary to quantify the noise properties of every spaxel. We initially subtract the QSO continuum and the sky background here, following a procedure similar to that used for the DLA spectral region. Then, we perform two additional steps: continuum subtraction and emission line masking.

After completion of these steps, the data cubes are background- and QSO continuum-subtracted within the DLA spectral regions, and the data cubes are background-, QSO continuum-, source continuum-, and source emission line-subtracted within the field spectral regions. We reiterate that processing of the field spectral regions is necessary for characterizing the spectral noise, which is critical for quantifying the line flux signal-to-noise ratio and the line detection completeness (see Section~\ref{3.3}).

\subsection{Emission line identification}
\label{3.2}

We describe here the line search algorithm used to identify emission features in each data cube. The purpose of this algorithm is to identify emission lines in large volumes of data, which led us to adopt a Levenberg-Marquardt least-squares (i.e. $\chi^2$) minimization approach. We fitted the spectrum of every spaxel in the DLA spectral region with a Gaussian emission line profile with four free parameters: the line flux at peak ($f_0$), the line centroid ($\lambda_0$), the line width ($\sigma_0$), and the continuum flux ($f_1$). Formally, the observed spectrum $f_\lambda$ was fitted with the function
\begin{equation}
\label{line_equation}
f_\lambda (\lambda) = f_0 e^{{(\lambda - \lambda_0)^2}/{2\sigma_0^2}} + f_1 \,.
\end{equation}

We adopted the following constraints on the allowed values of the parameters of the fit: (i)~the absolute value of the line flux at peak could not exceed the maximum absolute value of the observed $f_\lambda$ by a factor larger than four, (ii)~the line width was bounded between 0.75~\AA\ and 3~\AA\ (i.e., between the approximate KCWI resolution for the chosen configurations and a maximum FWHM line width of $500$~km~s$^{-1}$), and (iii)~the line centroid had to be within the observed spectrum. No explicit constraints were set for the value of the continuum flux.

We ran our line fitting algorithm $N$ times per spectrum, where $N$ is the total number of spectral pixels. For iteration number $i$, the algorithm was executed with specific initial conditions: 
\begin{equation}
f_0^i = f_{\lambda_i},~ 
\lambda_0^i = \lambda_i,~ 
\sigma_0^i = 2\,\mbox{\AA},~
\mbox{and } f_1^i = 0,
\end{equation}
i.e., all iterations have initial conditions that match an observed pixel ($\lambda_i, f_{\lambda_i}$). This ensures that all reasonable emission lines throughout the spectrum (i.e., all local minima in $\chi^2$) are searched for. Then, the global minima for the line centroid and the line width of every spaxel are set by the solution with the largest value for $f_0$ that satisfies reduced $\chi^2 < 1.5$ (i.e., the brightest solution that fits the data reasonably well). We refer to the associated solution for the line centroid and the line width for any spaxel as $\lambda_0 = \lambda_0^*$ and $\sigma_0 = \sigma_0^*$, respectively.

Having found the line centroid and width, we estimate the total flux of the identified feature. To this end, we simply integrate the region of the spectrum that encloses 2$\sigma$ of the total line flux and then apply a correction factor of $(2\sigma)^{-1}\approx1.05$. In equation form, the single spaxel line flux $F_0^*$ is computed as
\begin{gather}
\label{eq_flux}
F_0^* = \frac{1}{2\sigma}\sum_{\lambda_0^* - 2\sigma_0^*}^{\lambda_0^* + 2\sigma_0^*}f_{\lambda} \Delta \lambda.
\end{gather}

As a result of this process, every spaxel has an associated value for $F_0^*$, $\lambda_0^*$, and $\sigma_0^*$. We reiterate that these steps only serve the purpose of identifying the brightest emission feature, with the final estimation of the line flux being performed after we formally determine the signal-to-noise ratio of each feature (Section~\ref{3.3}).

\subsection{\lya\ signal-to-noise ratio and flux}
\label{3.3}

Next, we estimate the signal-to-noise of the $F_0^*$ measured in every spaxel. For each data cube, we created $N = 200$ simulated cubes that match the noise properties of the original cube. This was done using the variance in the field spectral regions (i.e., at wavelengths away from the DLA redshift). We then inserted the model QSO continuum into each simulated cube and subjected each simulated cube to all the processing steps outlined in Sections~\ref{3.1} and \ref{3.2}. The simulated cubes were then used to obtain a random line flux distribution for every spaxel, yielding a spaxel-dependent signal-to-noise ratio ($S/N$). The process was repeated with $N = 600$ simulated cubes for every spaxel that contains a bright line in the original cube, allowing for accurate accounting of the random line flux distribution and signal-to-noise ratio.

We found the signal-to-noise ratio to be a good metric for identifying emission lines, with the threshold $S/N \geq 4$ sufficient for distinguishing statistically-significant features in the cubes. However, a simple $S/N$ criterion can wrongly identify non-Gaussian artifacts in the cubes as significant emission features. We hence employed two additional signal-to-noise definitions: based on the variance in the cube at the peak wavelength of the emission feature ($S/N_{\rm cube}$) and based on the error on $F_0^*$ from error propagation in Equation~\ref{eq_flux} ($S/N_{\rm spec}$). In addition to the original $S/N$ threshold, we also imposed the following criteria: $S/N_{\rm cube}>3$ (to remove sky subtraction residuals) and $S/N_{\rm spec}>1.5$ (to remove spurious features arising from outliers in individual exposures). 

We then remeasured the fluxes of all significant detections. This is necessary because the fluxes measured with Equation~(\ref{eq_flux}) do not exclude continuum emission and because the original flux estimate does not account for the possibility of extended emission. Therefore, we visually inspected all the adjacent spaxels of all significant features, identified cases of extended emission, and co-added the spectra when necessary. After co-addition, the underlying continuum was fitted and subtracted out, and the emission line fluxes were measured following Equation~(\ref{eq_flux}). The line fluxes reported in this paper correspond to these corrected, spatially co-added values. 

\begin{figure}[t]
\centering 
\includegraphics[width=3.2in]{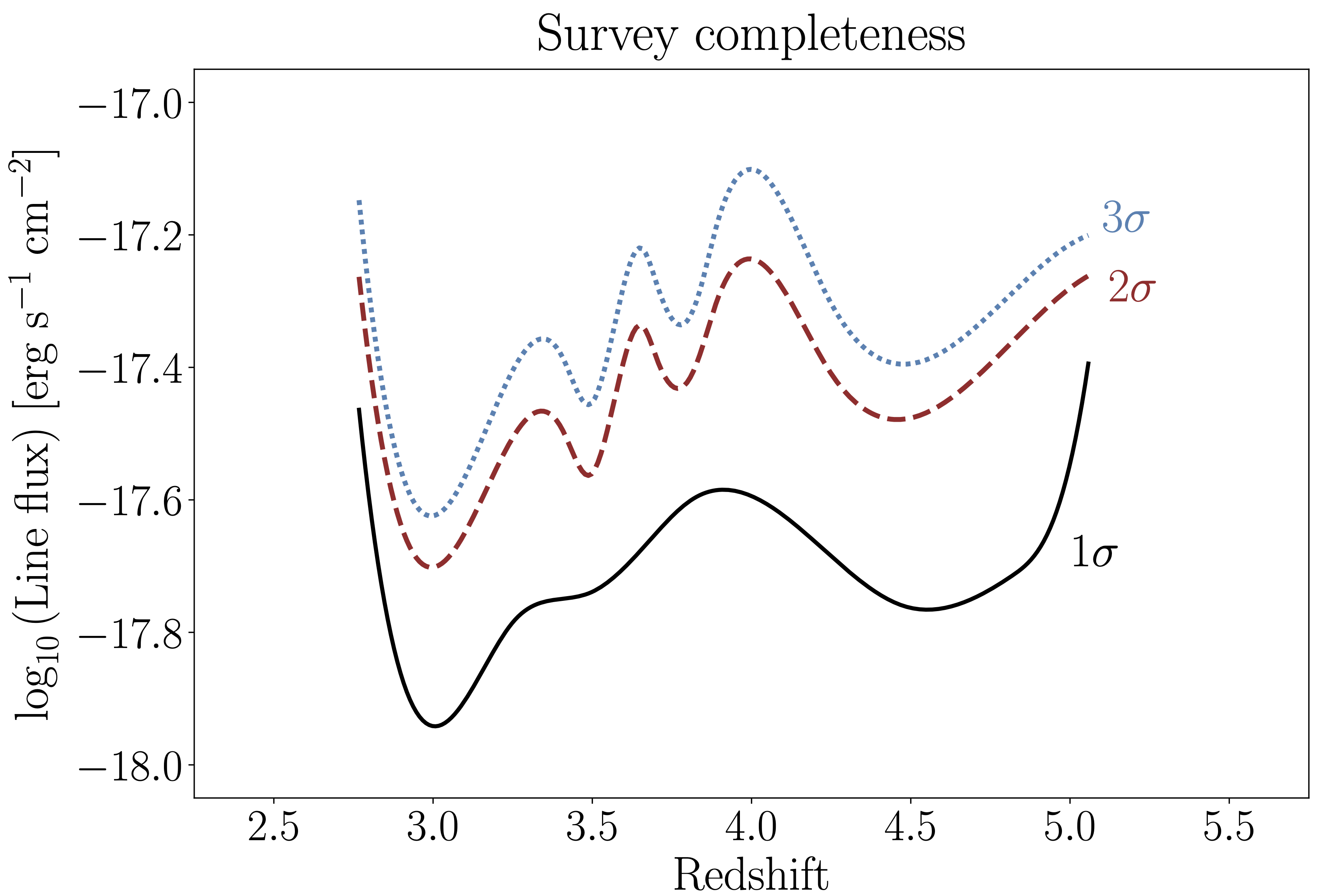}\\
\vspace{-0.5em}
\caption{Line detection completeness of KAGG as a function of redshift. Plotted are three completeness levels: $1\sigma$ (black line), $2\sigma$ (red dashed line), and $3\sigma$ (yellow dotted line). These curves were obtained by smoothing the completeness dependence on redshift for the 36 DLAs in the survey. Note that this figure does not account for sightline dependent Galactic extinction. Throughout the paper, non-detection upper limits for individual targets account for Galactic extinction and correspond to 2$\sigma$ completeness.}
\label{fig_completeness}
\end{figure}

Lastly, we estimated the line flux sensitivity limit of each cube. To this end, we inserted 5,000 simulated emission lines in the DLA spectral region at random spaxels and with varying line fluxes at peak, line centroids, and line widths. We then measured all three signal-to-noise metrics for each inserted line and determined the detection completeness as a function of line flux given our signal-to-noise ratio thresholds. This completeness function is shown in Figure~\ref{fig_completeness}. The upper limits to the line flux reported in the paper correspond to detection rates of $2\sigma\approx 0.95$, i.e., these line flux upper limits indicate the values at which we recover lines with sufficient significance 95\% of the time. 

On average, the KAGG survey achieves slightly fainter line fluxes than the MAGG survey. In KAGG, we reach 90\% completeness at line fluxes of $\approx 4 \times 10^{-18}$~erg~s$^{-1}$~cm$^{-2}$, whereas data from the MAGG survey reaches the same completeness at fluxes of $\approx 6 \times 10^{-18}$~erg~s$^{-1}$~cm$^{-2}$ (\citealt{lofthouse2023}). Because these completeness estimates depend on spectral sensitivity, methodology, and line significance thresholds, this difference in completeness may not necessarily imply that the Keck/KCWI spectra are more sensitive than the VLT/MUSE spectra. Regarding flux calibration, the MAGG survey is more precise than the KAGG survey (see Section~\ref{2.2}).

All line flux measurements and upper limits were corrected for extinction in the Milky Way using the \citet{fitzpatrick1999} extinction curve, \citet{schlegel1998} dust-reddeding maps, and the \citet{schlafly-finkbeiner2011} reddening to $A_V$ conversion. This correction was only applied to the inferred line flux values; all spectra shown in the manuscript have not been corrected for extinction or dust-reddening. The velocity offsets between the \lya\ emission lines and the absorption redshifts that we report in this paper were measured from the best fit \lya\ line centroids (see Equation \ref{line_equation}). 

\subsection{Interloper contamination}
\label{3.4}

Finally, we identified and excluded from the analysis emission lines that do not correspond to \lya\ emission at the redshift of the DLA, but rather to optical emission lines from galaxies at lower redshifts. Possible line contaminants include the [O{\sc ii}] doublet, H$\gamma$, H$\beta$, [O{\sc iii}], N{\sc i}, and He{\sc i}. With the exception of [O{\sc ii}]$\lambda$3727\AA\ emitters, all other lines are easily identifiable by the presence of other emission lines across the rest-frame wavelength range $\lambda = 4000-6000$~\AA. We find that our detections include one H$\gamma$ emitter at $z\sim 0.6$, one H$\beta$ emitter at $z\sim 0.2$, two [O{\sc iii}] emitters at $z\sim 0.3-0.5$, and one N{\sc i} emitter at $z\sim 0.2$. The low number of H$\gamma$, H$\beta$, [O{\sc iii}], N{\sc i}, and He{\sc i} interlopers suggests that the number of [O{\sc ii}]$\lambda$3727\AA\ interlopers is also likely to be much lower than the number of higher-redshift \lya\ emitters. In any case, we remove four significant detections that are likely to be [O{\sc ii}]$\lambda$3727\AA\ emitters at $z\sim 0.4-0.9$ due to their bright continua, their large angular sizes, and their double-peaked emission consistent with the [O{\sc ii}]$\lambda$3727\AA\ doublet profile.

\section{KAGG \lya\ emitters} 
\label{4}

In this section, we describe our \lya\ emission detections. We convert their emission redshifts to velocities relative to the DLA redshift, which we use to define different \lya\ emitter samples: DLA \lya\ emitters, Group \lya\ emitters, and Large scale \lya\ emitters. Then, we discuss our observations of DLA fields that have also been observed by other surveys and observatories.

\subsection{\lya\ emission velocities relative to the DLA redshift}
\label{4.1}

\begin{figure*}[h!]
\vspace{0em}
\centering 
\includegraphics[width=7.1in]{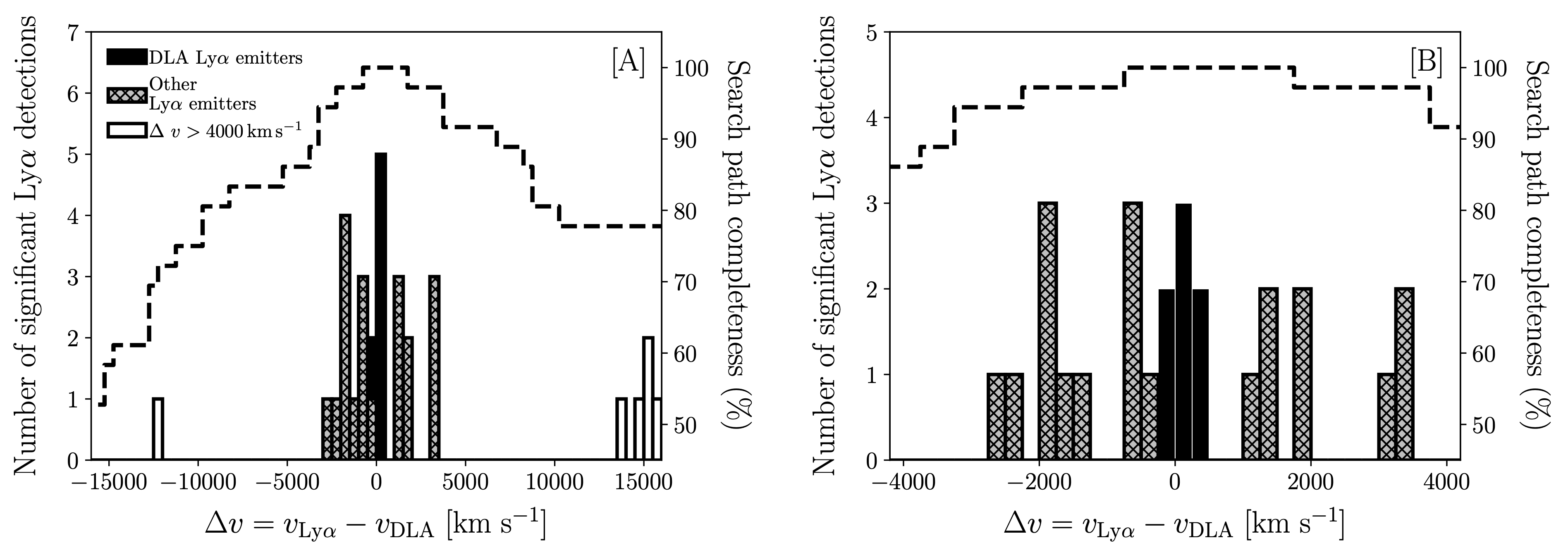}\\
\caption{Histogram of significant \lya\ emitters in velocity relative to the DLA redshift. The left panel shows the complete distribution, whereas the right panel is zoomed in and the data has been more finely binned to show only \lya\ emitters within $\Delta v < 4,000$~km~s$^{-1}$ of the DLA. The search path of KAGG as a function of $\Delta v$ is plotted as the black dashed line. Because the search path is highly complete ($\gtrsim90\%$) within $|\Delta v| \lesssim 4000$~km~s$^{-1}$, we adopt this value as our sample cutoff. For the rest of the paper, we define DLA \lya\ emitters as the 7 galaxies with $-400 <\Delta v~[\rm km~s^{-1}]< 800$ (black histogram). DLA \lya\ emitters plausibly correspond to the primary emission counterparts of DLAs.}
\label{fig_velocity}
\end{figure*}

We define the search path completeness of the KAGG survey as the fraction of the DLA sample for which our data are sensitive to \lya\ emission, which is dependent on the wavelength coverage of the observations and on the redshift of the DLAs. This search path completeness is shown as a function of the velocity offset relative to the DLA redshift (hereafter, $\Delta v$) as dashed lines in Figure~\ref{fig_velocity}. As apparent in panel~[B] of this figure, the completeness is highest at low $\Delta v$, remaining $> 90$\% within $\Delta v \lesssim 4000$~km~s$^{-1}$. Panel~[A] shows that the completeness decreases toward higher $\Delta v$, but it still remains $\gtrsim 50$\% within $\Delta v \lesssim 15,000$~km~s$^{-1}$.

\begin{figure*}[h!]
\vspace{0em}
\centering 
\includegraphics[width=7.1in]{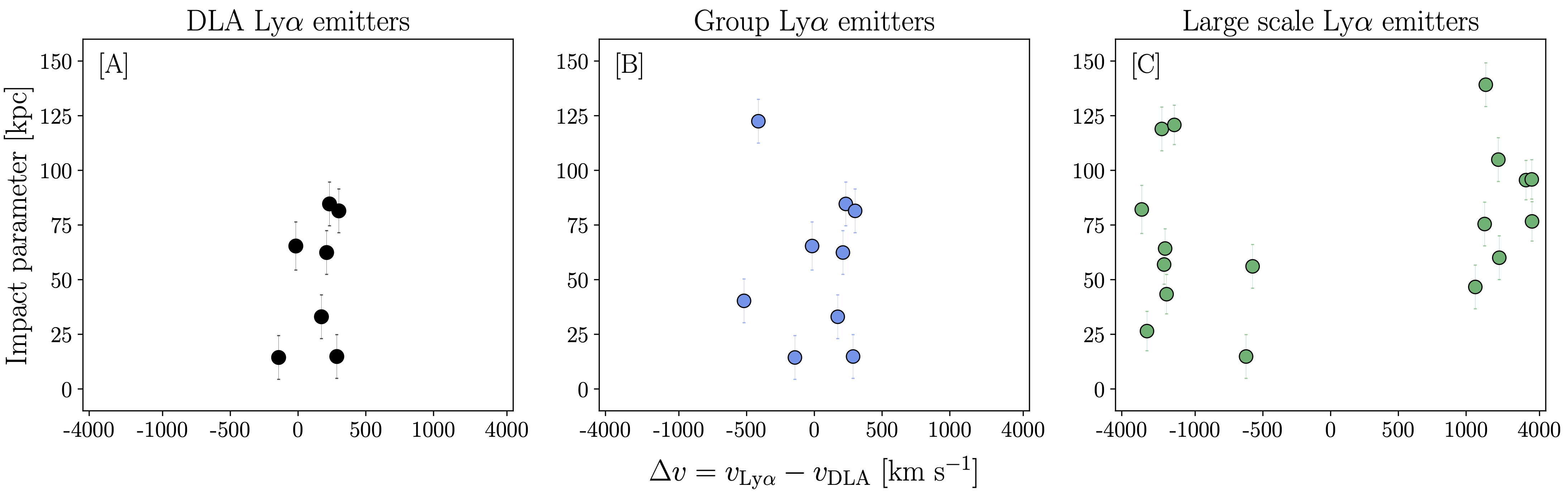}\\
\caption{The three subsamples into which we divide our \lya\ emitter detections. Each panel shows the distribution of a different subsample in velocity offset relative to the DLA redshift and impact parameter. Panel~[A]: DLA \lya\ emitters are defined to be within $-400 < \Delta v < 800$~km~s$^{-1}$ and plausibly correspond to the DLA galaxy. This subsample includes 7 detections. Panel~[B]: Group \lya\ emitters are selected according to the velocity offset criterion $-550 < \Delta v < 950$~km~s$^{-1}$. This selection is designed to include any galaxies that may belong in the same galaxy group as the DLA, and therefore includes all DLA \lya\ emitters. The Group \lya\ emitter subsample includes 9 detections. Panel~[C]: Large scale \lya\ emitters are selected to be at larger $\Delta v$ than Group \lya\ emitters and within $\Delta v < 4000$~km~s$^{-1}$. This velocity criterion is equivalent to a proper distance selection of $0.3\lesssim d~[\rm Mpc]\lesssim 10$, and thus Large scale \lya\ emitters probe galaxies in the same large-scale structures as the DLA. This subsample is composed of 17 detections.}
\label{fig_subsamples}
\end{figure*}

Our line search and identification procedure described in the previous section yielded the detection of 32 significant \lya\ emitters out to $\Delta v \approx 15,000$~km~s$^{-1}$. Their $\Delta v$ distribution is shown in Figure~\ref{fig_velocity}[A]. It is clear that there are no \lya\ emission lines identified at velocity offsets of $\approx 4,000-12,000$~km~s$^{-1}$. The 6 \lya\ emitters identified at larger velocity offsets ($\Delta v \approx 12,000-15,000$~km~s$^{-1}$) are likely to be field galaxies unassociated with the DLAs. We exclude them from our final \lya\ emission sample, but we list their properties in Appendix~\ref{appA}. We divide the remaining sample of 26 \lya\ emitters into three not mutually exclusive categories: DLA \lya\ emitters, Group \lya\ emitters, and Large-scale \lya\ emitters. The criteria to define each subsample is presented in the following subsections (Sections \ref{4.2}, \ref{4.3}, and \ref{4.4}). The distribution of each subsample in velocity offset and impact parameter is presented in Figure \ref{fig_subsamples}.

\subsection{DLA \lya\ emitters}
\label{4.2}

The velocity offset between the galaxy and the absorber is critical for the identification of the emission counterparts of DLAs. This is particularly relevant for searches in \lya\ emission because the redshifts of the galaxy and of the \lya\ emission line may differ due to \lya\ resonant scattering. Specifically, the \lya\ emission redshift is typically redward of the systemic redshift by a few hundred km~s$^{-1}$ \citep[e.g.][]{shapley2003,kulas2012,verhamme2018,cassata2020}. For example, analysis of 53 galaxies from the ALPINE survey of main-sequence galaxies at $z \approx 4.4-6$ found the distribution of velocity offsets between \lya\ and \cii\ emission to have a median of $\approx 200$~km~s$^{-1}$ and a spread of $\approx \pm 500$~km~s$^{-1}$ \citep{cassata2020}. \citet{verhamme2018} found similar velocity offsets ($\lesssim 800$~km~s$^{-1}$) between \lya\ and C{\sc iii}]$\lambda\lambda$1907,1909\AA\ emission for a sample of 13 galaxies at $z \approx 2.9-3.8$. Based on these two studies, we will assume that the median velocity offset between the \lya\ emission redshift and the galaxy systemic redshift is $\approx 200$~km~s$^{-1}$ with a spread of $\pm 600$~km~s$^{-1}$.

\begin{figure*}[ht!]
\centering 
\includegraphics[width=6.8in]{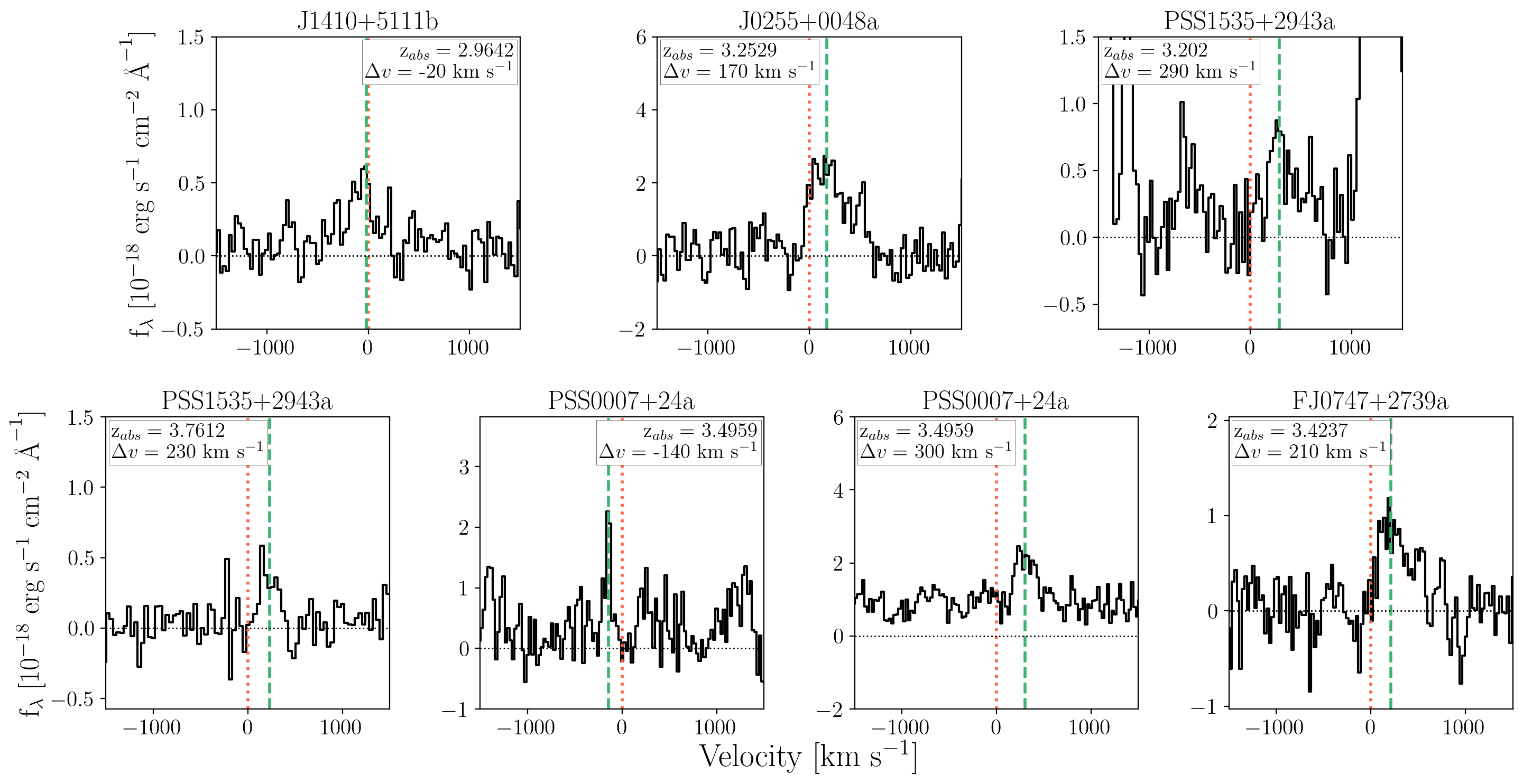}\\
\vspace{-0.9em}
\caption{Spatially integrated spectra of the 7 DLA \lya\ emitters detected in the KAGG survey. The dashed vertical lines in each panel indicate $\Delta v = 0$ (i.e., the \lya\ wavelength at the DLA redshift; red) and the emission line centroid (green). The QSO sightline identifiers and the DLA redshifts are indicated at the top of each subplot. We note that we identify two DLA \lya\ emitters associated with the DLA at $z\approx 3.5$ along PSS0007+24. We also note that the spikes at $|\Delta v| \gtrsim 1000$~km~s$^{-1}$ in the spectrum of PSS1535+2943a correspond to QSO emission near this small impact parameter \lya\ emitter. Also apparent in the PSS1535+2943a spectrum is a second \lya\ emitter at $\Delta v \approx - 620$~km~$s^{-1}$ that belongs in our Large-scale \lya\ emitter subsample due to its large negative velocity offset (Section \ref{4.4} and Appendix \ref{appA}).}
\vspace{-0.7em}
\label{fig_lya_lines}
\end{figure*}

\begin{figure*}[hb!]
\vspace{-0.75em}
\centering 
\includegraphics[width=7in]{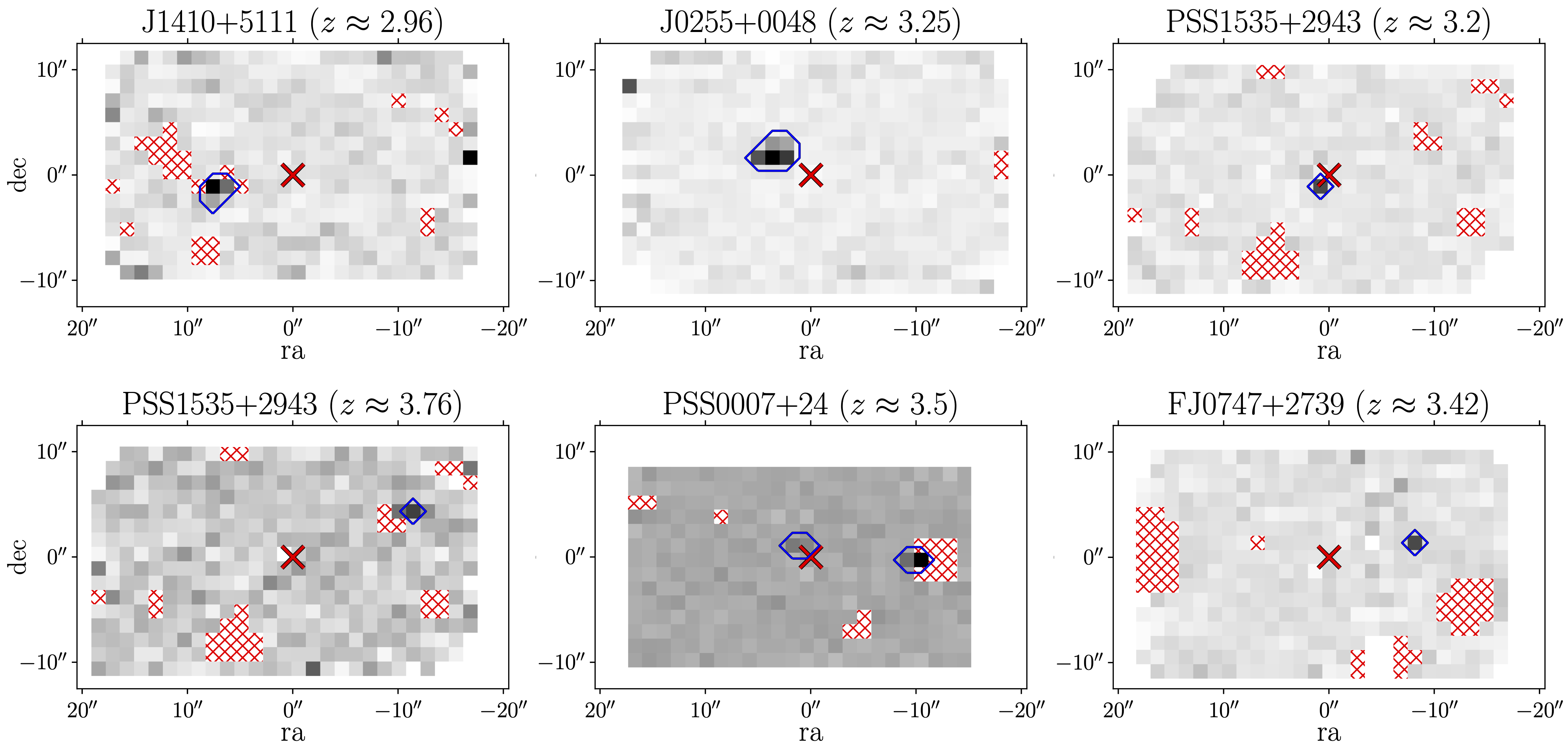}
\vspace{-0.75em}
\caption{KCWI images of the 6 DLAs featuring 7 DLA \lya\ emitters. These images were constructed by integrating the KCWI datacubes within $\pm100$~km~s$^{-1}$ of the DLA redshifts, except for the outlined, colored spaxels that feature significant line emission: the colormap at these spaxels corresponds to the KCWI datacube centered at the wavelength of the emission line and integrated within $\pm 3\sigma$ of the measured line width. The red crosses shows the location of the QSO and the hatched regions show the location of continuum and/or line emission in the foreground. We note that the search for \lya\ emission was conducted in the whole cube, and thus spaxels with foreground emission are masked only for visualization.}
\label{fig_lya_images}
\end{figure*}

While the DLA redshift could be offset from the galaxy systemic redshift, especially if the absorption arises in the CGM \citep[e.g.][]{neeleman2025}, the redshifts of the DLA and its galaxy counterpart should, on average, match. Indeed, the $\Delta v$ distribution of the 26 significant \lya\ emitters within $\Delta v < 4000$~km~s$^{-1}$ of the DLA redshift (Figure~\ref{fig_velocity}~[B]) has a median of $\approx 250$~km~s$^{-1}$, consistent with our choice of $200$~km~s$^{-1}$ as the median offset between the \lya\ and galaxy redshifts. Hence, we will use the criterion $-400 < \Delta v < 800$~km~s$^{-1}$ to identify the galaxy counterparts of our target DLAs in \lya\ emission. Seven \lya\ emitters satisfy this criterion, and we will henceforth refer to them as DLA~\lya\ emitters. Their velocity offsets and impact parameters are show in panel~[A] of Figure \ref{fig_subsamples}. Their properties are summarized in Table~\ref{table_DLA_lya_emitters} and their spectra are shown in Figure~\ref{fig_lya_lines}. We constructed narrowband images of \lya\ emission for these 7 DLA \lya\ emitters, and we show them in Figure~\ref{fig_lya_images}.

Implementation of the same criteria to select DLA \lya\ emitters from the MAGG survey results in the inclusion of \lya\ emitters at very large impact parameters ($b \approx 100-300$~kpc). Such large impact parameters exceed the largest virial radii ($r_{vir}$) at these redshifts by factors as large as $\approx 3$, and hence these \lya\ emitters are very unlikely to correspond to the DLA galaxy. Therefore, it is clear that an additional selection criterion involving impact parameter is required to select DLA \lya\ emitters from MAGG. Earlier studies have found that the covering fraction of DLAs decreases with galactocentric distance, from $f_{\rm cov} \approx 1$ at $r=0$ to $f_{\rm cov} \lesssim 0.15$ at $r= r_{vir}$ \citep{rahmati2015,stern2021}. Thus, $b \leq r_{vir}$ is a conservative limit on the maximum physical separation between galaxies and high \NHI\ gas. The most massive halos at $z\approx 4$ have halo masses of $M_h\approx 10^{13}$~M$_{\odot}$, corresponding to $r_{vir}\approx 100$~kpc. Thus, we can use the criteria $-400 < \Delta v < 800$~km~s$^{-1}$ and $b \leq 100$~kpc to more generally select DLA \lya\ emitter candidates. These criteria result in 4 MAGG DLA \lya\ emitters, and their properties are listed in Table~\ref{table_DLA_lya_emitters}.

\startlongtable 
\begin{deluxetable*}{cccccccc} 
\tabletypesize{\normalsize} 
\vspace{-0.5em}
\tablecaption{Properties of the 11 DLA \lya\ emitters included in our study. The first seven are from the KAGG survey, and the last four are from the MAGG survey. The columns of the table are (1)~detection identifier, (2)~DLA redshift, (3, 4)~the J2000 coordinates of the \lya\ emission, (5)~the velocity offset $\Delta v$ between the \lya\ redshift and the DLA redshift in km~s$^{-1}$, (6)~the \lya\ flux in $10^{-18}$~erg~s$^{-1}$~cm$^{-2}$, (7)~the impact parameter to the QSO sightline at the DLA redshift in kpc, and (8)~the sample (KAGG or MAGG).
\vspace{-0.5em}
} 
\label{table_DLA_lya_emitters} 
\setlength{\tabcolsep}{0.09in} 
\tablehead{
\colhead{Identifier} & \colhead{z$_{abs}$} & \colhead{ra}  & \colhead{dec} & \colhead{vel. offset} & \colhead{\lya\ flux} & \colhead{b} & \colhead{sample of} \vspace{-0.3em} \\
\colhead{} & \colhead{} & \colhead{[hours]}  & \colhead{[degrees]} & \colhead{[km~s$^{-1}$]} & \colhead{[$10^{-18}$~erg~s$^{-1}$~cm$^{-2}$]} & \colhead{[kpc]} & \colhead{origin}} 
\startdata 
PSS0007+24a & 3.4959 & 00:07:38.77 & 24:17:25.80 & $-140$ & $4.0 \pm 2$ & $14 \pm 10$ & KAGG  \\
PSS0007+24b & 3.4959 & 00:07:37.85 & 24:17:24.30 & $300$ & $7.9 \pm 4$ & $81 \pm 10$ & KAGG  \\
J0255+0048a & 3.2529 & 02:55:18.81 & 00:48:49.00 & $170$ & $19.6 \pm 9$ & $33 \pm 10$ & KAGG  \\
FJ0747+2739a & 3.4237 & 07:47:10.55 & 27:39:04.80 & $210$ & $10.9 \pm 6$ & $62 \pm 10$ & KAGG \\
J1410+5111a & 2.9642 & 14:10:31.41 & 51:11:12.60 & $-20$ & $1.7 \pm 1$ & $65 \pm 11$ & KAGG \\
PSS1535+2943a & 3.202 & 15:35:53.88 & 29:43:12.70 & $290$ & $3.0 \pm 2$ & $15 \pm 10$ & KAGG \\
PSS1535+2943b & 3.7612 & 15:35:52.92 & 29:43:18.20 & $230$ & $2.0 \pm 1$ & $85 \pm 10$ & KAGG \\
\hline
J0851+2332a  & $3.5297$ & 08:51:43.60 & 23:32:12.00 & $262$ & $10.0 \pm 1$ & $24$  & MAGG \\
J1111-0804a & $3.6077$ & 11:11:13.75 & -08:04:11.05 & $57$ & $8.5 \pm 1$ & $67$  & MAGG  \\
J2315+1456a & $3.2732$ & 23:15:44.39 & 14:56:01.40 & $244$ & $3.9 \pm 1$ & $97$  & MAGG  \\
J2334-0908a & $3.0572$ & 23:34:45.64 & -09:08:07.48 & $212$ & $15.1 \pm 2$ & $98$  & MAGG  \\
\hline
\hline
\enddata
\tablecomments{The quoted errors on the KAGG line fluxes include uncertainties in the flux calibration. Hence, these errors are significantly larger than the line detection uncertainty.}
\end{deluxetable*}

\newpage

\subsection{Group \lya\ emitters}
\label{4.3}

It is plausible that some of the DLA \lya\ emitters actually correspond to neighboring galaxies of the DLA galaxy. For instance, \citet{oyarzun2024} showed two examples of \lya\ emitters within $\approx 500$~km~s$^{-1}$ and $b<100~$kpc of $z\approx 2$ DLAs that do not correspond to the DLA galaxy, which was identified in CO emission and undetected in \lya\ emission. Hence, one can adopt the more agnostic approach of treating DLA \lya\ emitters as Group \lya\ emitters. Since we are no longer interested in primary associations to select this sample, we can relax our original relative velocity and impact parameter criteria. Because the search paths of KAGG and MAGG are larger in relative velocity than in impact parameter, it is more sensible to let the impact parameter coverage dictate the sample selection cuts. MAGG has the largest impact parameter of the two surveys, with a field of view that extends as far as $b\lesssim 300$~kpc. According to the Hubble flow, a line of sight distance of $b\lesssim 300$~kpc is equivalent to relative velocities of $\Delta v \lesssim 150$~km~s$^{-1}$ at $z\approx 4$. Thus, we add this $\Delta v$ value to our original DLA \lya\ emitter $\Delta v$ selection. This results in the criteria $b\lesssim 300$~kpc and $-550 < \Delta v < 950$~km~s$^{-1}$ to select Group \lya\ emitters. 

Implementation of these criteria in KAGG yields 9 Group \lya\ emitters. We note that all 7 DLA \lya\ emitters are included in this category, and therefore the difference between the two subsamples is only two \lya\ emitters. The velocity offsets and impact parameters of this subsample are shown in Figure \ref{fig_subsamples}~[B]. The Group \lya\ emitter selection criteria are significantly more impactful for selecting \lya\ emitters in the MAGG survey due to their far larger field of view. Application of these criteria in MAGG yields 26 MAGG Group \lya\ emitters, a significant increase over the 4 MAGG DLA \lya\ emitters. The properties of the 35 Group \lya\ emitters are presented in Appendix \ref{appA}.

\subsection{Large-scale \lya\ emitters}
\label{4.4}

There are 17 additional \lya\ emitters in KAGG that are found at larger velocity offsets from the DLA redshifts than $-550 < \Delta v < 950$~km~s$^{-1}$. These objects probably arise from galaxies in the same large scale structures as the DLAs. We will refer to this sample as Large-scale \lya\ emitters and list their properties in Appendix \ref{appA}. We note that this sample only includes \lya\ emitters from the KAGG survey. As with the other two \lya\ emitter subsamples, we plot their velocity offsets and impact parameters in Figure~\ref{fig_subsamples}.

\subsection{Searches for \lya\ emission in previously characterized DLA fields}
\label{4.5}

 \begin{figure*}[t!]
\vspace{0em}
\centering 
\includegraphics[width=7.1in]{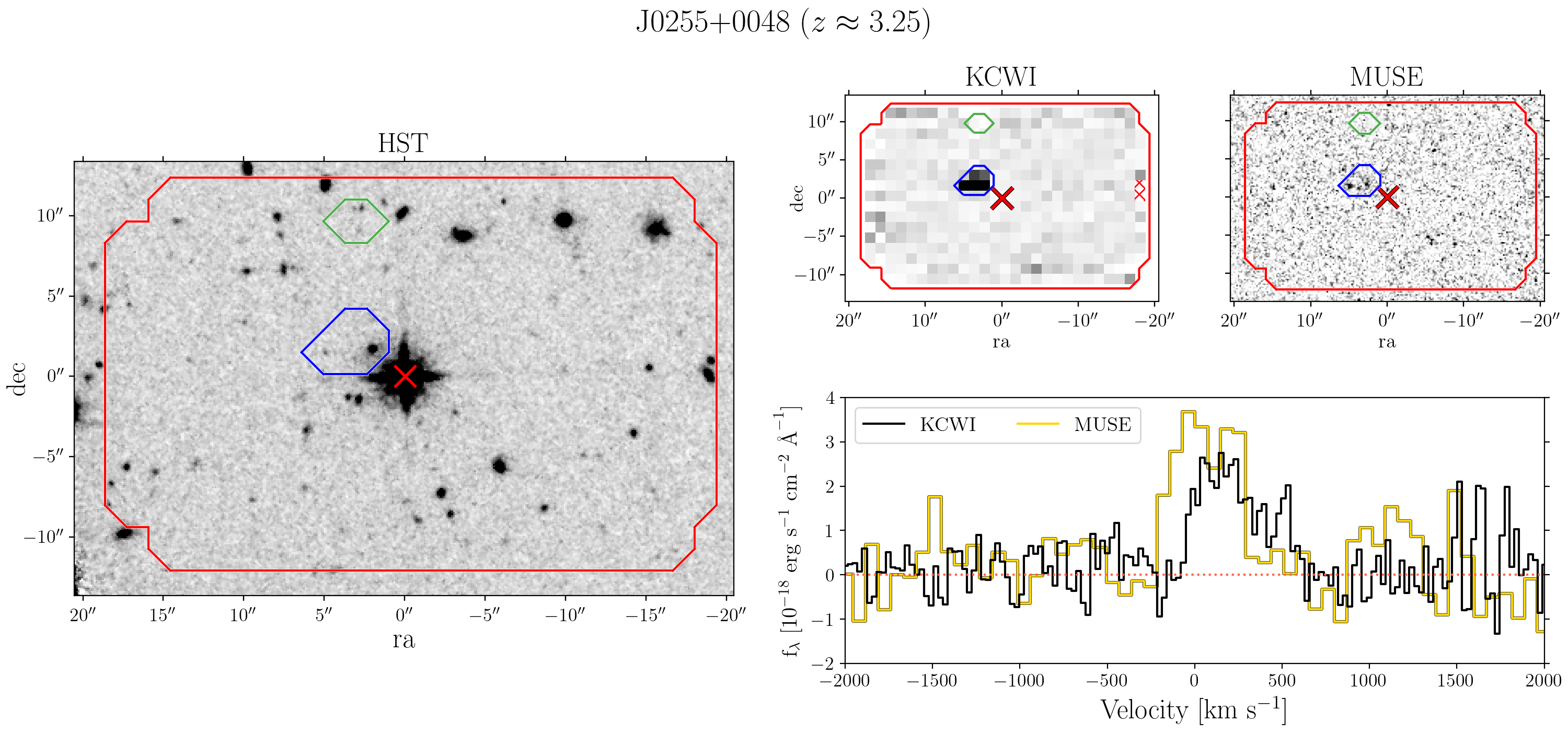}
\vspace{-1.4em}
\caption{Overview of the data for the DLA at $z\approx 3.25$ along the sightline of QSO J0255+0048. The top left panel shows HST imaging of this field in the F160W band, with the QSO location represented by the red cross. The top right panels show the KCWI and MUSE datacubes integrated within $\pm100$~km~s$^{-1}$ of the DLA redshift, with exception of spaxels featuring line emission: here, the integrated datacubes were computed at the wavelength of the emission line and within $\pm 3\sigma$ of the measured line width. We identify a significant \lya\ emitter at an impact parameter of $b\approx30$~kpc (blue contours), which was also identified in observations with VLT/MUSE (\citealt{fumagalli2017,mackenzie2019,pruto2024}). The bottom right panel shows the spatially integrated spectra obtained with KCWI (black) and MUSE (yellow) for this source. We do not detect in KAGG the tentative \lya\ emitter at $b\approx 84$~kpc (green contours) reported by MAGG (\citealt{mackenzie2019}).}
\label{fig_J0255}
\end{figure*}

A number of our target DLA fields have been searched for redshifted \cii\ emission with ALMA \citep{neeleman2025} or for \lya\ emission in the MAGG survey \citep{lofthouse2023}. Remarkably, we do not detect \lya\ emission from {\it any} of the known \cii\ emitters in our target DLA fields, including  
(i)~the \cii\ emitter at $z_{abs} = 3.80$ in the field of J1201+2117, (ii)~the two \cii\ emitters at $z_{abs} = 4.13$ in the field of J1054+1633, (iii)~the \cii\ emitter at $z_{abs} = 4.31$ in the field of J1626+2751, or (iv)~the \cii\ emitter at $z_{abs} = 4.39$ in the field of J0834+2140. There is a similarity between this result and the non-detection of \lya\ emission from DLA galaxies at $z\approx 2$ originally detected by their CO emission (\citealt{oyarzun2024}). \citet{oyarzun2024} concluded that DLA CO emitters at $z\approx 2$ are faint in \lya\ because of their high stellar and molecular gas masses, as well as their high dust extinctions. This explanation may not be sufficient to explain the low \lya\ luminosities of the \cii-emitting galaxies at $z \approx 4$, which have lower masses than the $z \approx 2$ CO emitters \citep[e.g.][]{kanekar2020,kaur2021} and show good agreement between their NUV-based and far-infrared-based SFRs, indicating low or moderate dust extinctions \citep{kaur2021}. 

As mentioned in Section \ref{2.1}, two of our QSO sightlines have been previously observed with VLT/MUSE as part of the MAGG survey and its precursor studies \citep{fumagalli2017,mackenzie2019,lofthouse2023}. We found no significant \lya\ emitters in the KCWI data cube for PSS0209+05, which has DLAs at $z \approx 3.67$ and $z\approx 3.86$. This is consistent with the results by \citet{lofthouse2023}, who found no significant \lya\ emitters within 150~kpc of either DLA  in their MUSE data cubes. While \citet{lofthouse2023} report the detection of a \lya\ emitter at $z\approx 3.86$, this object lies outside the KCWI field of view at a larger impact parameter ($b\approx211$~kpc). The second sightline that has been observed with VLT/MUSE is J0255+0048, which features a DLA at $z\approx 3.25$ \citep{fumagalli2017,mackenzie2019}. We identified one significant \lya\ emitter at $b\approx 30$~kpc in this field (see Figure~\ref{fig_J0255}), which is consistent with the VLT/MUSE results. However, we measure a slightly lower \lya\ flux ($19.6 \pm 9 \times 10^{-18}$~erg~s$^{-1}$~cm$^{-2}$) than the $31 \pm 1 \times 10^{-18}$~erg~s$^{-1}$~cm$^{-2}$ measured by  \citet{mackenzie2019}. \citet{mackenzie2019} also identify two tentative \lya\ emitters in this field, one at $b\approx 84$~kpc that we do not confirm (Figure \ref{fig_J0255}) and another at $b\approx 259$~kpc that lies outside of the KCWI field of view. Throughout this paper, we will use the KAGG \lya\ line flux for the DLA galaxy at $z \approx 3.25$ towards J0255+0048. 

\section{DLA \lya\ emitters} 
\label{5}

In this section, we analyze DLA \lya\ emitters, i.e., the \lya\ emitters that plausibly correspond to the DLA galaxy. We dissect their properties and then discuss implications for the population of DLA galaxies.

\subsection{The properties of DLA \lya\ emitters}
\label{5.1}

\begin{figure*}[h!]
\centering 
\includegraphics[width=7in]{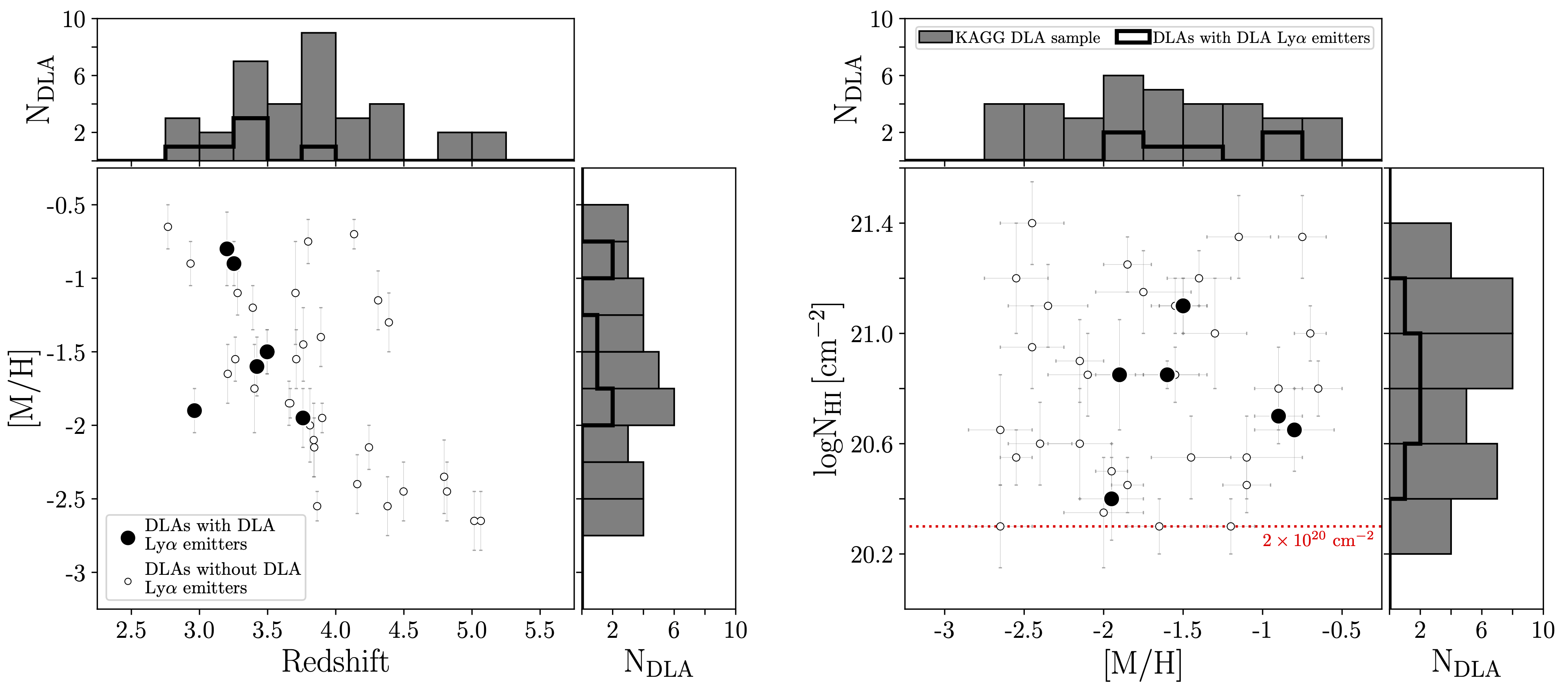}
\vspace{-0.5em}
\caption{The properties of the DLAs in KAGG that feature DLA \lya\ emitters. The redshifts and metallicities are plotted on the left panel, while the metallicities and \NHI\ are plotted on the right panel. The different symbols represent DLAs with DLA \lya\ emitter detections (black circles) and without DLA \lya\ emitter detections (white open circles). The red dotted line shows the DLA \NHI\ limit. The insets show the redshift, metallicity, and \NHI\ distributions for the full DLA sample (gray) and for DLAs with DLA \lya\ emitter detections (black). We only detect DLA \lya\ emitters at $z<4$.}
\label{fig_detections_met_nh1}
\end{figure*}

Figure \ref{fig_detections_met_nh1} shows the redshifts, metallicities, and \NHI\ values of the DLAs featuring DLA \lya\ emitter detections. We note that all significant DLA \lya\ emitters are found in the fields of low redshift ($z<4$), intermediate metallicity ($<-2<$~[M/H]~$<-0.5$) DLAs. The absence of DLA \lya\ emitters at higher redshift and lower DLA metallicities is noteworthy, with no significant detections in the low-metallicity, high-redshift half of the 36-DLA sample. Because of the degeneracy between redshift and DLA metallicity, we cannot distinguish which of these two parameters is driving this difference in detection rates. 

\begin{figure*}[hb!]
\centering 
\includegraphics[width=7.1in]{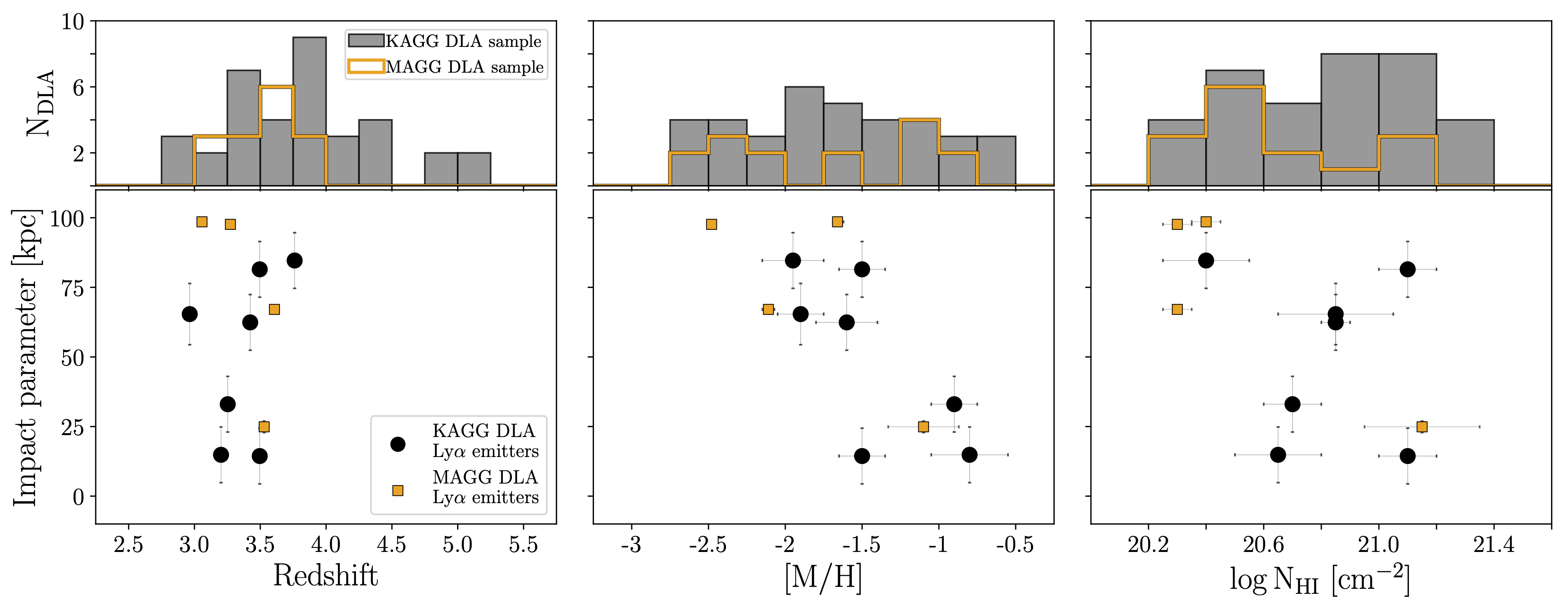}
\vspace{-1.7em}
\caption{Top panels: Histograms of the redshift, metallicity, and \NHI\ distributions of the DLA samples of KAGG and MAGG. The KAGG DLA sample is represented by the gray, filled histograms and the MAGG DLA sample is represented by the yellow lines. Bottom panels: Impact parameters of DLA \lya\ galaxies as a function of DLA redshift, metallicity, and \NHI. The black points represent DLA \lya\ emitters from KAGG and the yellow squares represent DLA \lya\ emitters from MAGG. Our DLA \lya\ galaxy detection rate is clearly higher in the lower redshift half of the DLA sample ($z<4$). We identify tentative evidence of an anti-correlation between DLA metallicity and DLA \lya\ impact parameter (1.8$\sigma$ significance).}
\label{fig_impact_param}
\end{figure*}

Figure \ref{fig_impact_param} shows how the impact parameters of DLA \lya\ emitters in KAGG and MAGG depend on redshift, DLA metallicity, and \NHI. For comparison, the corresponding distributions of the parent KAGG and MAGG DLA samples are also included in the top panels of this figure. As shown in Figure \ref{fig_detections_met_nh1}, DLA \lya\ emitters in KAGG are in the $z=3-4$ range, with no detections in DLAs at $z>4$. DLA \lya\ emitters in MAGG are also in the same redshift range, although this is by design because the redshift distribution of the MAGG DLA sample is limited to $z\lesssim 4$. 

In the middle panel of Figure \ref{fig_impact_param}, we identify a tentative anti-correlation between DLA metallicity and the impact parameters of DLA \lya\ emitters. It appears that DLA \lya\ emitters associated with $\rm [M/H]\approx -1$ DLAs are confined to $b\lesssim 50$~kpc, whereas DLA \lya\ emitters in the fields of $\rm [M/H] \lesssim -1.5$ DLAs are at larger impact parameters ($b\gtrsim 50$~kpc). A Pearson correlation test associates a $>2\sigma$ significance to this anti-correlation. However, this test does not take into account non-detections or differences between the impact parameter coverages of KAGG and MAGG. To better quantify the significance of this anti-correlation, we conduct Monte Carlo simulations of the KAGG and MAGG observations. Under the assumption that no correlation between [M/H] and impact parameter exists, and if we assume that the data for all DLAs have roughly similar detection limits, each spaxel in KAGG and MAGG has approximately the same probability of featuring a \lya\ emitter. By simulating this setup $N= 500$ times, we obtain $N$ random instances of the $b = b(\rm [M/H])$ relation. We then characterize the strength of each relation by fitting first order polynomials, yielding the random distribution of $b = b(\rm [M/H])$ slopes. The observed slope in the middle panel of Figure \ref{fig_impact_param} is steeper than $\approx 93\%$ of the simulated instances, and therefore we conclude that impact parameter decreases with DLA metallicity with 1.8$\sigma$ significance. 

 \begin{figure*}[ht!]
\centering 
\includegraphics[width=7.1in]{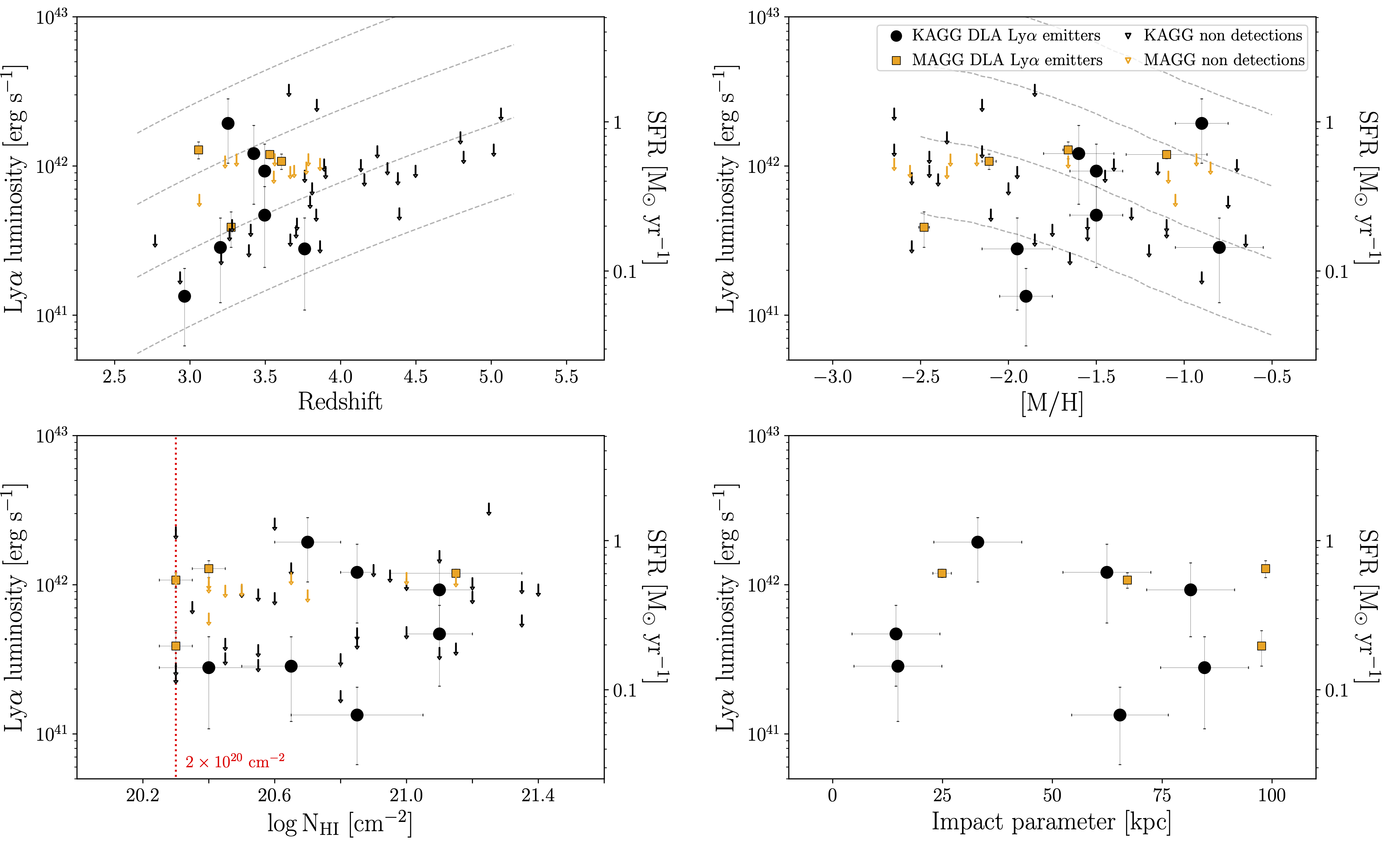}
\vspace{-1.2em}
\caption{The dependence of \lya\ luminosity of DLA \lya\ emitters on redshift, DLA metallicity, \NHI, and impact parameter. Shown are detections in KAGG (black points) and in MAGG (yellow squares). KAGG and MAGG non-detections of DLA \lya\ emitters are shown as black and yellow downward pointing arrows, respectively. The increase in luminosity distance with redshift results in a correlation between \lya\ luminosity and redshift (top left) and a weak anti-correlation between \lya\ luminosity and DLA metallicity (top right). The black dashed lines show the magnitude of the luminosity distance effect based on the sensitivity of our data. The red line shows the \NHI\ limit that separates DLAs from lower \NHI\ absorbers.}
\label{fig_lya_luminosity}
\end{figure*}

The right panel of Figure \ref{fig_impact_param} reveals that the impact parameters of DLA \lya\ emitters show no significant dependence on \NHI. It is also apparent that most of the \lya\ associations identified in KAGG are at higher \NHI\ than the associations established in MAGG. This difference is driven by differences between the \NHI\ distributions of the DLA samples (see top panels of Figure \ref{fig_sample}). In fact, the \NHI\ distributions of KAGG and MAGG DLA \lya\ emitters are both consistent with being randomly drawn from the \NHI\ distributions of their respective parent DLA samples.

Figure \ref{fig_lya_luminosity} shows how the \lya\ luminosities of DLA \lya\ emitters in KAGG and MAGG vary with redshift, DLA metallicity, \NHI, and impact parameter. We also show on the right side of the y-axes the star-formation rate (SFR) associated with the measured \lya\ luminosities. This assumes the H$\alpha$ SFR calibration of \citet{kennicutt2012}, a dust-free \lya\ to H$\alpha$ luminosity ratio of 8.7, and a \citep{chabrier2003} initial mass function \citep[e.g.][]{matthee2016,oyarzun2017,sobral2019}. However, we warn that these \lya\ SFRs suffer from large uncertainties because the escape fraction of \lya\ radiation can widely vary from galaxy to galaxy (e.g., \citealt{matthee2016}). These variations in the \lya\ escape fraction are largely driven by differences in the dust extinction and \hi\ column density of the host galaxy, which we cannot currently measure.

The slightly deeper line flux depth of KAGG over MAGG (Section \ref{3.3}) is apparent in the first panel of Figure \ref{fig_lya_luminosity}, which shows \lya\ luminosity as a function of redshift. We can also see in this panel that that there is a correlation between \lya\ luminosity sensitivity and redshift in KAGG. Given the increase in luminosity distance with redshift, this correlation is expected. This implies that our \lya\ luminosity sensitivity is also dependent on DLA metallicity, since low metallicity DLAs tend to be at high redshifts (\citealt{rafelski2012,rafelski2014,wisz2026}; see Figure \ref{fig_detections_met_nh1}). To illustrate the magnitude of these variations, we plot curves of iso-sensitivity as a function of redshift and DLA metallicity in the top panels of Figure \ref{fig_lya_luminosity} (black dashed lines). These curves are derived from the combination between the wavelength dependence of our \lya\ line flux sensitivity and the most recent fit to the DLA redshift-metallicity relation (\citealt{wisz2026}). The third and fourth panels of Figure \ref{fig_lya_luminosity} shows that there is no significant correlation between DLA \lya\ emitter luminosity and \NHI\ or between DLA \lya\ emitter luminosity and impact parameter. 

\subsection{Implications for the population of DLA galaxies}
\label{5.2}

ALMA mm-wave CO studies of high-metallicity DLAs ([M/H]$\gtrsim -0.5$) at $z\approx 2$ have shown that most of such DLAs probe \hi\ gas in the ISMs (or inner CGMs) of massive galaxies (M$_*>10^{10}$~M$_{\odot}$) detected by their CO emission \citep{neeleman2018,kanekar2020,kaur2022b,kaur2022a,kaur2025}. At these stellar masses, the \lya\ escape fraction can be low due to resonant scattering and absorption of \lya\ radiation by dust and/or high \NHI\ gas \citep{oyarzun2016}. For instance, \citet{oyarzun2024} searched for \lya\ emission in the fields of 4 CO emitters associated with metal-rich ($-1<$~[M/H]~$<0$) DLAs at $z\approx2$ and did not detect any emission at the galaxies' location. In three cases, they do not detect any \lya\ emission within $b\lesssim 50$~kpc of the DLAs, whereas in one case they detect two \lya\ emitters at larger impact parameters ($b\approx 50-70$~kpc) than the CO emitter ($b\approx 15$~kpc). 

In KAGG, we find a similar disconnect between samples of DLA galaxies detected in \lya\ emission and in the sub-mm via \cii\ emission. As described in Section \ref{4.2}, we do not detect \lya\ emission from any of the five \cii\ emitters detected in KAGG DLA fields. However, galaxy samples identified by their CO emission at $z\approx 2$ and by their \cii\ emission at $z\approx 4$ show some differences. While CO emitters have stellar masses exceeding M$_* \gtrsim 10 ^{10}$~M$_{\odot}$, \cii\ emitters at $z\approx 4$ can reach lower stellar masses (M$_* \gtrsim 10 ^{9}$~M$_{\odot}$). The agreement between FIR-SFRs and UV-SFRs of \cii\ emitters at $z\approx 4$ \citep{kaur2021} also indicates that their dust extinctions are, on average, significantly lower than the dust extinctions of CO emitters at $z\approx 2$. Further analysis is needed to determine whether these moderate amounts of dust are still sufficient to severely reduce the \lya\ luminosities of DLA galaxies. Also possible is that high \hi\ column density gas in the ISM of \cii\ emitters is scattering \lya\ photons along the line of sight, reducing their \lya\ escape fractions and luminosities.

We find success at detecting DLA \lya\ emitters in the intermediate DLA metallicity regime ($-1.5~\lesssim$~[M/H]$~\lesssim -0.5$; Figure \ref{fig_detections_met_nh1}), in similar fashion to slit spectroscopy surveys at $z\approx 2$ (\citealt{krogager2017}). Particularly, 4 DLA \lya\ emitters (3 in KAGG and 1 in MAGG) within $b\lesssim 50$~kpc are associated with [M/H]$~\sim -1$ DLAs. These impact parameters are consistent with the impact parameters of DLA \cii\ emitters at $z\approx 4$. Due to the remarkably good match between the redshift of the \cii\ emission and the redshift of the DLA ($\Delta v \lesssim 100$~km~s$^{-1}$), these \cii\ emitters are almost certainly the DLA galaxy. Unfortunately, a similar argument cannot be made for our sample of \lya\ emitters, since the \lya\ emission redshift is not a sufficiently accurate probe of the systemic redshift of the galaxy due to the highly complex \lya\ photon escape kinematics (Section \ref{4.1}). Therefore, we can only tentatively claim that these 4 \lya\ emitters around DLAs with [M/H]$~\approx -1$ correspond to the DLA galaxies.

Our DLA \lya\ emitters at [M/H]$~\sim -2$ have noticeably larger impact parameters ($b\approx 50-100$~kpc) than the \lya\ emitters around [M/H]$~\sim -1$ DLAs. Greater impact parameters at lower DLA metallicities are consistent with observations that link high CGM metallicity with low galactocentric distances (e.g. \citealt{nielsen2013,mendez-hernandez2022}). This trend is modulated by a combination of physical processes: how the ejection of metals onto the CGM is driven by outflows and how the reaccretion of CGM gas is modulated by the gravitational potentials of dark matter halos (\citealt{barai2013,cook2025}). In this context, the physical separation between the galaxy and the \hi\ gas reservoir probed by the DLA may drive variations in DLA metallicity. 

Alternatively, [M/H]$~\sim -2$ DLAs with \lya\ emitter detections at $b=50-100$~kpc could be associated with galaxies that we have not detected. In fact, the majority of our DLAs at these metallicities feature no DLA \lya\ emitter detections at any impact parameter (Figure \ref{fig_detections_met_nh1}), further suggesting that DLA \lya\ emitter detections $b=50-100$~kpc away from [M/H]$~\sim -2$ DLAs may not correspond to the DLA galaxy. Motivation for the DLA galaxy being at $b\lesssim 50$~kpc originates in theoretical work arguing that high \NHI\ gas is confined to the inner CGM of galaxies (e.g. \citealt{rahmati2015,rhodin2019,stern2021}). The primary candidates for such hosts are faint galaxies with low intrinsic star-formation rates. Such galaxies could be challenging to detect in \lya\ emission (e.g., KAGG and MAGG), CO emission (\citealt{kanekar2018,kaur2021}), and \cii\ emission (\citealt{neeleman2017,neeleman2025}). Because the SEDs of DLA \lya\ emitters are indicative of stellar masses of M$_*\approx 10^{8}- 10^{10}$~M$_{\odot}$ and intrinsic star formation rates of SFR~$\approx 1-100$~M$_{\odot}$~yr$^{-1}$ (\citealt{rhodin2021}), any missing primary emission counterparts of metal-poor DLAs would likely have M$_* <10^{8}$~M$_{\odot}$ and SFR~$<1$~M$_{\odot}$~yr$^{-1}$. These galaxies are unlikely to significantly enrich their surrounding \hi\ gas reservoirs, and thus are sensible candidates for hosting [M/H]$~\sim -2$ DLAs.

An important variable of the baryon cycle that we have yet to discuss is cosmic time. As apparent in Figures \ref{fig_lya_luminosity}, there is no clear trend in how the luminosity or the impact parameters of \lya\ emitters vary with redshift. However, we note a strong difference in the detection rate of DLA \lya\ emitters between $z<4$ and $z>4$ (Figure \ref{fig_impact_param}). While some of this difference could be driven by how our sensitivity decreases with redshift, we point out that the number of Large-scale \lya\ emitters remains high up to $z\approx 5$ (see Section \ref{6}). Thus, we cannot rule out that the number of \lya\ emitters at low $\Delta v$ and/or low impact parameter intrinsically decreases at $z>4$. 

An increase in the average impact parameter and/or $\Delta v$ of DLA galaxies with redshift would qualitatively agree with model predictions, particularly from the FIRE simulation (\citealt{hopkins2014,hopkins2018}). Their galaxy formation models predict that the fraction of high \NHI\ gas found at large galactocentric distances increases at early times (\citealt{stern2021}), implying a correlation between DLA impact parameter and redshift. In addition, they find that CGM turbulence and motion --- which are connected to $\Delta v$ --- are strongly redshift dependent. At early times, the velocity distribution of CGM gas is broad and can significantly shift from that of the galaxy. As halos grow more massive than $\sim 10^{12}$M$_{\odot}$, the CGM virializes and the velocity offsets between the galaxy and CGM decrease (\citealt{stern2021b}).

Samples of sub-mm DLA galaxies could distinguish between all of these scenarios. The impact parameter distribution of sub-mm DLA galaxies appears to differ significantly between $z\approx 2$ and $z\approx 4$: while $\approx 70\%$ CO emitters at $z\sim 2$ found to date are at $b<20$~kpc (\citealt{kanekar2018,kaur2022b}), only $\approx30\%$ of \cii\ emitters at $z\sim 4$ are found within this same impact parameter range (\citealt{neeleman2017,neeleman2025}). Larger samples are needed to confirm that this evolution is statistically significant, to rule out biases between CO and \cii\ selected samples, and to determine if there is any difference between the $\Delta v$ distributions of these two samples.

\section{Group and Large-scale \lya\ emitters} 
\label{6}

In the previous section, we discussed the properties of DLA galaxies by interpreting DLA \lya\ emitters as the primary emission counterparts of DLAs. However, it is also plausible that some of these DLA \lya\ emitters actually correspond to companions of undetected galaxies that give rise to the DLA. In this section, we consider the more general scenario of Group \lya\ emitters and Large-scale \lya\ emitters. We study their properties and interpret them as probes of the environments of DLAs.

\subsection{The properties of Group \lya\ emitters}
\label{6.1}

\begin{figure*}[h!]
\centering 
\includegraphics[width=7.1in]{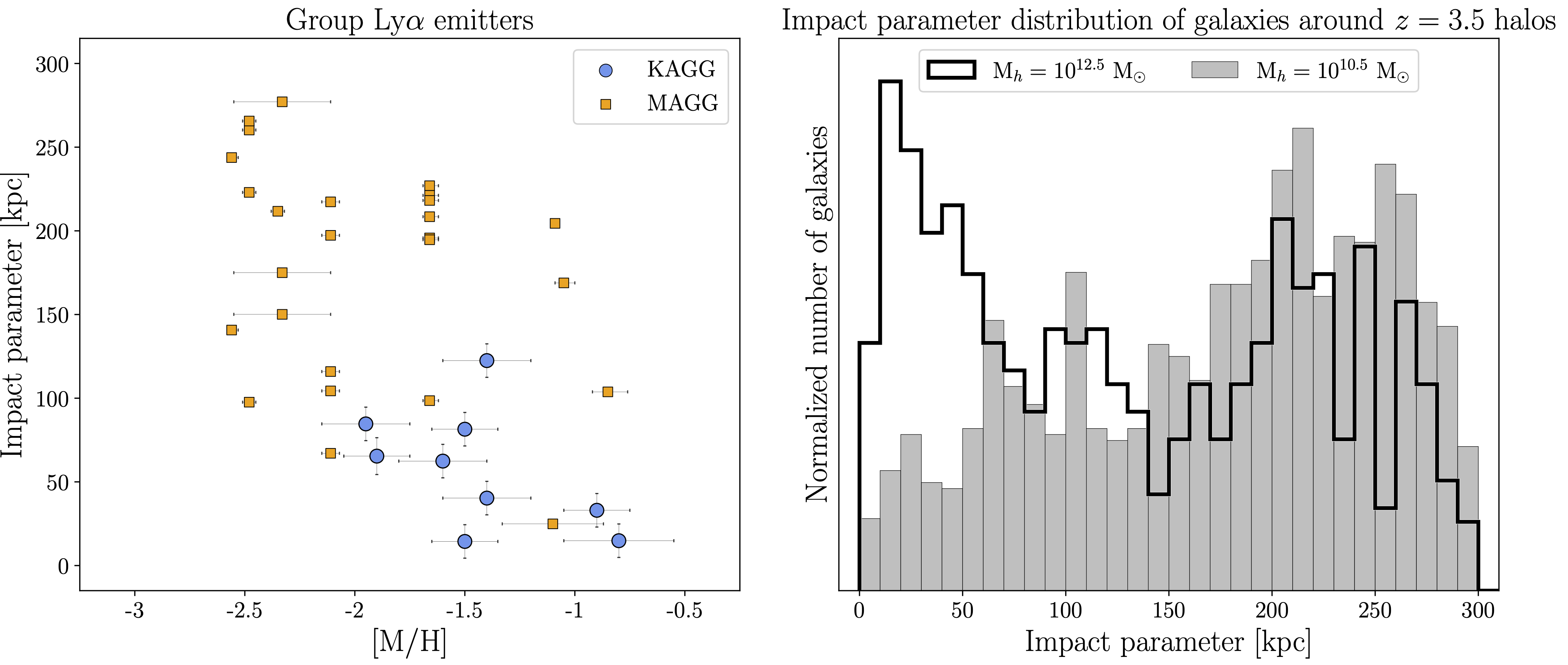}
\vspace{-0.7em}
\caption{Left: Impact parameters of Group \lya\ emitters in KAGG (blue circles) and in MAGG (yellow squares) as a function of DLA metallicity. Group \lya\ emitters are selected to be within impact parameters and velocity offsets equivalent to $b<300$~kpc. Under the assumption that none of these \lya\ emitters correspond to the DLA galaxy, this figure shows the impact parameter distribution of the \lya\ emitting neighbors of DLA galaxies. We find an anti-correlation between DLA metallicity and Group \lya\ emitter impact parameter at 2.3$\sigma$ significance. A distribution that is skewed toward larger impact parameters is consistent with low DLA galaxy halo masses. Right: To show this, we plot on the right panel the normalized impact parameter distribution of halo neighbors (M$_{h}>10^{10}$~M$_{\odot}$) in the Universe Machine empirical model ($z\approx 3.5$). These halo neighbor impact parameter distributions were measured around  M$_{h}=10^{12.5}$~M$_{\odot}$ host halos (black, open histogram) and M$_{h}=10^{10.5}$~M$_{\odot}$ host halos (gray, filled histogram). As halo mass decreases, the impact parameter distribution of neighboring halos shifts toward larger impact parameters.}
\label{fig_large_scale_impact_param}
\end{figure*}

We start by characterizing how the impact parameters of Group \lya\ emitters vary with redshift, DLA metallicity, and \NHI. The standout result --- shown in Figure \ref{fig_large_scale_impact_param} --- is an apparent anti-correlation between DLA metallicity and Group \lya\ emitter impact parameter. This result is similar to, and not independent of, the anti-correlation between DLA metallicity and DLA \lya\ emitter impact parameter that we reported in Section \ref{5.1} and Figure \ref{fig_impact_param}. Similarly to our analysis in Section \ref{5.1}, we conduct Monte Carlo simulations of this anti-correlation. We find it to have a significance of 2.3$\sigma$, slightly higher than that for DLA \lya\ emitters.

This variation in the typical impact parameter of Group \lya\ emitters with DLA metallicity may be an indication of differences in galaxy clustering with DLA metallicity. To explore this, we study how the impact parameter distribution of galaxies around halos vary with host halo mass. Explicitly, we analyze the CANDELS catalogs generated with the Universe Machine empirical model of galaxy formation (\citealt{behroozi2019}), and we proceed to select halos within the redshift bin $z = 3.4-3.6$ and with halo masses M$_h > 10^{10}$~M$_{\odot}$. With this redshift and halo mass selections, we ensure that the Universe Machine catalogs and our observations are as well matched as possible. Then, we compute the average impact parameter distribution of galaxies around halos of two different masses: M$_h = 10^{10.5}$~M$_{\odot}$ and M$_h = 10^{12.5}$~M$_{\odot}$. The results are plotted in the right panel of Figure \ref{fig_large_scale_impact_param}.

Differences between the impact parameter distributions of galaxies around halos of different masses are apparent. For halos of mass M$_h = 10^{12.5}$~M$_{\odot}$ --- the most massive halos at $z\approx 3.5$ --- there are two peaks in the impact parameter distribution, at $b\approx 0-50$~kpc and a lower peak at $\approx 200-300$~kpc. On the other hand, most neighboring galaxies around M$_h = 10^{10.5}$~M$_{\odot}$ halos are found at $b\approx 200-300$~kpc. Upon comparison with the left panel of Figure \ref{fig_large_scale_impact_param}, we can conclude that variations in the impact parameters of Group \lya\ emitters with DLA metallicity are consistent with variations in host halo mass with DLA metallicity. 

\begin{figure*}[h!]
\centering 
\vspace{-0.7em}
\includegraphics[width=7.1in]{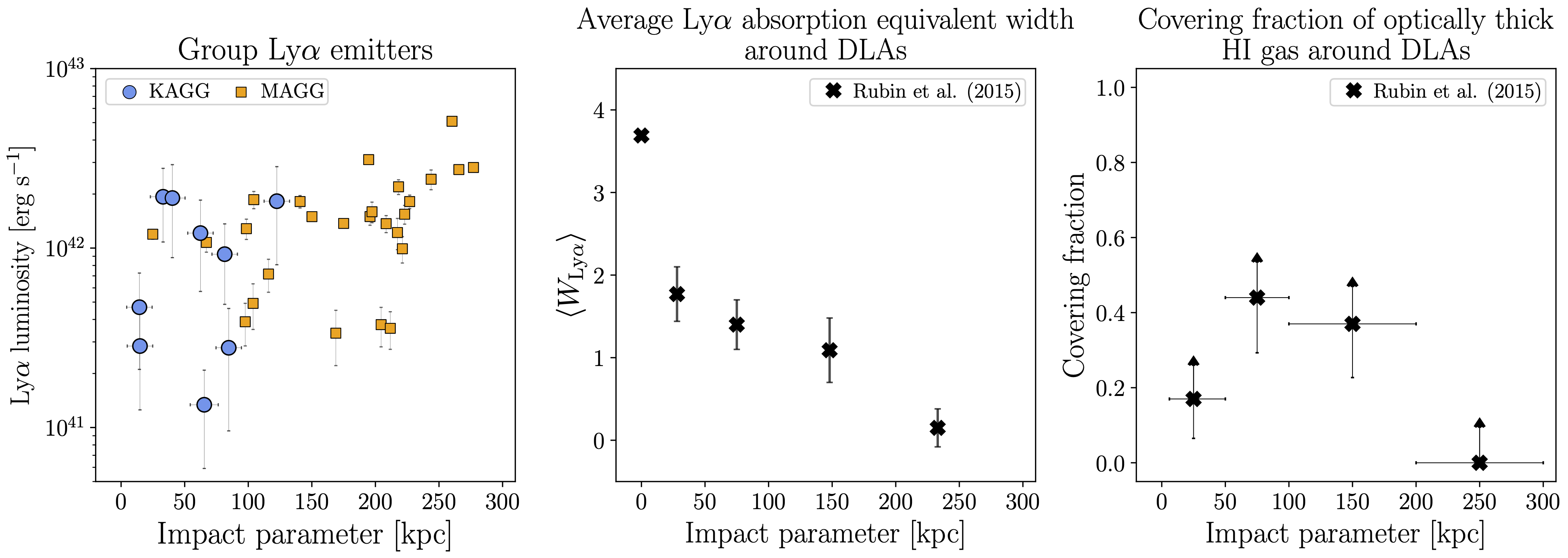}
\caption{Left: \lya\ luminosities of Group \lya\ emitters in KAGG (blue circles) and in MAGG (yellow squares) as a function of impact parameter. These \lya\ emitters were selected to at $z < 300$~kpc, and to have velocity offsets equivalent to $b<300$~kpc. We find that Group \lya\ luminosity and impact parameter are correlated (2.3$\sigma$ significance; Appendix \ref{appB}). This correlation is consistent with a decrease in the covering fraction of dust and high \NHI\ gas with impact parameter. Middle: Support for this interpretation can be found in studies of spatially coincident \hi\ absorption from QSO pairs (\citealt{rubin2015}). The figure shows that the \lya\ absorption equivalent width decreases with increasing DLA impact parameter. Right: Lower limits on the covering fraction of optically thick \hi\ gas around DLAs as a function of impact parameter, also from \citet{rubin2015}.}
\label{fig_large_scale_luminosities}
\end{figure*}

We also characterize how the \lya\ luminosities of Group \lya\ emitters vary with different properties. There are hints of a correlation between \lya\ luminosity and impact parameter, which we show in Figure \ref{fig_large_scale_luminosities}. We conduct a detailed Bayesian analysis of this trend, which accounts for the measured \lya\ luminosities and their errors, non-detection upper limits, and the different fields of view of KAGG and MAGG. This analysis is presented in Appendix \ref{appB} and indicates that this correlation is significant at the 2.3$\sigma$ level. 

\newpage

Other observations are consistent with the faint \lya\ luminosities of low impact parameter associations. For example, \citet{krogager2017} and \citet{rhodin2018} utilized slit spectroscopy to search for \lya\ emission within $b\lesssim 40$~kpc of DLAs at $z < 3$, finding \lya\ emitters that are as faint as those at $b \lesssim 40$~kpc in KAGG in $\gtrsim50\%$ of their DLA sample ($L_{\mbox{\small\lya}} = 10^{41}-10^{42}$~erg~s$^{-1}$; Figure \ref{fig_large_scale_luminosities}). 

Figure~\ref{fig_large_scale_luminosities} suggests that changes in the \lya\ luminosity with impact parameter are apparent for $b \gtrsim 100$~kpc. Therefore, it is interesting to consider whether there are large-scale effects that may hinder the escape of \lya\ radiation around DLAs. The middle panel of Figure~\ref{fig_large_scale_luminosities} shows the variation of the average \lya\ absorption equivalent width ($<W_{\rm Ly\alpha}>$) with DLA impact parameter for QSO pairs \citep{rubin2015}. The figure indicates that QSO sightlines through DLA environments feature an excess of \lya\ absorption compared to random sightlines. Further, \citet{rubin2015} estimate the covering fraction of optically thick gas (\NHI$~> 10^{17.2}$~cm$^{-2}$) within $\approx 150$~kpc of DLAs and found it to be at least $\gtrsim 30\%$ (right panel of Figure~\ref{fig_large_scale_luminosities}). We discuss in Section \ref{6.3} how this spatially coincident high-\NHI\ gas could be, at least in part, responsible for the decrease in \lya\ luminosity toward lower DLA impact parameters.

\subsection{The properties of Large-scale \lya\ emitters}
\label{6.2}

\begin{figure*}[hb!]
\centering 
\includegraphics[width=7.1in]{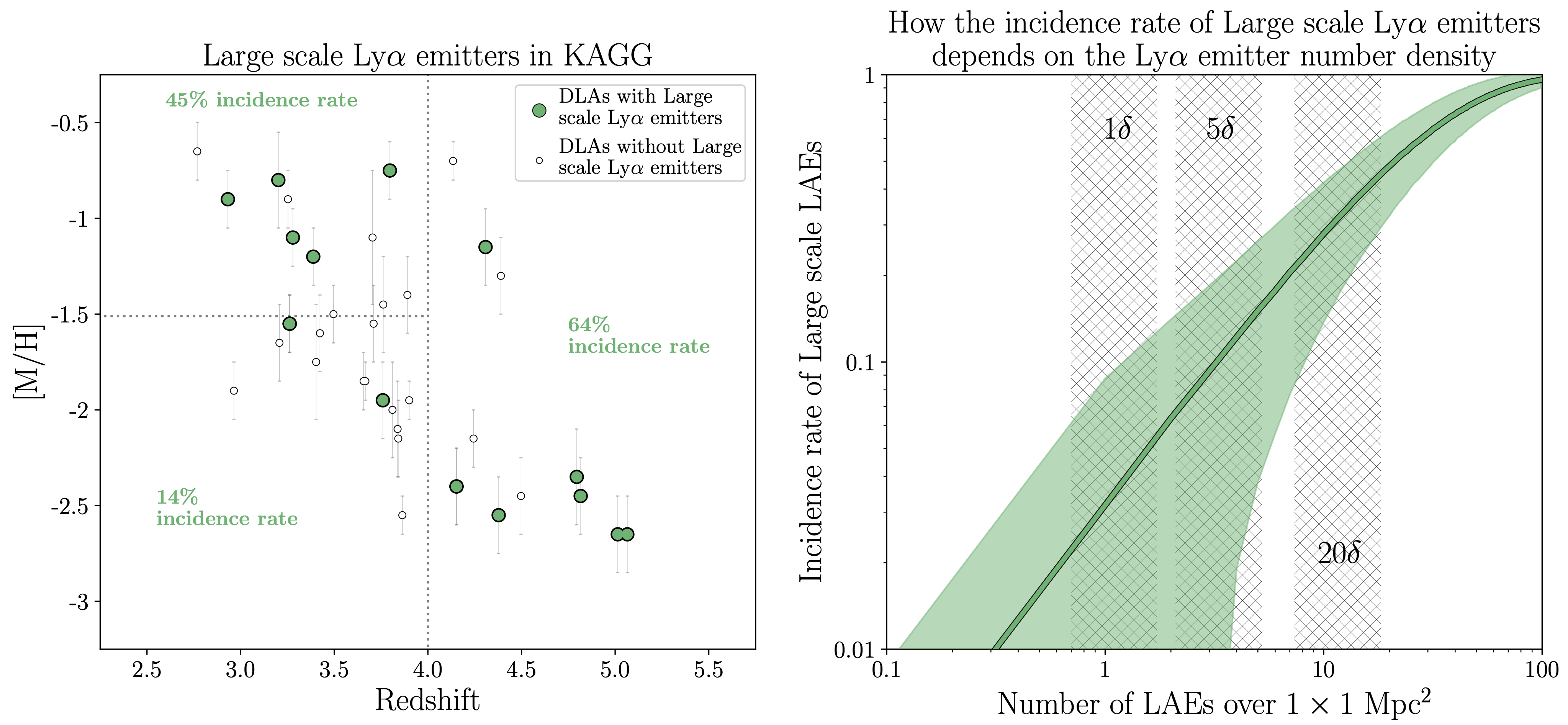}
\caption{Left: The incidence of Large-scale \lya\ emitters around DLAs from the KAGG survey as a function of redshift and DLA metallicity. DLAs with Large-scale \lya\ emitter detections are shown as filled green circles, while those without Large-scale \lya\ emitters are shown as open white circles. We divide the DLA sample into three quadrants according to redshift and DLA metallicity. At $z<4$, the incidence rate of DLAs with Large-scale \lya\ emitters is higher at high DLA metallicities ($\approx 45\%$) than at low DLA metallicities ($\approx 14\%$). Right: Simulations of the incidence rate of Large-scale \lya\ emitters in KAGG as a function of the \lya\ emitter surface density over $1 \times 1$~Mpc$^2$. The mean relation is shown as the green solid line and the $68$\% confidence interval shown as the green shaded region. Large-scale \lya\ emitter incidence rates exceeding $20\%$ are consistent with originating in galaxy over-densities ($>5\delta$).}
\label{fig_large_scale_incidence}
\end{figure*}

Large scale \lya\ emitters correspond to galaxies at larger velocity offsets than DLA \lya\ emitters and Group \lya\ emitters. With a relative velocity selection that is equivalent to $0.3\lesssim b~[\rm Mpc] \lesssim 10$, such \lya\ emitters trace galaxies within the same large scale structures as the DLAs (Section \ref{4.4}). In Figure \ref{fig_large_scale_incidence}, we show how the detection rate of Large-scale \lya\ emitters depends on DLA metallicity and redshift. There is at least one Large-scale \lya\ emitter associated with 14 of the 36 DLAs, which corresponds to an incidence rate of $\approx 39\%$. To inspect whether this incidence rate depends on DLA metallicity or redshift, we separate the sample into three bins with roughly equal number of DLAs (see Figure~\ref{fig_large_scale_incidence}). While $\approx45\%$ of intermediate metallicity DLAs at $z<4$ feature Large-scale \lya\ emitters, this incidence rate drops to $\approx 14\%$ for metal-poor DLAs at the same redshift. The incidence rate of Large-scale \lya\ emitters is the highest in the $z>4$ bin ($\approx64\%$). 

\newpage

To interpret these results, we simulate how the incidence rate of Large-scale \lya\ emitters over the KCWI field of view depends on the surface density of \lya\ emitters over a larger area. These simulations were conducted by randomly placing galaxies over areas of $1\times1~$Mpc$^{2}$ and then measuring the surface density of \lya\ emitters within a central region of $230 \times 140$~kpc$^{2}$ (i.e., the KCWI field of view at $z\approx 4$). Even for protoclusters at these redshifts --- which typically feature $\lesssim 10$ \lya\ emitters per Mpc$^{2}$ \citep{kuiper0211,liang2018,laishram2026} --- the expected number of Large-scale \lya\ emitters within the KCWI field of view is $\lesssim 1$. 

From these simulations, we recover the incidence rate of Large-scale \lya\ emitters in samples of 10 fields, i.e., the number of fields in each DLA subsample (left panel of Figure \ref{fig_large_scale_incidence}). We plot the median and error on the simulated incidence rate in the right panel of Figure \ref{fig_large_scale_incidence}. Our measurements of the incidence rate from DLA samples are consistent with a wide range of \lya\ emitter surface densities, from 1 to 20 \lya\ emitters per Mpc$^{2}$. To contextualize these \lya\ emitter surface densities, we estimated galaxy over-densities defined as 
\begin{equation}
\label{delta}
\delta = \frac{\rho_{gal}}{<\rho_{gal}>} - 1, 
\end{equation}
where $\rho_{gal}$ corresponds to the number density of galaxies and $<\rho_{gal}>$ corresponds to the mean number density of galaxies (\citealt{alpaslan-tinker2020,oyarzun2024b}). We derive $<\rho_{gal}>$ at $z\approx 4$ by integrating the galaxy stellar mass function down to M$_* = 10^{9}$~M$_{\odot}$ from the best-fit Schechter function derived by \citet{weaver2023}. To convert these galaxy number densities into galaxy surface densities, we adopt a line of sight distance of $10~$pMpc, which matches the search path $\Delta v \lesssim 4,000$~km~s$^{-1}$ that we adopted to identify Large-scale \lya\ emitters in KAGG. Finally, we convert these galaxy surface densities into \lya\ emitter surface densities by adopting a \lya\ emitter fraction of $10-50\%$ at these redshifts \citep{oyarzun2017}, where this uncertainty drives the error on $\delta$. The corresponding ranges of $\delta$ are plotted as shaded areas in Figure \ref{fig_large_scale_incidence}. 

From this analysis, the $\gtrsim 50\%$ incidence rate of Large-scale \lya\ emitters in intermediate metallicity DLAs at $z<4$ is consistent with galaxy over-densities $>5\delta$. On the other hand, the $\approx 14\%$ incidence rate of Large-scale \lya\ emitters around metal-poor DLAs at $z<4$ can be explained by both the field and by low over-densities ($1-5\delta$). We discuss the implications of these results in Section~\ref{6.3}.    

\subsection{Implications for the environments of DLAs}
\label{6.3}

Searches for line emission around DLAs --- as well as lower \NHI\ absorbers --- often reveal \lya\ emitters at the absorption redshift. While in some cases these identifications may well correspond to the galaxy associated with the DLA gas \citep[e.g.,][]{fumagalli2017,nielsen2022}, the frequent identification of (sometimes several) \lya\ emitters at large offsets in both velocity  ($|\Delta v| > 500$~km~s$^{-1}$) and impact parameter ($b\gtrsim 100~$kpc) indicates that DLAs cannot always be directly linked to \lya\ emitters (e.g. \citealt{lofthouse2023}, this work). Thus, it is reasonable to interpret Group and Large-scale \lya\ emitters as emission probes of the environments of DLAs. 

The velocity distribution of \lya\ emitters relative to the redshift of the DLA ($\Delta v$) can be used to constrain these environments. We start by quantifying the incidence rate of \lya\ emitters at arbitrary velocities, i.e., galaxies unassociated to the DLA. Figure \ref{fig_velocity} shows that we detect only 6 galaxies at $|\Delta v| = 5,000 - 16,000$~km~s$^{-1}$, indicating a low incidence rate of field \lya\ emitters. Instead, we identify a clear abundance of \lya\ emitters at velocities corresponding to large over-densities ($|\Delta v| = 1000 - 4000$~km~s$^{-1}$; Figure \ref{fig_velocity}). Further, there is at least one Large-scale \lya\ emitter in $\approx 40\%$ of our DLAs (Figure~\ref{fig_large_scale_incidence}). Taking into account that \lya\ emission is typically offset by $\lesssim 600$~km~s$^{-1}$ relative to the systemic velocity of the galaxy (e.g. \citealt{cassata2020}), velocities of $\Delta v = 1000- 4000$~km~s$^{-1}$ at $z\approx 4$ correspond to comoving sizes of $\Delta d \approx 5 - 40$~cMpc or proper sizes of $d \approx 1 - 10$~Mpc. These scales are consistent with cosmic overdensities at $z>2$ (e.g., \citealt{toshikawa2020,tornotti2025}), hinting that a significant fraction of DLAs at $z=3-5$ may reside in these structures. Indeed, our analysis in Section \ref{6.2} indicates that the incidence rate of Large-scale \lya\ emitters at the high-metallicity end of the distribution is consistent with $>5\delta$ over-densities. 

We also find evidence of a connection between host halo mass and DLA metallicity in our analysis of the impact parameter of Group \lya\ emitters (Section \ref{6.1}). As apparent in Figure \ref{fig_large_scale_impact_param}, the impact parameters of Group \lya\ emitters are, on average, smaller at high DLA metallicities than at low DLA metallicites. Based on our analysis of the CANDELS catalogs of galaxy formation generated with the Universe Machine empirical model (\citealt{behroozi2019}), these differences in the average impact parameter are consistent with differences in the mass of the DLA host halo. 

The host environments of DLAs can be independently inferred from studies of DLA clustering (e.g., \citealt{cooke2006,barnes2014,alonso2018}). \citet{font-ribera2012} found that QSOs in SDSS/BOSS (\citealt{dawson2013}) featuring DLAs at $z\sim 2-3$ are highly biased, where bias is a quantitative measure of the clustering of a population relative to the underlying dark-matter density field (\citealt{sheth1999,wechsler2018}). Furthermore, stacking of \lya\ emitter spectra from the HETDEX survey (\citealt{gebhardt2021,hill2021}) reveals that the \lya\ absorption troughs of stacked \lya\ emitter spectra at $z\approx 2-3$ are more pronounced in high density regions of the Universe (\citealt{weiss2024}). These results suggest that the host halos of DLAs are, on average, at the high end of the halo mass function (M$_h \gtrsim 10^{12}$~M$_{\odot}$).

In a different study, \citet{perez-rafols2018} found that the connection between DLAs and high mass halos is more nuanced: while DLAs with strong metal absorption indeed tend to reside in massive halos, DLAs with weak metal absorption are less biased and more consistent with residing in lower mass halos (M$_h < 10^{11}$~M$_{\odot}$). This finding is in qualitative agreement with our findings utilizing Group and Large-scale \lya\ emitters that link high mass halos with high DLA metallicities at these redshifts (Sections \ref{6.1} and \ref{6.2}).

This link between halo mass and DLA metallicity is also supported by kinematic analyses of DLAs that utilize $\Delta v_{90}$, the velocity interval encompassing 90\% of the total optical depth from a low-ion transition metal absorber (\citealt{prochaska-wolfe1997}). \citet{prochaska2008b} found a strong correlation between the equivalent width of [Si\,II]$\,\lambda1526$ absorption and $\Delta v_{90}$, suggesting a correlation between DLA metallicity and halo mass (see also \citealt{neeleman2013}). Theoretical work have since reproduced this correlation and reinforced the interpretation that intermediate-high metallicity DLAs tend to originate in massive host halos (e.g., \citealt{bird2014,bird2015}). 

We can also gain insights into the environments of DLAs from the apparent correlation between the \lya\ luminosities and the impact parameters of Group \lya\ emitters. Studies of \lya\ emission in galactic environments (e.g., \citealt{laursen2009,blanc2011,atek2014,oyarzun2017,smith2019b,weiss2021}) indicate that \lya\ line brightness is primarily dictated by the production, resonant scattering, and absorption of \lya\ photons, with effects such as the IGM transmission at $z\sim 3-5$ and the slope of the stellar initial mass function making only second-order contributions \citep{verhamme2006,verhamme2008,verhamme2012,dijkstra2012,gronke2016,gronke2016b,sobral2019}. Therefore, one could consider two explanations for why the \lya\ luminosities of Group \lya\ emitters are lower at small impact parameters (Figure \ref{fig_large_scale_luminosities}): low \lya\ photon productions (i.e., due to low SFRs) and/or low \lya\ escape fractions (i.e., due to high \NHI\ or high dust extinction).

In our first scenario, the production of \lya\ photons is intrinsically low near DLAs. However, this is contrary to observations that link high \hi\ surface densities with high SFR surface densities \citep{kennicutt1998,kennicutt2012}. The second scenario is that high \NHI\ gas within $b\lesssim100$~kpc of the DLAs is resonantly scattering \lya\ radiation. The gas responsible for the scattering would not be the DLA clouds themselves, which we know have characteristic sizes and coherence scales of $\lesssim10$~kpc \citep{faerman-werk2023,lopez2024,berg2025b,berg2025,shaban2025}. Rather, we posit that star-forming galaxies within $b\lesssim100$~kpc of DLAs may have environments that are rich in \HI\ and dust, facilitating the scattering and absorption of \lya\ photons. We find motivation for this scenario in the fact that the covering fraction of optically thick \hi\ gas remains significant within $b\lesssim 100$~kpc of DLAs (Section \ref{6.1} and Figure \ref{fig_large_scale_luminosities}; \citealt{rauch2002,ellison2007,prochaska2013,rubin2015}). This association between DLA environments and high \NHI\ star-forming galaxies is in qualitative agreement with our previous analyses that link metal-rich DLAs at $z\sim 3-5$ with galaxy overdensities.

An alternative observable to \lya\ luminosity is the rest-frame \lya\ equivalent width ($W_{\rm Ly\alpha}$), i.e., the \lya\ line flux divided by the rest-frame UV continuum. Unlike differences in the \lya\ luminosity, differences in $W_{\rm Ly\alpha}$ are more straightforward to associate with variations in galaxy properties: the shape of the $W_{\rm Ly\alpha}$ distribution depends on stellar mass, as well as proxies for \NHI\ and dust extinction (e.g., \citealt{oyarzun2016,oyarzun2017}). Unfortunately, because only very deep IFU observations constrain the rest-frame UV continua of faint galaxies at $z\gtrsim2$ (e.g., \citealt{revalski2024}), constraining the $W_{\rm Ly\alpha}$ of DLA galaxies often requires ancillary imaging. The convenience of $W_{\rm Ly\alpha}$ over \lya\ luminosity, as well as being able to constrain galaxy morphology, highlights the benefit of rest-frame UV imaging for samples of \lya\ DLA galaxies. Unfortunately, only one DLA \lya\ emitter in KAGG is in a field that features rest-frame UV imaging (Figure \ref{fig_J0255}).

An alternative view of the correlation between \lya\ luminosity and impact parameter is that bright \lya\ emitters --- such as those detected by MAGG at large impact parameters --- are absent in the vicinity of DLAs. One could associate this absence with an observational effect. The field of view --- and therefore the cosmic volume --- of the search for \lya\ emitters increases with impact parameter. As a result, the likelihood of finding bright \lya\ emitters increases with impact parameter. The magnitude of this effect depends on the steepness of the \lya\ luminosity function, which is unconstrained for DLA galaxies. Regardless of whether the correlation between \lya\ luminosity and impact parameter is driven by the physics of the baryon cycle or by observational effects, it appears that luminosity-limited searches for \lya\ emitters around DLAs --- such as KAGG and MAGG --- are bound to identify more and brighter galaxies at large impact parameters than at low impact parameters.

\section{Summary} 
\label{7}

DLAs enable studies of the baryon cycle at high redshift because they probe the physical properties of high \NHI\ gas in absorption. Integral field spectrographs allow one to connect DLAs with their parent and companion galaxies, a goal that has been elusive for decades. In this paper, we introduce KAGG (KCWI Analysis of Gas around Galaxies), a survey with Keck/KCWI designed to search for \lya\ emission around DLAs at $z\sim 3-5$. KAGG exploits the broad wavelength coverage of KCWI to search for \lya\ emitters in 17 QSO fields featuring 36 DLAs. The DLA sample was constructed solely on the presence of multiple DLAs along QSO sightlines, and thus KAGG provides \lya\ emission constraints on the galaxy counterparts of DLAs at $z\sim 3-5$ without any metallicity pre-selection. The observations allow us to identify \lya\ emitters down to faint \lya\ line luminosities of $L_{\mbox{\small\lya}}\approx10^{41}$~erg~s$^{-1}$. Because we deployed KCWI in its large slicer configuration, we are sensitive to galaxies over large fields of view ($\approx 240\times 150$~kpc) with spaxel sizes of $\approx 10$~kpc. We developed algorithms to search for emission lines, diagnose emission line signal-to-noise, and quantify our line detection completeness. Our findings are as follows:

1. We identify 26 significant \lya\ emitters within $\Delta v = 4,000$~km~s$^{-1}$ of the DLA redshifts. This velocity range is broad and encompasses different types of DLA associations. Based on studies of \lya\ emission redshifts relative to galaxy systemic redshifts, we first select all \lya\ emitters whose redshifts are  consistent with the DLA redshift and which lie at impact parameters within 100~kpc of the DLA sightline. This selection yields 7 detections, which we classify as DLA \lya\ emitters and represent our primary DLA galaxy candidates.

2. These 7 DLA \lya\ emitters span a wide impact parameter range ($b = 0-100$~kpc) and are mostly associated with intermediate-high metallicity DLAs ($-1.5\lesssim$~[M/H]~$\lesssim-0.5$). These results confirm previous findings that the success of \lya\ emission searches at low impact parameters peaks at these metallicities (e.g., \citealt{krogager2017}).

3. Based on a composite sample of 11 DLA \lya\ emitters from the KAGG and MAGG \citep[MUSE Analysis of Gas around Galaxies; ][]{mackenzie2019,lofthouse2023} surveys, we constrain how the properties of DLA \lya\ emitters vary with DLA properties. We find tentative (1.8$\sigma$) evidence of an anti-correlation between DLA metallicity and DLA \lya\ emitter impact parameter. While DLA \lya\ emitters around [M/H]~$\approx -1$ DLAs lie at low impact parameters ($b<50~$kpc), DLA \lya\ emitters around [M/H]~$\approx -2$ DLAs are all found at $b>50~$kpc. One possible explanation for this trend is that metal enrichment of the CGM is more efficient at lower galactocentric distances. Alternatively, [M/H]~$\approx -2$ DLAs could be associated with low stellar mass (M$\lesssim 10^{8}$~M$_{\odot}$) galaxies at $b<50~$kpc that are below our detection limits.

4. There are five \cii\ emitters in our DLA sample, none of which we detect in \lya\ emission. We conclude that a significant fraction of the DLA galaxy population at $z\approx 4$ is faint in \lya\ emission, presumably due to \lya\ photon scattering by high \NHI\ gas and/or absorption by dust grains. 

5. We have no significant DLA \lya\ emitter detections at low metallicities ([M/H]$~\lesssim-2$) or at the highest redshifts ($z>4$). This low detection rate could be caused by a decrease in \lya\ luminosity sensitivity with redshift. Alternatively, metal-poor DLAs at $z\gtrsim 4$ could mostly arise from faint galaxies with low stellar masses and low star-formation rates (M$_* <10^{8}$~M$_{\odot}$ and SFR~$<1$~M$_{\odot}$~yr$^{-1}$). 

6. We define a second subsample called Group \lya\ emitters. They are selected to lie at $b\lesssim 300$~kpc of the DLA location and redshift, i.e. to lie within the group environment of the DLA. Our sample of 35 Group \lya\ emitters includes the 11 DLA \lya\ emitters from KAGG and MAGG, 2 additional \lya\ emitters at larger velocity offsets from KAGG, and 22 additional \lya\ emitters from MAGG.

7. We find that the impact parameters of Group \lya\ emitters decrease with increasing DLA metallicity at $2.3\sigma$ significance. Group \lya\ emitters around DLAs with [M/H]~$\gtrsim-1.5$ typically have $b\approx 0-150$~kpc, whereas Group \lya\ emitters around DLAs with [M/H]~$\lesssim -2$ tend to be at $b\approx 100-300$~kpc. This difference in impact parameter is consistent with variations in halo mass, according to a comparison with the Universe Machine empirical model. We interpret this result as a signature that DLA metallicity is correlated with the halo mass of the DLA galaxy. This conclusion is in agreement with results from DLA clustering studies linking metal-enriched DLAs with higher bias factors (\citealt{perez-rafols2018}).

8. We also identify an increase in Group \lya\ emitter luminosity with Group \lya\ emitter impact parameter. We conduct a Bayesian analysis of this correlation and find it to have $2.3\sigma$ significance (Appendix \ref{appB}). Studies of coincident \lya\ absorption in QSO pairs \citep{rubin2015} have yielded evidence that DLAs reside in extended ($b\lesssim100$~kpc) high \NHI\ environments, which may lead to lower \lya\ escape fractions at impact parameters of even $b \approx 100$~kpc. Thus, we propose that galaxies within $b\lesssim100$~kpc of DLAs are more likely to scatter and absorb \lya\ photons. 

9. We define a third \lya\ emitter subsample named Large-scale \lya\ emitters. This subsample encompasses all remaining galaxies in KAGG that are found at larger velocity offsets~$\lesssim 4000$~km~s$^{-1}$, corresponding to proper distances of $0.3\lesssim d~[\rm Mpc]\lesssim 10$ at $z\approx 4$. We find that the incidence rate of Large-scale \lya\ emitters peaks at high DLA metallicities ([M/H]~$\geq-1.5$) and at high redshifts ($z>4$). We model the incidence rate of Large-scale \lya\ emitters in our survey, and we conclude that intermediate-high metallicity DLAs ([M/H]$~\gtrsim -1.5$) reside in galaxy over-densities ($>5\delta$). This result provides further evidence for a link between DLA metallicity and halo mass that is independent of our analysis of Group \lya\ emitters.

\begin{acknowledgements}
We thank the anonymous reviewer for their constructive comments regarding this manuscript. We acknowledge support from the National Science Foundation under grant No. 2107989. MR also acknowledges support from the STScI's Director's Discretionary Research Funding (PIDs D0101.90306 \& D0101.90362). NK acknowledges support from the Department of Atomic Energy, under projects 12-R\&D-TFR-5.02-0700 and RTI4017 ``Next Generation Instrumentation for Radio Astronomy'', and also from the Anusandhan National Research Foundation of the Department of Science and Technology, India, via a J. C. Bose Fellowship (JCB/2023/000030). This work made use of Astropy\footnote{\url{http://www.astropy.org}}, a community-developed core Python package and an ecosystem of tools and resources for astronomy \citep{astropy:2013, astropy:2018, astropy:2022}.
\end{acknowledgements}

\appendix

\section{Group and Large scale \lya\ emitters}
\label{appA}

Throughout this work, we have distinguished between DLA \lya\ emitters, Group \lya\ emitters, and Large-scale \lya\ emitters. The properties of DLA \lya\ emitters in KAGG and MAGG are presented in Table \ref{table_DLA_lya_emitters}, which is part of the main text. Here, we present the properties of Group \lya\ emitters and Large-scale \lya\ emitters. The former are in Table \ref{table_group_lya_emitters} and the latter in Table \ref{table_large_scale_lya_emitters}. Lastly, we present in Table \ref{table_excluded_lya_emitters} the properties of 6 line emitters that we excluded from our analysis due to their large velocity offsets ($\approx 4,000-12,000$~km~s$^{-1}$) relative to the DLA redshifts.

\startlongtable
\begin{deluxetable*}{cccccccc}
\tabletypesize{\normalsize} 
\vspace{-0.5em}
\tablecaption{Properties of the 35 Group \lya\ emitters in the KAGG and MAGG surveys. The columns of the table are (1)~the detection identifier, (2)~the DLA redshift, (3, 4)~the J2000 coordinates of the \lya\ emission, (5)~the velocity offset $\Delta v$ between the \lya\ redshift and the DLA redshift in km~s$^{-1}$ (6)~the \lya\ flux in $10^{-18}$~erg~s$^{-1}$~cm$^{-2}$, (7)~the impact parameter to the QSO sightline at the DLA redshift in kpc, and (8)~the sample (KAGG or MAGG).}
\label{table_group_lya_emitters} 
\setlength{\tabcolsep}{0.08in} 
\tablehead{
\colhead{Identifier} & \colhead{z$_{abs}$} & \colhead{ra}  & \colhead{dec} & \colhead{vel. offset} & \colhead{\lya\ flux} & \colhead{b} & \colhead{sample of} \vspace{-0.3em} \\
\colhead{} & \colhead{} & \colhead{[hours]}  & \colhead{[degrees]} & \colhead{[km~s$^{-1}$]} & \colhead{[$10^{-18}$~erg~s$^{-1}$~cm$^{-2}$]} & \colhead{[kpc]} & \colhead{origin}} 
\startdata 
PSS0007+24a & 3.4959 & 00:07:38.77 & 24:17:25.80 & $-140$ & $4.0 \pm 2$ & $14 \pm 10$ & KAGG \\
PSS0007+24b & 3.4959 & 00:07:37.85 & 24:17:24.30 & $300$ & $7.9 \pm 4$ & $81 \pm 10$ & KAGG \\
J0255+0048a & 3.2529 & 02:55:18.81 & 00:48:49.00 & $170$ & $19.6 \pm 9$ & $33 \pm 10$ & KAGG \\
FJ0747+2739a & 3.4237 & 07:47:10.55 & 27:39:04.80 & $210$ & $10.9 \pm 6$ & $62 \pm 10$ & KAGG\\
PC0953+47a & 3.8907 & 09:56:25.07 & 47:34:48.00 & $-520$ & $12.6 \pm 6$ & $40 \pm 10$ & KAGG \\
PC0953+47b & 3.8907 & 09:56:23.69 & 47:34:51.00 & $-410$ & $12.1 \pm 7$ & $122 \pm 10$ & KAGG \\
J1410+5111a & 2.9642 & 14:10:31.41 & 51:11:12.60 & $-20$ & $1.7 \pm 1$ & $65 \pm 11$ & KAGG \\
PSS1535+2943a & 3.2020 & 15:35:53.88 & 29:43:12.70 & $290$ & $3.0 \pm 2$ & $15 \pm 10$ & KAGG \\
PSS1535+2943b & 3.7612 & 15:35:52.92 & 29:43:18.20 & $230$ & $2.0 \pm 1$ & $85 \pm 10$ & KAGG \\
\hline
J0133+0400a & $3.6916$ & 01:33:39.69 & 04:00:43.25 & $291$ & $13.6 \pm 1$ & $140$  & MAGG \\
J0133+0400b & $3.6916$ & 01:33:38.32 & 04:00:46.16 & $711$ & $18.2 \pm 2$ & $243$  & MAGG \\
J0133+0400c & $3.7724$ & 01:33:40.691 & 04:01:12.92 & $713$ & $3.5 \pm 1$ & $103$  & MAGG \\
J0209+0517a & $3.6659$ & 02:09:43.40 & 05:16:51.55 & $-83$ & $2.7 \pm 1$ & $211$  & MAGG \\
J0334-1612a & $3.5566$ & 03:34:14.78 & -16:12:25.05 & $131$ & $3.1 \pm 1$ & $204$  & MAGG \\
J0339-0133a & $3.0622$ & 03:39:02.31 & -01:33:26.71 & $653$ & $3.9 \pm 1$ & $168$  & MAGG  \\
J0851+2332a  & $3.5297$ & 08:51:43.60 & 23:32:12.00 & $262$ & $10.0 \pm 1$ & $24$  & MAGG \\
J1111-0804a & $3.6077$ & 11:11:13.75 & -08:04:11.05 & $57$ & $8.5 \pm 1$ & $67$  & MAGG  \\
J1111-0804b & $3.6077$ & 11:11:13.23 & -08:04:13.34 & $291$ & $14.8 \pm 2$ & $104$  & MAGG  \\
J1111-0804c & $3.6077$ & 11:11:12.91 & -08:04:10.38 & $307$ & $5.7 \pm 1$ & $115$  & MAGG  \\
J1111-0804d & $3.6077$ & 11:11:13.37 & -08:04:27.89 & $175$ & $12.6 \pm 2$ & $197$  & MAGG  \\
J1111-0804e & $3.6077$ & 11:11:13.25 & -08:04:30.18 & $212$ & $9.7 \pm 2$ & $217$  & MAGG  \\
J1220+0921a & $3.3090$ & 12:20:21.80 & 09:21:17.00 & $-279$ & $14.6 \pm 1$ & $150$  & MAGG  \\
J1220+0921b & $3.3090$ & 12:20:20.30 & 09:21:52.00 & $229$ & $13.4 \pm 1$ & $175$  & MAGG \\
J1220+0921c & $3.3090$ & 12:20:23.10 & 09:21:10.00 & $370$ & $27.4 \pm 1$ & $277$  & MAGG \\
J2315+1456a & $3.2732$ & 23:15:44.39 & 14:56:01.40 & $244$ & $3.9 \pm 1$ & $97$  & MAGG  \\
J2315+1456b & $3.2732$ & 23:15:45.46 & 14:55:56.08 & $221$ & $15.4 \pm 2$ & $222$  & MAGG  \\
J2315+1456c & $3.2732$ & 23:15:44.90 & 14:55:38.24 & $288$ & $50.9 \pm 2$ & $260$  & MAGG  \\
J2315+1456d & $3.2732$ & 23:15:45.31 & 14:56:30.15 & $251$ & $27.4 \pm 2$ & $265$  & MAGG  \\
J2334-0908a & $3.0572$ & 23:34:45.64 & -09:08:07.48 & $212$ & $15.1 \pm 2$ & $98$  & MAGG  \\
J2334-0908b & $3.0572$ & 23:34:48.08 & -09:08:08.97 & $223$ & $36.6 \pm 2$ & $194$  & MAGG  \\
J2334-0908c & $3.0572$ & 23:34:45.44 & -09:07:52.05 & $284$ & $17.6 \pm 2$ & $195$  & MAGG  \\
J2334-0908d & $3.0572$ & 23:34:44.83 & -09:07:59.96 & $158$ & $16.1 \pm 2$ & $208$  & MAGG  \\
J2334-0908e & $3.0572$ & 23:34:45.19 & -09:07:51.34 & $-340$ & $25.8 \pm 2$ & $218$  & MAGG  \\
J2334-0908f & $3.0572$ & 23:34:48.15 & -09:08:00.53 & $87$ & $11.7 \pm 2$ & $221$  & MAGG  \\
J2334-0908g & $3.0572$ & 23:34:47.18 & -09:07:45.77 & $817$ & $21.4 \pm 2$ & $226$  & MAGG  \\
\enddata
\tablecomments{The quoted errors on the KAGG line fluxes include uncertainties in the flux calibration. Hence, these errors are significantly larger than the line detection uncertainty.}
\end{deluxetable*}

\begin{deluxetable*}{cccccccc}
\tabletypesize{\normalsize} 
\tablecaption{Properties of the 17 Large-scale \lya\ emitters in the KAGG survey. The columns of the table are (1)~the detection identifier, (2)~the DLA redshift, (3, 4)~the J2000 coordinates of the \lya\ emission, (5)~the velocity offset $\Delta v$ between the \lya\ redshift and the DLA redshift in km~s$^{-1}$ (6)~the \lya\ flux in $10^{-18}$~erg~s$^{-1}$~cm$^{-2}$, (7)~the impact parameter to the QSO sightline at the DLA redshift in kpc, and (8)~the sample (KAGG or MAGG).}
\label{table_large_scale_lya_emitters} 
\setlength{\tabcolsep}{0.05in} 
\tablehead{
\colhead{Identifier} & \colhead{z$_{abs}$} & \colhead{ra}  & \colhead{dec} & \colhead{vel. offset} & \colhead{\lya\ flux} & \colhead{b} & \colhead{sample of} \vspace{-0.3em} \\
\colhead{} & \colhead{} & \colhead{[hours]}  & \colhead{[degrees]} & \colhead{[km~s$^{-1}$]} & \colhead{[$10^{-18}$~erg~s$^{-1}$~cm$^{-2}$]} & \colhead{[kpc]} & \colhead{origin}} 
\startdata 
J0929+2825a & 3.2627 & 09:29:14.27 & 28:25:34.60 & $1190$ & $18.0 \pm 8$ & $47 \pm 10$ & KAGG\\
J0929+2825b & 3.2627 & 09:29:13.36 & 28:25:25.10 & $-1880$ & $5.3 \pm 2$ & $119 \pm 10$ & KAGG \\
J0929+2825c & 3.2627 & 09:29:15.83 & 28:25:25.10 & $1450$ & $24.0 \pm 10$ & $139 \pm 10$ & KAGG\\
PSS0957+33a & 3.2796 & 09:57:44.63 & 33:08:13.30 & $-580$ & $5.6 \pm 3$ & $56 \pm 10$ & KAGG\\
J1054+1633a & 4.8166 & 10:54:45.81 & 16:33:30.90 & $-1800$ & $3.2 \pm 2$ & $57 \pm 9$ & KAGG\\
J1132+1209a & 4.3802 & 11:32:45.34 & 12:08:54.70 & $-1480$ & $20.3 \pm 9$ & $121 \pm 9$  & KAGG\\
J1132+1209b & 5.0165 & 11:32:45.44 & 12:09:01.70 & $3450$ & $2.8 \pm 2$ & $96 \pm 9$ & KAGG\\
J1201+2117a & 3.7975 & 12:01:11.06 & 21:18:08.10 & $1840$ & $7.4 \pm 4$ & $105 \pm 10$ & KAGG\\
J1201+2117b & 4.1578 & 12:01:09.88 & 21:18:08.10 & $3470$ & $6.4 \pm 3$ & $77 \pm 9$  & KAGG\\
J1201+2117c & 4.1578 & 12:01:09.31 & 21:17:59.60 & $3100$ & $4.6 \pm 2$ & $96 \pm 9$ & KAGG\\
J1202+3235a & 4.7955 & 12:02:07.31 & 32:35:47.30 & $-1760$ & $12.5 \pm 6$ & $64 \pm 9$ &  KAGG\\
J1202+3235b & 5.0647 & 12:02:08.29 & 32:35:39.30 & $-1720$ & $55.0 \pm 22$ & $43 \pm 9$ &  KAGG\\
J1410+5111b & 2.9344 & 14:10:31.57 & 51:11:18.10 & $-2740$ & $4.0 \pm 2$ & $82 \pm 11$ & KAGG\\
PSS1535+2943c & 3.2020 & 15:35:53.88 & 29:43:12.70 & $-620$ & $4.0 \pm 2$ & $15 \pm 10$ & KAGG\\
PSS1535+2943d & 3.7612 & 15:35:53.03 & 29:43:18.20 & $1420$ & $4.5 \pm 2$ & $75 \pm 10$ & KAGG\\
J1626+2751a & 4.3110 & 16:26:26.27 & 27:51:29.00 & $-2480$ & $1.7 \pm 1$ & $26 \pm 9$ &  KAGG\\
PSS1802+5616a & 3.3912 & 18:02:48.06 & 56:16:46.00 & $1870$ & $11.5 \pm 5$ & $60 \pm 10$ &  KAGG\\
\enddata
\tablecomments{The quoted errors on the KAGG line fluxes include uncertainties in the flux calibration. Hence, these errors are significantly larger than the line detection uncertainty.}
\end{deluxetable*}

\begin{deluxetable*}{cccccccc}
\tabletypesize{\normalsize} 
\tablecaption{Properties of the 6 \lya\ emitters excluded from the analysis due to their large velocity offsets relative to the DLA redshifts. The columns of the table are (1)~the detection identifier, (2)~the DLA redshift, (3, 4)~the J2000 coordinates of the \lya\ emission relative to the DLA, (5)~the velocity offset $\Delta v$ between the \lya\ redshift and the DLA redshift in km~s$^{-1}$ (6)~the \lya\ flux in $10^{-18}$~erg~s$^{-1}$~cm$^{-2}$, (7)~the impact parameter to the QSO sightline at the DLA redshift in kpc, and (8)~the sample (KAGG or MAGG).}
\label{table_excluded_lya_emitters} 
\setlength{\tabcolsep}{0.05in} 
\tablehead{
\colhead{Identifier} & \colhead{z$_{abs}$} & \colhead{ra}  & \colhead{dec} & \colhead{vel. offset} & \colhead{\lya\ flux} & \colhead{b} & \colhead{sample of} \vspace{-0.3em} \\
\colhead{} & \colhead{} & \colhead{[hours]}  & \colhead{[degrees]} & \colhead{[km~s$^{-1}$]} & \colhead{[$10^{-18}$~erg~s$^{-1}$~cm$^{-2}$]} & \colhead{[kpc]} & \colhead{origin}} 
\startdata 
FJ0747+2739b & 3.9004 & 07:47:11.26 & 27:39:00.80 & $15,390$ & $74.0 \pm 31$ & $22 \pm 10$ & KAGG\\
FJ0747+2739c & 3.9004 & 07:47:11.34 & 27:39:04.80 & $15,420$ & $79.0 \pm 33$ & $22 \pm 10$ & KAGG\\
PC0953+47c & 4.2439 & 09:56:23.44 & 47:34:38.50 & $15,510$ & $44.5 \pm 22$ & $126 \pm 9$ & KAGG \\
J1132+1209c & 4.3802 & 11:32:45.82 & 12:09:11.20 & $13,770$ & $9.6 \pm 4$ & $92 \pm 9$ & KAGG \\
PSS1535+2943e & 3.2020 & 15:35:54.68 & 29:43:12.70 & $14,830$ & $11.6 \pm 5$ & $95 \pm 10$ & KAGG\\
PSS1535+2943f & 3.2020 & 15:35:54.68 & 29:43:09.70 & $-12,100$ & $7.3 \pm 3$ & $100 \pm 10$ & KAGG
\enddata
\tablecomments{The quoted errors on the KAGG line fluxes include uncertainties in the flux calibration. Hence, these errors are significantly larger than the line detection uncertainty.}
\end{deluxetable*}

\newpage

\section{Bayesian analysis of the dependence of Group \lya\ luminosity on DLA properties}
\label{appB}

Interpreting any variations in \lya\ luminosity with redshift, DLA metallicity, \NHI, and impact parameter is challenging for multiple reasons: (i) it is a multivariate space, (ii)~our \lya\ luminosity sensitivity varies with redshift (and therefore with DLA metallicity) and (iii)~KAGG and MAGG have different impact parameter coverage. Thus, in order to properly constrain any correlations in \lya\ luminosity with DLA properties --- as well as establishing their significance --- we fit a model to the observed $L_{\mbox{\small\lya}}$.

Hereafter, we denote the dataset (KAGG and MAGG) as $\mathcal{D} = \{L_{\mbox{\small\lya}}^i, \delta L_{\mbox{\small\lya}}^i, \Delta L_{\mbox{\small\lya}}^i\}$. Here, $L_{\mbox{\small\lya}}^i$ corresponds to the observed luminosities, $\delta L_{\mbox{\small\lya}}^i$ represents the error on the luminosities, and $\Delta L_{\mbox{\small\lya}}^i$ signifies the $1\sigma$ luminosity upper limits. For KAGG, the $\Delta L_{\mbox{\small\lya}}^i$ were computed from the measured 2$\sigma$ detection limits (Section \ref{3.2}) for every target. For MAGG, we adopted different 2$\sigma$ detection limits depending on the discovery paper: $L_{\mbox{\small\lya}} = 10^{42}$~erg~s$^{-1}$ for \citet[a line luminosity cut]{mackenzie2019} and $F_{\mbox{\small\lya}} = 10^{-17.2}$~erg~s$^{-1}$~cm$^{-2}$ for \citet[a line flux cut]{lofthouse2023}.

We denote our model $\mathcal{M} = \{P_0, \{\theta\}\}$. Here, $P_0$ quantifies the probability that a spaxel features \lya\ emission and $\{\theta\}$ captures all the parameters that fit the observed $L_{\mbox{\small\lya}} = L_{\mbox{\small\lya}}(z, \mbox{[M/H]}, \mbox{\NHI}, b \mid \{\theta\})$. In strict terms, $P_0$ corresponds to the probability that a $1\farcs 35$ spaxel ($\approx 10$~kpc at $z\sim 4$) features \lya\ emission within 500~km~s$^{-1}$ of the DLA redshift. We set the threshold between emission and no-emission as $L_{\mbox{\small\lya}} = 10^{41}$~erg~s$^{-1}$, which lies just below the KAGG and MAGG line detection limit. We can therefore write $P_0$ as 
\begin{equation}
P_0 = \mathbb{P}(L_{\mbox{\small\lya}} \geq 10^{41}~\mbox{erg~s}^{-1} \mid \{\theta\})
\end{equation}

The model $L_{\mbox{\small\lya}} = L_{\mbox{\small\lya}}(\theta)$ fits how \lya\ luminosities $L_{\mbox{\small\lya}} > 10^{41}~\mbox{erg~s}^{-1}$ vary with galaxy properties. We choose a linear model given by:
\begin{equation}
\begin{split}
L_{\mbox{\small\lya}} [1&0^{42}~\mbox{erg~s}^{-1}] = \mathcal{M}(z, \mbox{[M/H]}, \mbox{\NHI}, b) = \\
\theta_0 + \theta_{z} z  \, + \,  &\theta_{\mbox{[M/H]}} \,\mbox{[M/H]} +  \theta_{\mbox{\NHI}} \, \rm log \mbox{\NHI} + \theta_{b} \, b \, \, + \\ &\mathcal{N}(0, \theta_\Sigma^2),
\end{split}
\end{equation}
where $\theta_0$ is a normalization parameter, $\theta_i$ represents the linear dependence of the \lya\ luminosity on various properties, and $\theta_\Sigma$ corresponds to the standard deviation of normally distributed intrinsic scatter. This term encapsulates variations in the \lya\ luminosity that cannot be directly associated with the measured properties of the DLA (e.g., stochasticity in the galaxy star-formation histories, variations in the escape fraction of \lya\ photons with the line of sight, etc.).

We constrain our model by fitting all spaxels in KAGG and in MAGG. For the MAGG data, we created empty grids in celestial coordinates with $1\farcs 35$ spaxel resolutions, thus matching the KAGG data, and we populated these grids with the detected $v_{500}$ \lya\ emitters. For simplicity, we assumed that all sources (in KAGG and in MAGG) are contained within a single spaxel. We set the \lya\ luminosity upper limit of all spaxels with non-detections to the reported \lya\ luminosity upper limits, i.e., we assume no luminosity variation throughout the field of view of any target. 

We chose flat, bounded prior probability distributions for all parameters. Formally,

\begin{equation}
\begin{split}
&\mathbb{P}(P_0) \propto H(P_0 - 0)\,H(1-P_0), \\
&\mathbb{P}(\theta_i) \propto H(\theta_i - 100)\,H(100-\theta_i), \\
&\mathbb{P}(\theta_\Sigma)  \propto H(\theta_\Sigma - 0)\,H(10-\theta_\Sigma),
\end{split}
\end{equation}
where
\begin{equation}
H(x) =
\begin{cases}
0, & x < 0 \\
1, & x \ge 0
\end{cases}
\end{equation}
is the Heaviside function.  

For evaluating the likelihood function $\mathcal{L}(\mathcal{D} \mid \mathcal{M})$, we distinguished between spaxels with detections and non-detections. For spaxels with detections, the likelihood $\mathcal{L}_{\rm det}^i(\mathcal{D} \mid \mathcal{M})$ is composed of two terms: (i)~the probability of observing a line with luminosity $L_{\mbox{\small\lya}, \, i}$ and error $\delta L_{\mbox{\small\lya}, \, i}$ given the model and (ii)~the probability that the detection has $L_{\mbox{\small\lya}} < 10^{41}~\mbox{erg~s}^{-1}$. This yields the following expression for the likelihood of a spaxel with a detection: 
\begin{equation}
\begin{split}
\mathcal{L}_{\rm det}^i(\mathcal{D} \mid &\mathcal{M}) = \\P_0 \ \mathcal{N}(&L_{\mbox{\small\lya}, \,i} \mid L_{\mbox{\small\lya}}(\{\theta \}), \, \delta L_{\mbox{\small\lya}, \,i}^2 + {\theta_\Sigma}^2)\ + \\
(1 - P_0) &\int_{-\infty}^{10^{41}} \mathcal{N}(L_{\mbox{\small\lya}, \,i} \mid L_{\mbox{\small\lya}}, {\delta L_{\mbox{\small\lya}, \,i}^2}) \,dL_{\mbox{\small\lya}}
\end{split}
\end{equation}
where $\mathcal{N}(x \mid \mu, \sigma^2)$ is a normal distribution with mean $\mu$ and standard deviation $\sigma$ valued at $x$. 

For spaxels with non-detections, the likelihood $\mathcal{L}_{\rm non-det}^i(\mathcal{D} \mid \mathcal{M})$ is composed of (i)~the probability that a line with luminosity $ L_{\mbox{\small\lya}} = L_{\mbox{\small\lya}}(\theta)$ was undetected and (ii)~the probability that the line was not detected because it was fainter than the $ L_{\mbox{\small\lya}} < 10^{41}~\mbox{erg~s}^{-1}$ threshold. Formally,
\begin{equation}
\begin{split}
\mathcal{L}_{\rm non-det}^i(\mathcal{D} \mid &\mathcal{M}) = \\P_0 \ \mathcal{N}(& 0 \mid L_{\mbox{\small\lya}}(\{\theta \}), \, \Delta L_{\mbox{\small\lya}, \,i}^2 + {\theta_\Sigma}^2)\ + \\
(1 - P_0) &\int_{-\infty}^{10^{41}} \mathcal{N}(0 \mid L_{\mbox{\small\lya}}, {\Delta L_{\mbox{\small\lya}, \,i}^2}) \,dL_{\mbox{\small\lya}}
\end{split}
\end{equation}

Although our model allows for 7 parameters, we only constrained the posterior distributions of models with 5 free parameters at a time. The posterior distributions were constrained with the Markov Chain Monte Carlo code \texttt{emcee} (\citealt{emcee}). For 5-parameter posteriors, we found that a setup with 500~walkers, 350~burn-in steps, and 150~steps after burn-in led to converged, well-sampled solutions.  

 \begin{figure*}[b!]
\centering 
\includegraphics[width=7in]{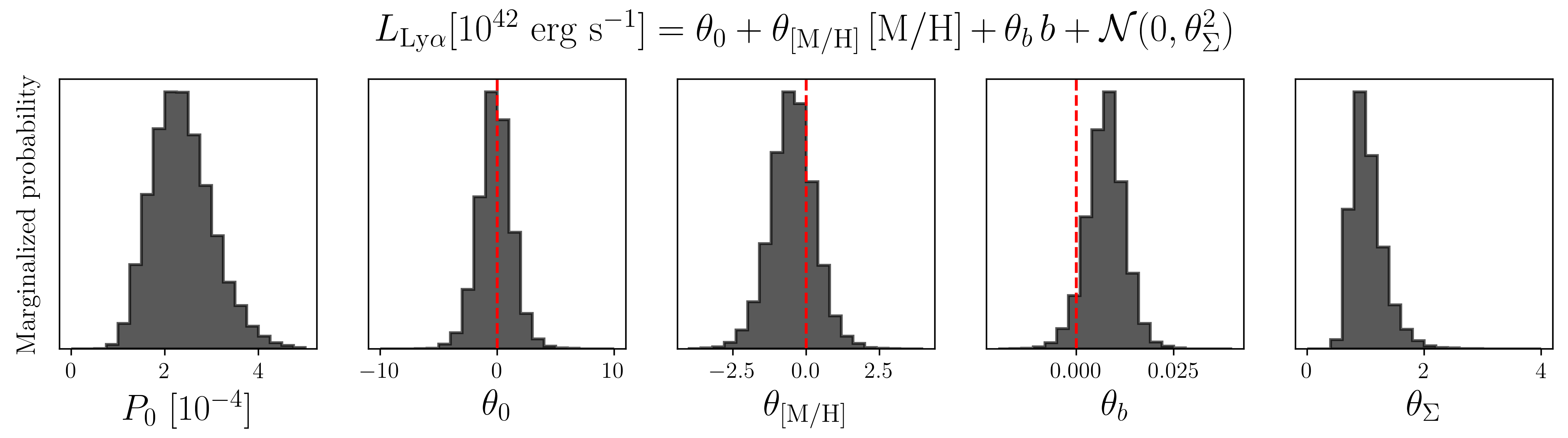}\\
\caption{Marginalized posterior distribution of the parameters of model $\mathcal{M} = \{P_0, \{\theta\}\}$ given the data from KAGG and MAGG. The parameter $P_0$ quantifies the probability that spaxels feature \lya\ emission with $L_{\mbox{\small\lya}} > 10^{41}$~erg~s$^{-1}$. The parameters $\theta_{\rm [M/H]}$ and $\theta_{b}$ quantify the linear dependence of the \lya\ luminosity on DLA metallicity and impact parameter, respectively. The normalization parameter is $\theta_0$ and the intrinsic scatter is $\theta_\Sigma$. The data indicate that the \lya\ luminosity of galaxies around DLAs (i)~are an univariate function of DLA properties, (ii) most strongly depend on impact parameter ($\approx 2\sigma$ significance), and (iii)~have an intrinsic scatter of $L_{\mbox{\small\lya}}\approx 10^{42}$~erg~s$^{-1}$.}
\label{fig_model}
\end{figure*}

We first fit the model assuming only variations in \lya\ luminosity with DLA metallicity and impact parameter, i.e., $L_{\mbox{\small\lya}} = L_{\mbox{\small\lya}}(\mbox{[M/H]}, b \mid \theta_0, \theta_{\mbox{[M/H]}}, \theta_b, \theta_\Sigma)$. As shown in Figure \ref{fig_model}, the emission probability per spaxel is low ($P_0 \lesssim 0.1\%$). The dataset is best reproduced by a model with no significant DLA metallicity component and a $1.9\sigma$ increase in the \lya\ luminosity with impact parameter. The posterior distribution also indicates the presence of significant intrinsic scatter ($\theta_\Sigma \approx 10^{42}$~erg~s$^{-1}$). Then, we fit models that account for variations in \lya\ luminosity with impact parameter and redshift, as well as with impact parameter and \NHI. We found the same result in all cases: the data are best reproduced by variations in the \lya\ luminosity with impact parameter ($\approx 1.9\sigma$ significance), and that the intrinsic scatter in the \lya\ luminosity is significant and with magnitude $\theta_\Sigma \approx 10^{42}$~erg~s$^{-1}$. Our interpretation for this large intrinsic scatter is that the \lya\ luminosities of DLA galaxies depend on several physical processes, including the nuances of the baryon cycle, the stochasticity of galaxy evolution, and the complexities of \lya\ radiation escape from galactic environments.

Our analysis in this section assumes that DLA metallicity, \NHI, and impact parameter are independent variables. This assumption is not entirely correct, since the impact parameters of \lya\ emitters are anti-correlated with DLA metallicity and \NHI\ (Section \ref{5.2}). Therefore, the appropriate conclusion from our modeling is that the \lya\ luminosities measured by KAGG and MAGG are best reproduced by a function that is univariate, with this single variable being impact parameter. To confirm this, we repeated our analysis by fitting univariate functions to the data, i.e., models with only 4 parameters: $\mathcal{M} = \{P_0, \theta_0, \theta_i, \theta_\Sigma \}$. For the impact parameter case ($\theta_i = \theta_b$), we detect a correlation with $L_{\mbox{\small\lya}}$ at the $2.3 \sigma$ level. For the DLA metallicity and \NHI\ cases (i.e., $\theta_i = \theta_{\rm [M/H]}$ and $\theta_i = \theta_{\mbox{\NHI}}$), we detect anti-correlations with $L_{\mbox{\small\lya}}$ at the $\approx 1.5 \sigma$ level.

\section{KAGG data}
\label{appC}
\begin{figure*}[h!]
\centering 
\includegraphics[width=\textwidth]{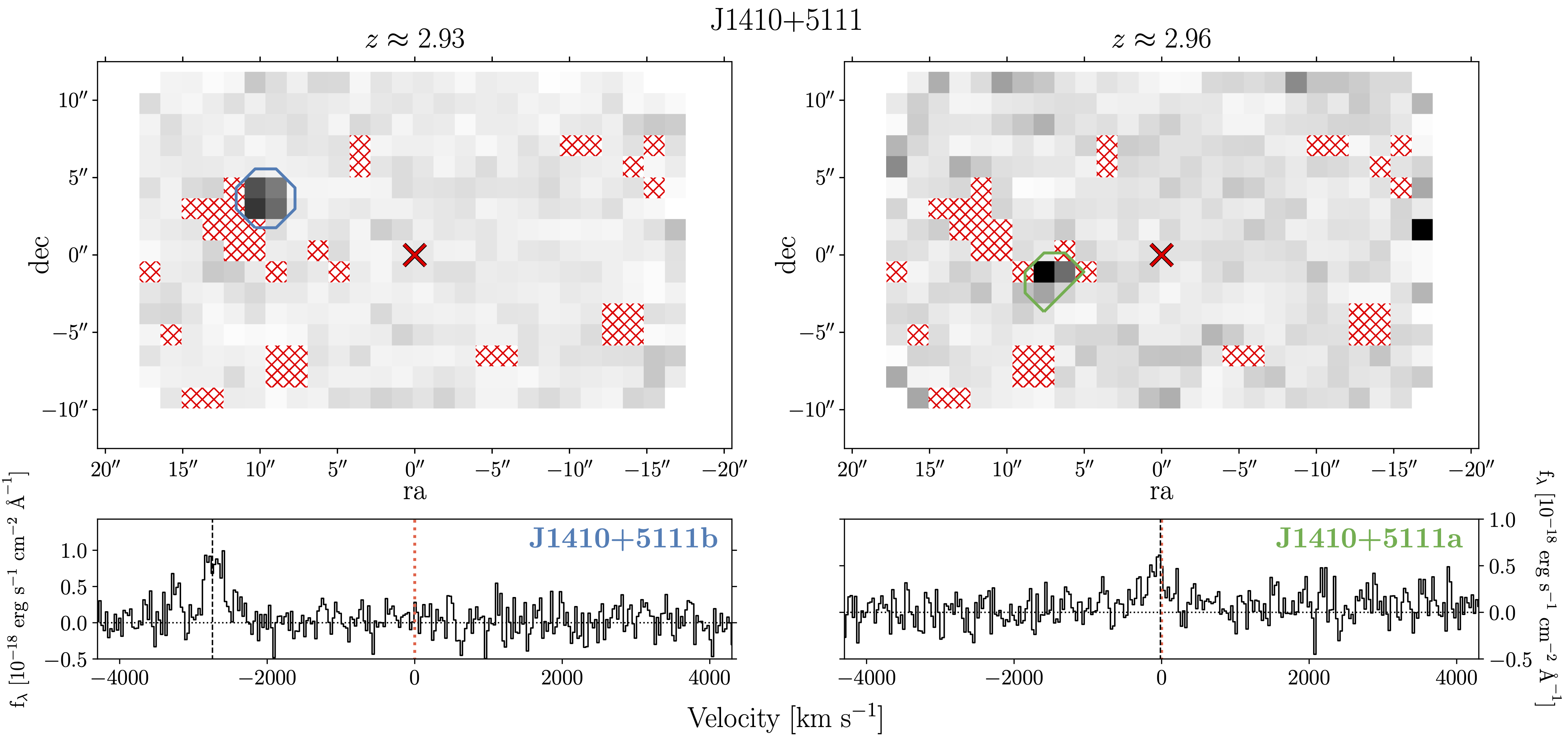}\\
\vspace{-0.5em}
\caption{KAGG data for the QSO sightline J1410+5111. See Appendix \ref{appC} for details.}
\label{fig_J1410}
\end{figure*}

\begin{figure*}[h!]
\centering 
\includegraphics[width=\textwidth]{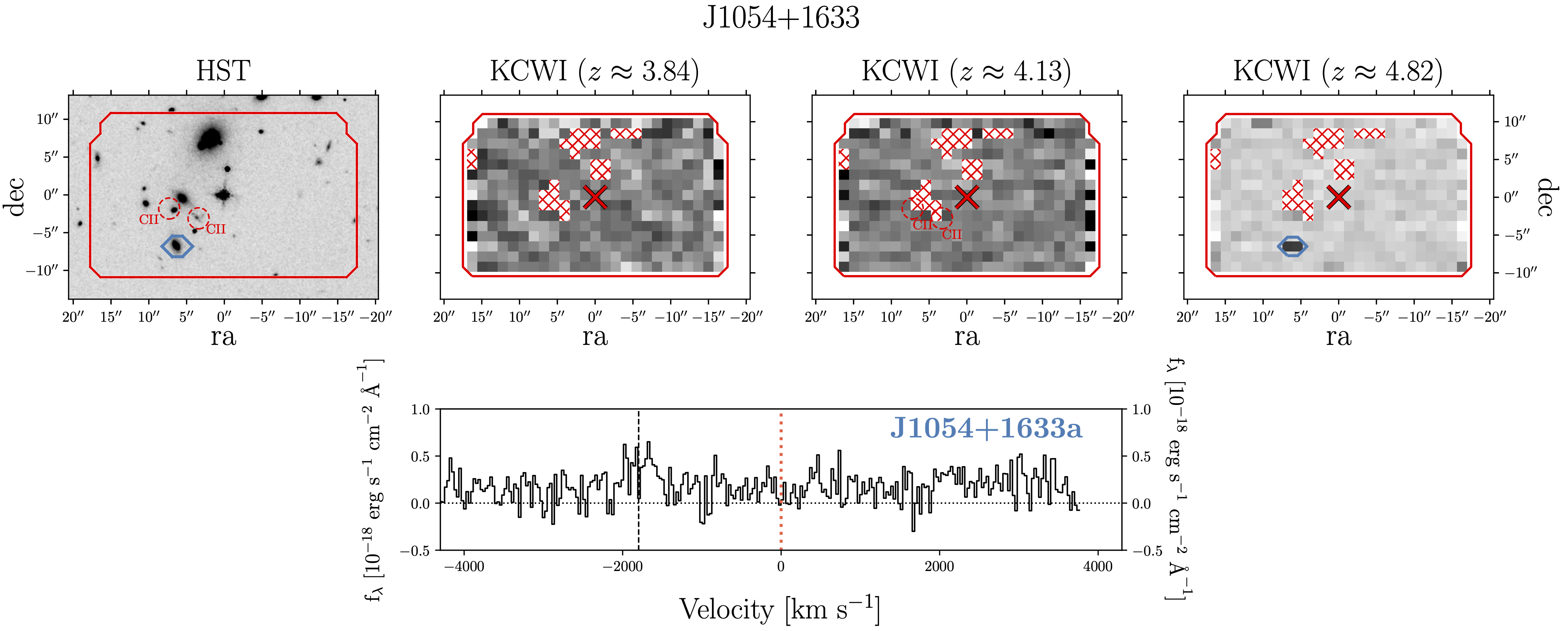}\\
\caption{KAGG data for the QSO sightline J1054+1633. See Appendix \ref{appC} for details.}
\vspace{-0.15in}
\label{fig_J1054}
\end{figure*}

\begin{figure*}[h!]
\centering 
\includegraphics[width=\textwidth]{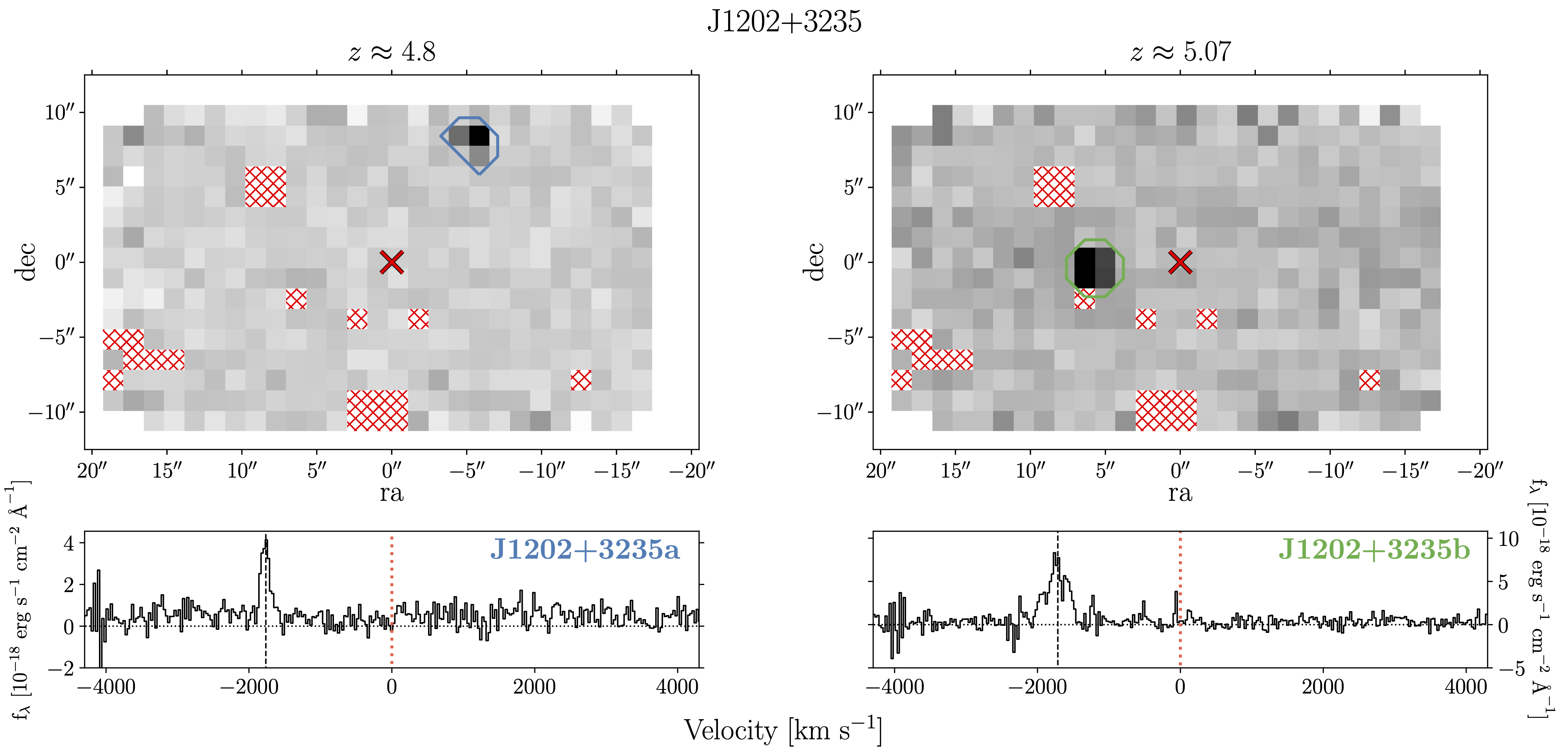}\\
\caption{KAGG data for the QSO sightline J1202+3235. See Appendix \ref{appC} for details.}
\vspace{-0.15in}
\label{fig_J1202}
\end{figure*}

We present in Figures \ref{fig_J1410} through \ref{fig_J1202} the KCWI images and spectra for all detections, with the exception of J0255+0048 (see Figure \ref{fig_J0255}). The top panels show the integrated KCWI datacubes within $\pm100$~km~s$^{-1}$ of the DLA absorption redshift for the non-detection spaxels or within the $\pm 3\sigma$ of the \lya\ emission line width for the detections. Indicated are also the QSO and DLA as the red cross, continuum and/or line emission unassociated with the DLA as hatched regions, and \cii\ emitters identified in the sub-millimeter as red circles. We also include in the top panels HST imaging in the HST/F140W or HST/F160W filters when available. The lower panels show the spectra of the respective \lya\ emitters. All the {\it HST} data used in this paper can be found in MAST: \dataset[10.17909/wkh6-0w77]{http://dx.doi.org/10.17909/wkh6-0w77}.

\newpage

 \begin{figure*}[h!]
\centering 
\includegraphics[width=0.98\textwidth]{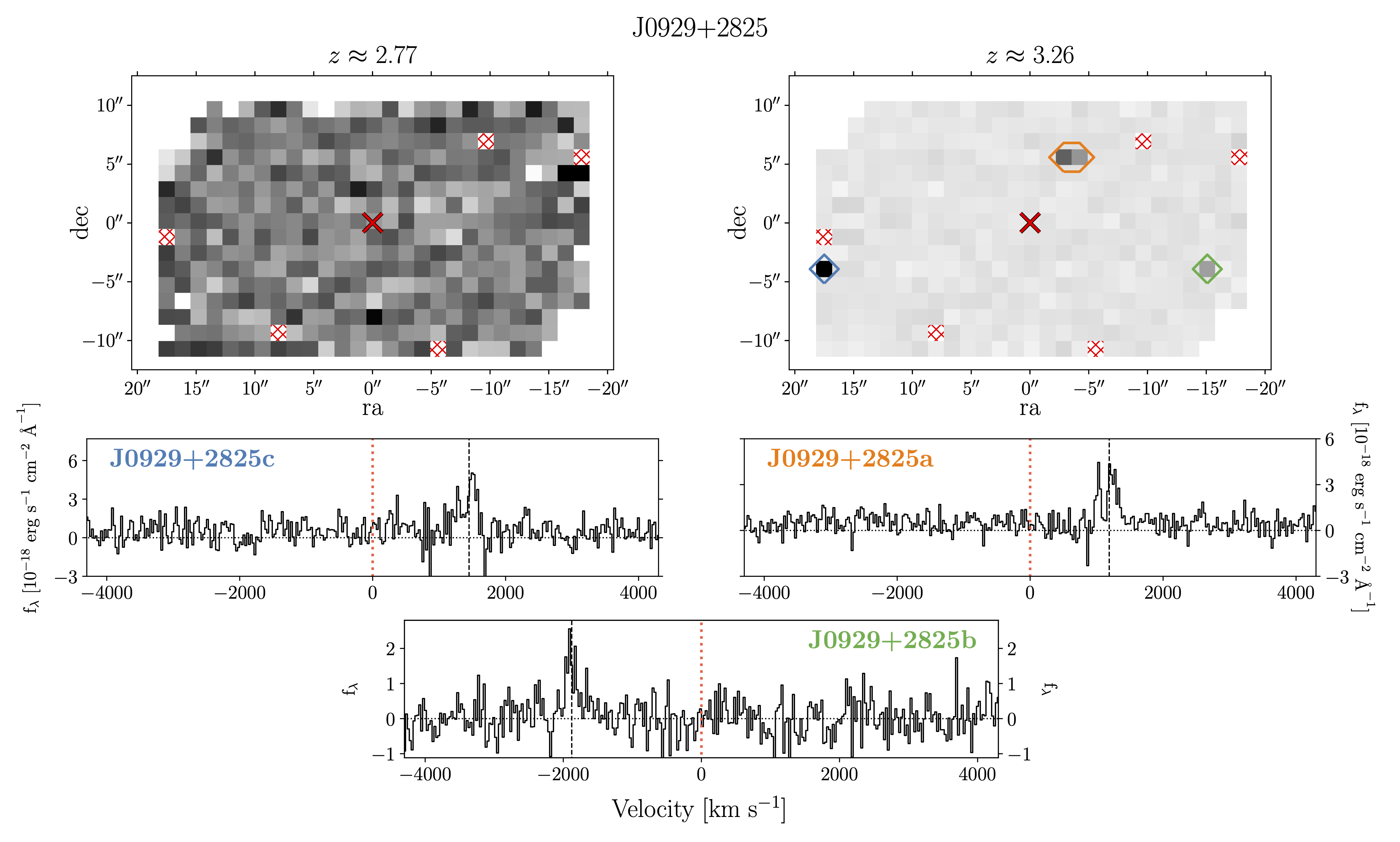}\\
\vspace{-1em}
\caption{KAGG data for the QSO sightline J0929+2825. See Appendix \ref{appC} for details.}
\vspace{-0.5em}
\end{figure*}

 \begin{figure*}[h!]
\centering 
\vspace{-1em}
\includegraphics[width=0.98\textwidth]{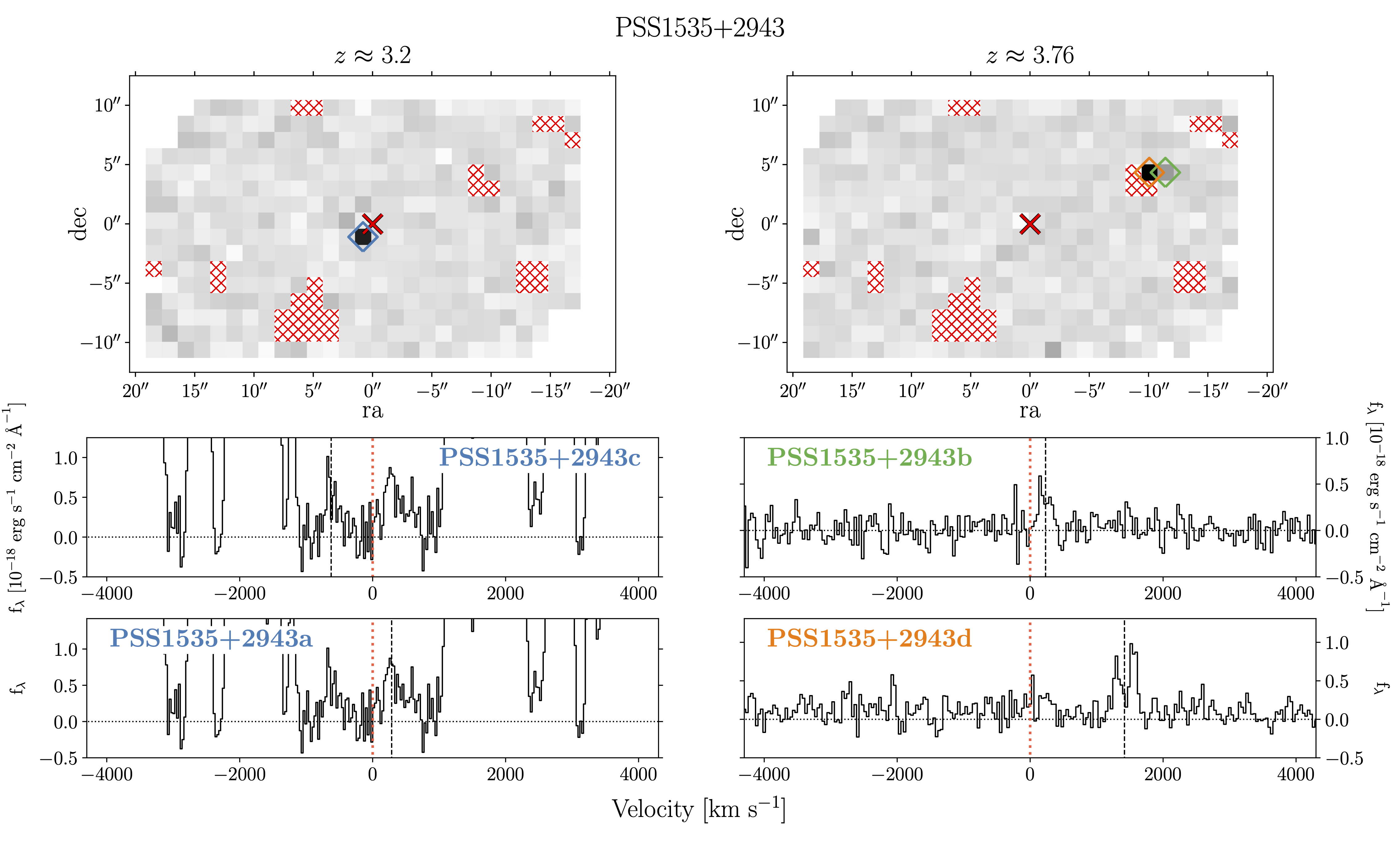}\\
\vspace{-1em}
\caption{KAGG data for the QSO sightline PSS1535+2943. See Appendix \ref{appC} for details.}
\end{figure*}

\newpage

 \begin{figure*}[h!]
\centering 
\includegraphics[width=\textwidth]{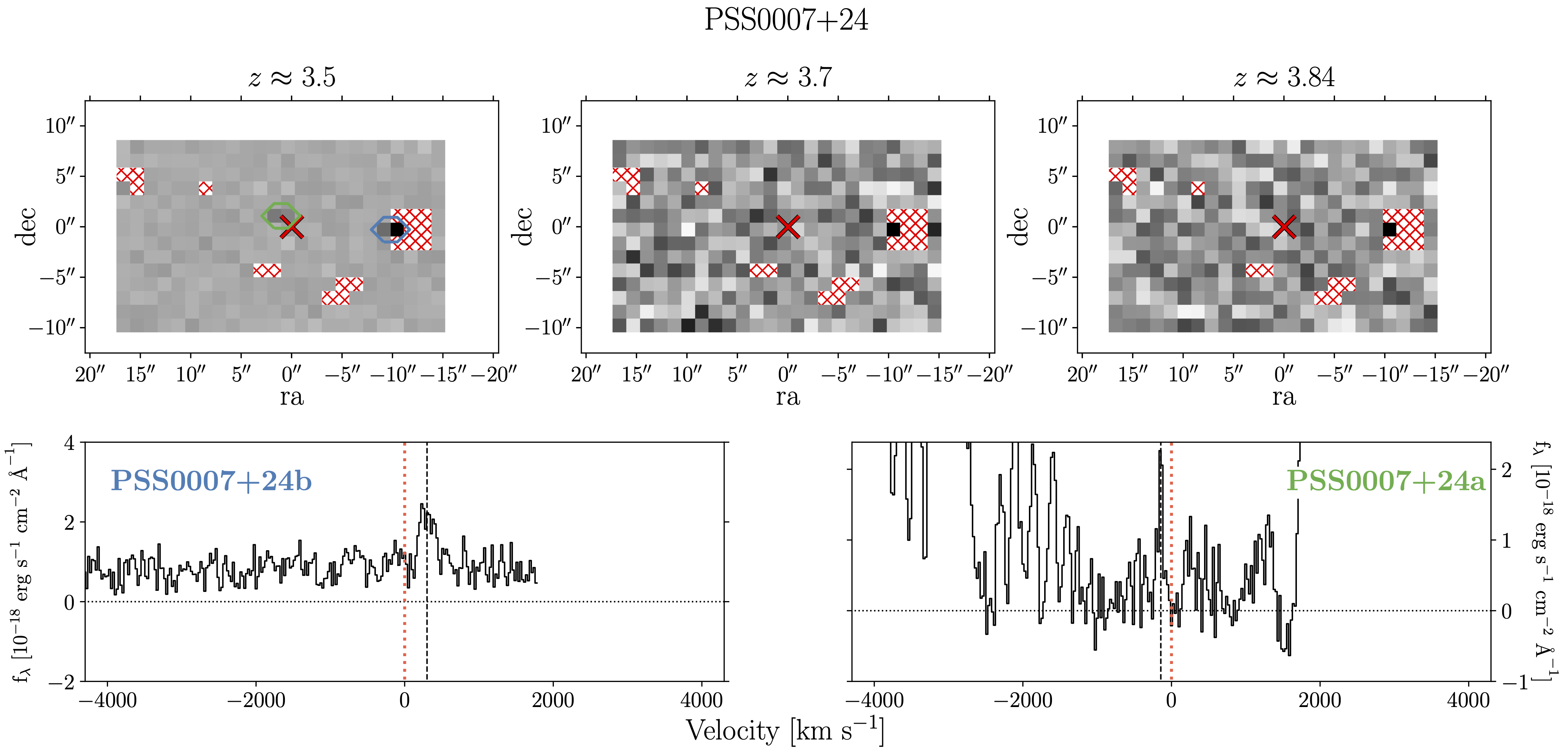}\\
\caption{KAGG data for the QSO sightline PSS0007+24. See Appendix \ref{appC} for details.}
\end{figure*}

 \begin{figure*}[h!]
\centering 
\includegraphics[width=\textwidth]{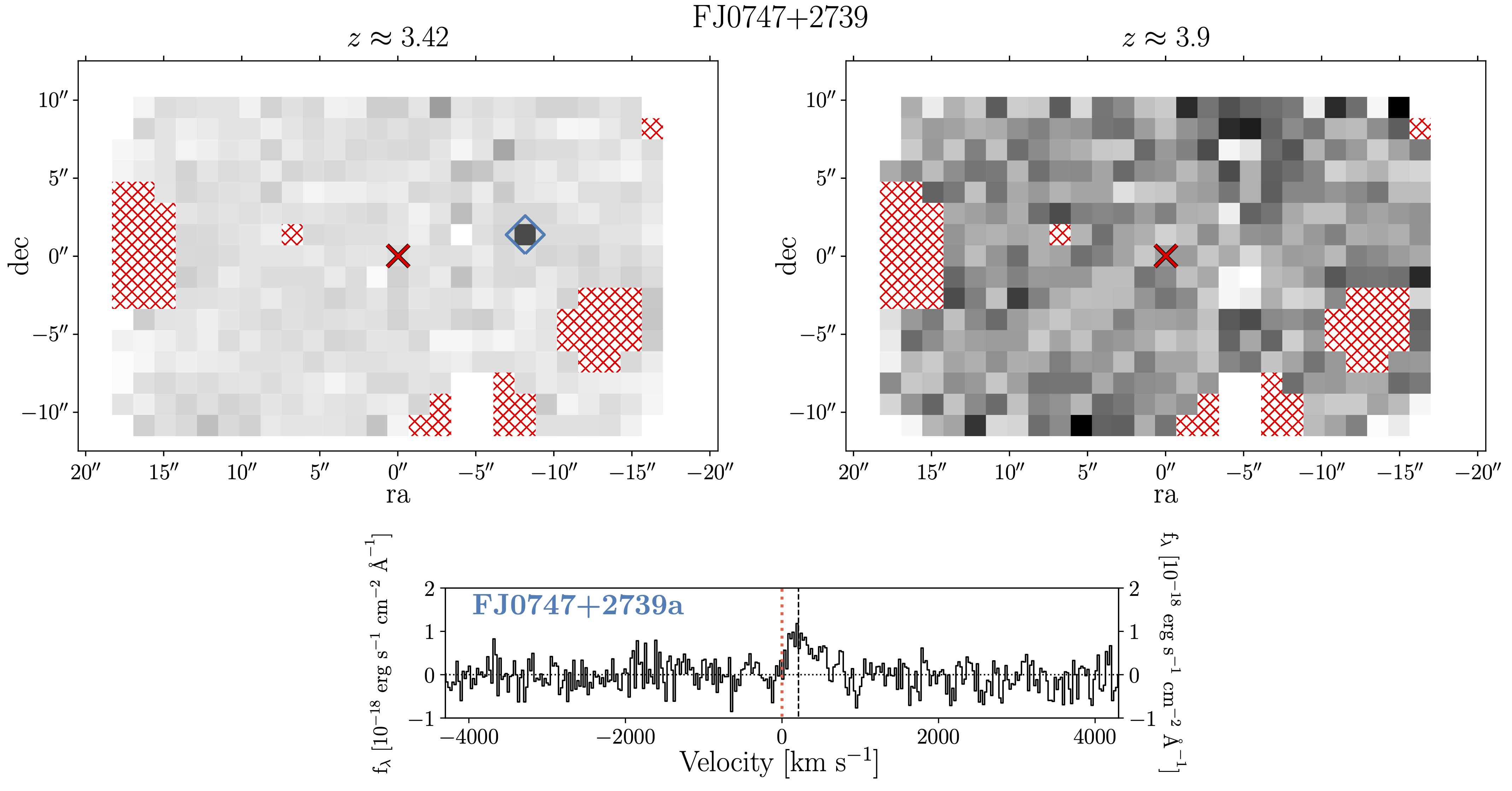}\\
\caption{KAGG data for the QSO sightline FJ0747+2739. See Appendix \ref{appC} for details.}
\end{figure*}

\newpage

 \begin{figure*}[h!]
\centering 
\includegraphics[width=\textwidth]{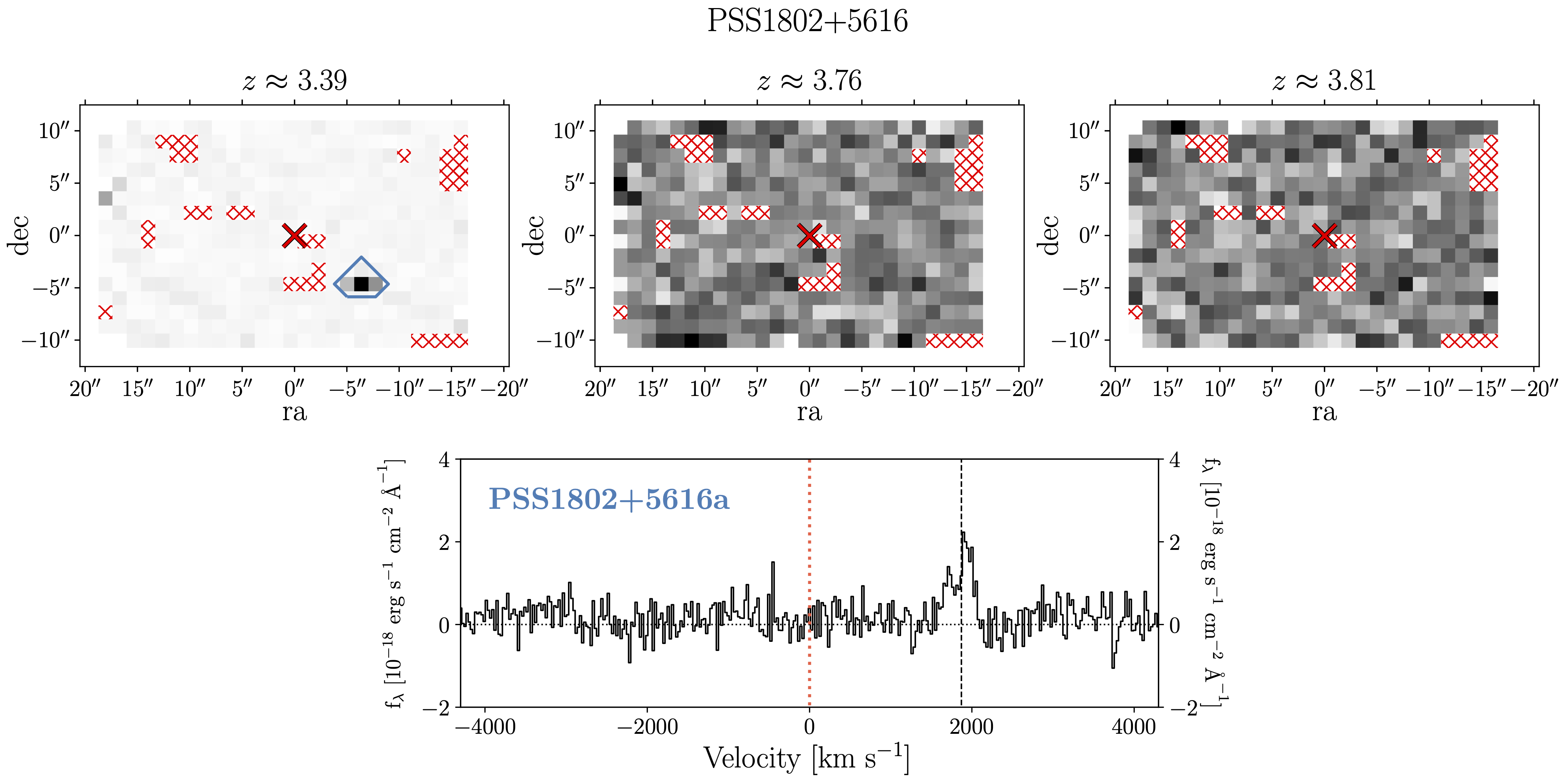}\\
\caption{KAGG data for the QSO sightline PSS1802+5616. See Appendix \ref{appC} for details.}
\end{figure*}

 \begin{figure*}[h!]
\centering 
\includegraphics[width=0.8\textwidth]{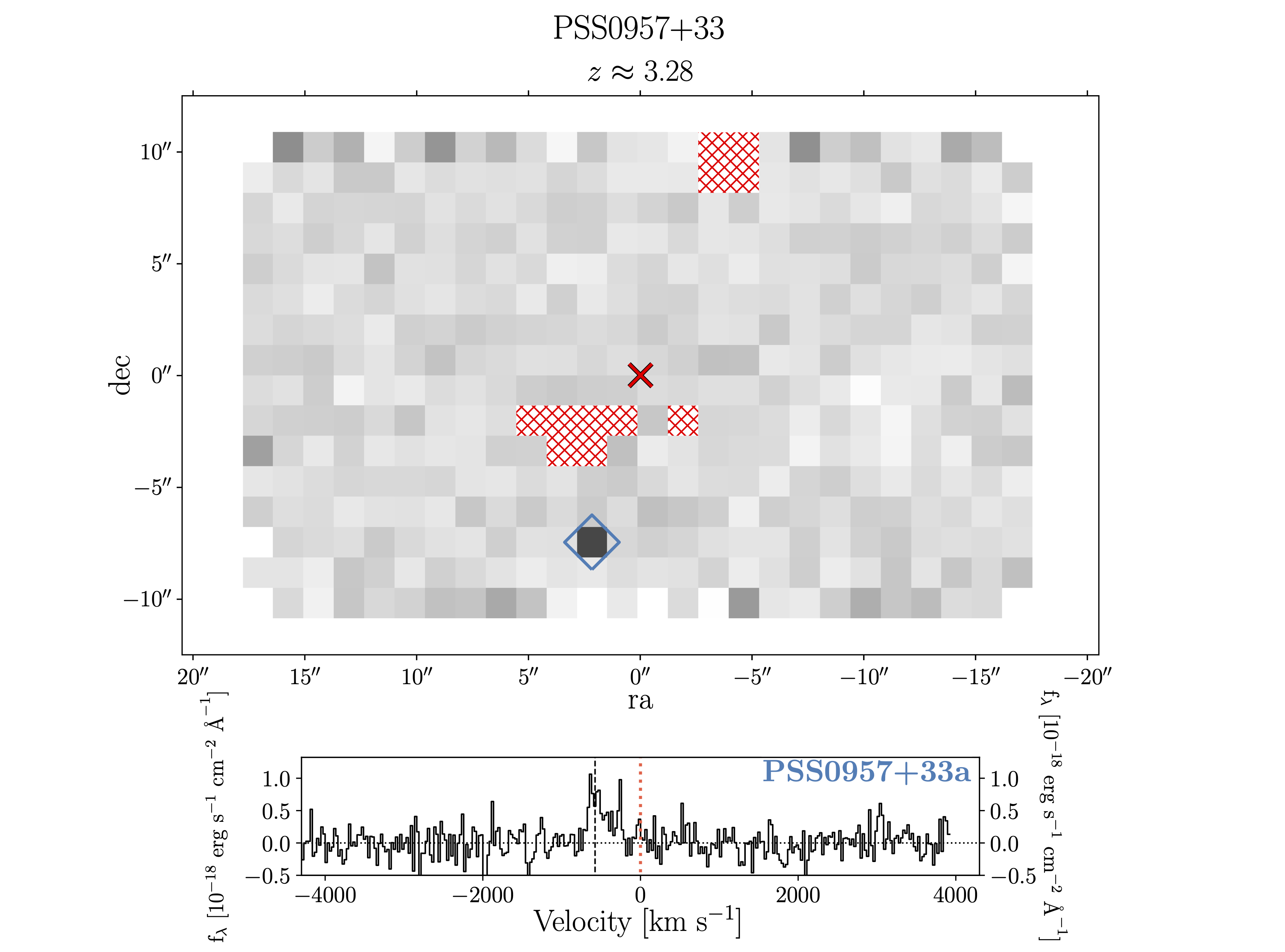}\\
\caption{KAGG data for the QSO sightline PSS0957+33. See Appendix \ref{appC} for details.}
\end{figure*}

\newpage

 \begin{figure*}[h!]
\centering 
\includegraphics[width=\textwidth]{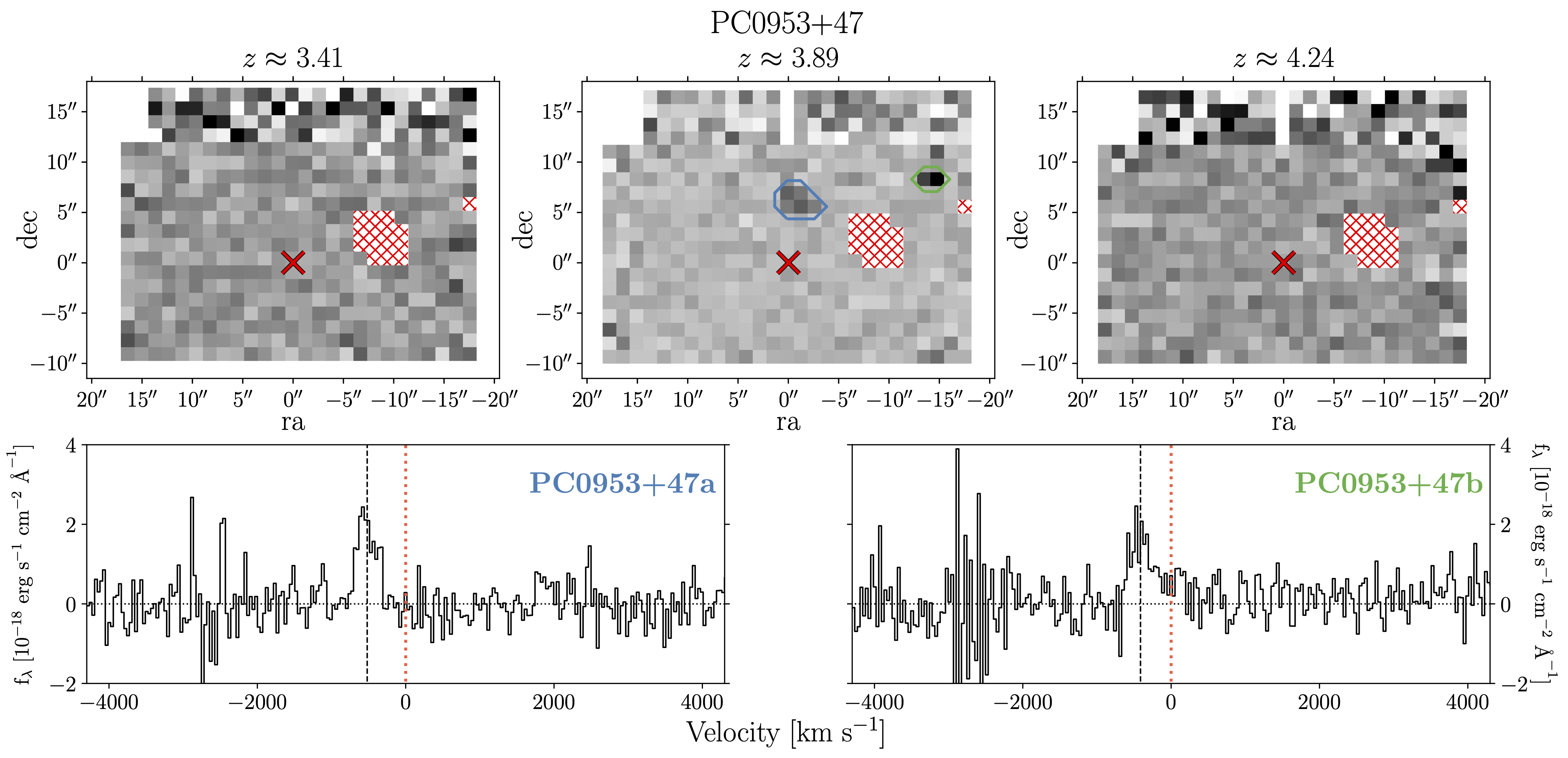}\\
\caption{KAGG data for the QSO sightline PC0953+47. See Appendix \ref{appC} for details.}
\end{figure*}

 \begin{figure*}[h!]
\centering 
\includegraphics[width=\textwidth]{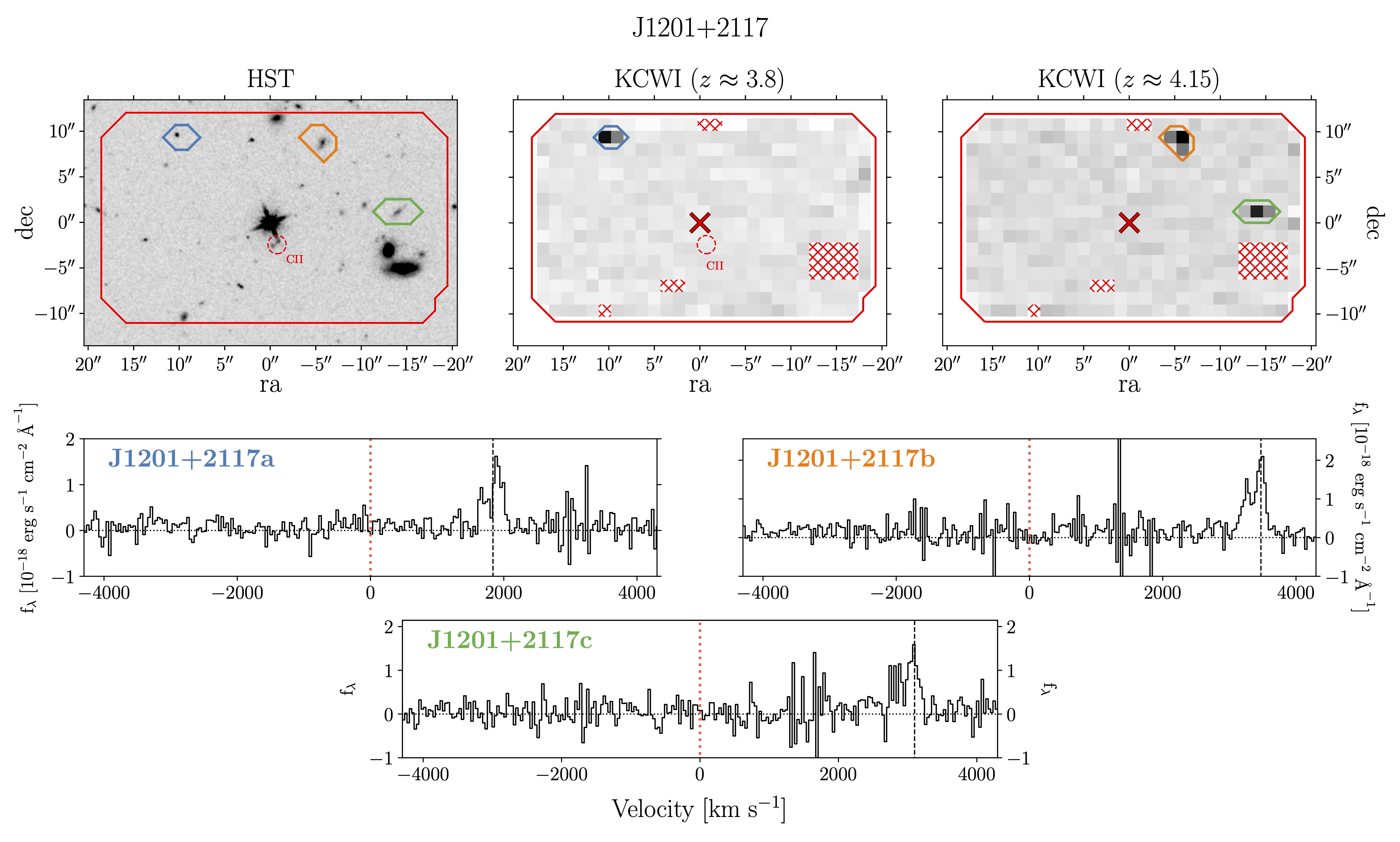}\\
\caption{KAGG data for the QSO sightline J1201+2117. See Appendix \ref{appC} for details.}
\end{figure*}

\newpage

 \begin{figure*}[h!]
\centering 
\includegraphics[width=\textwidth]{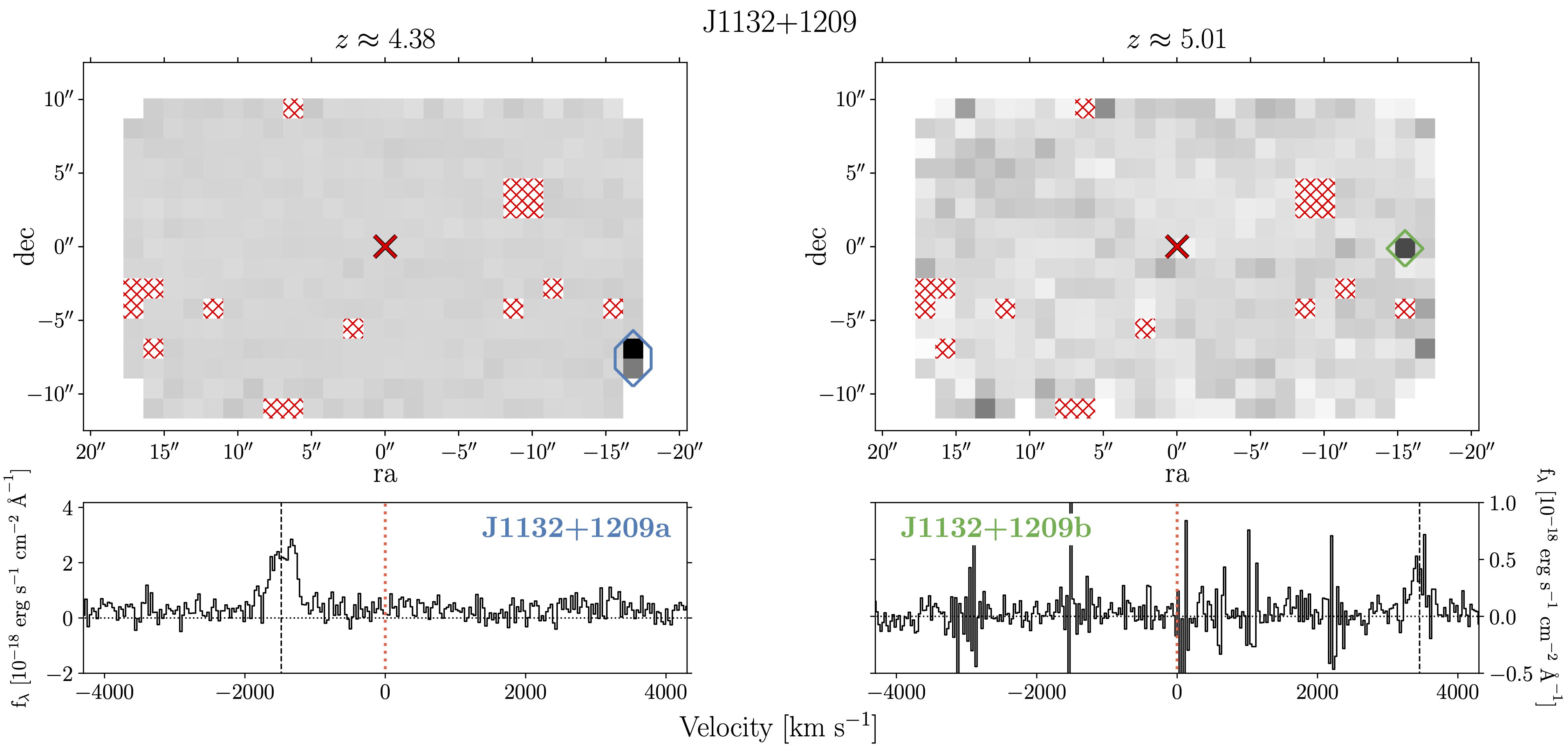}\\
\caption{KAGG data for the QSO sightline J1132+1209. See Appendix \ref{appC} for details.}
\end{figure*}

 \begin{figure*}[h!]
\centering 
\includegraphics[width=\textwidth]{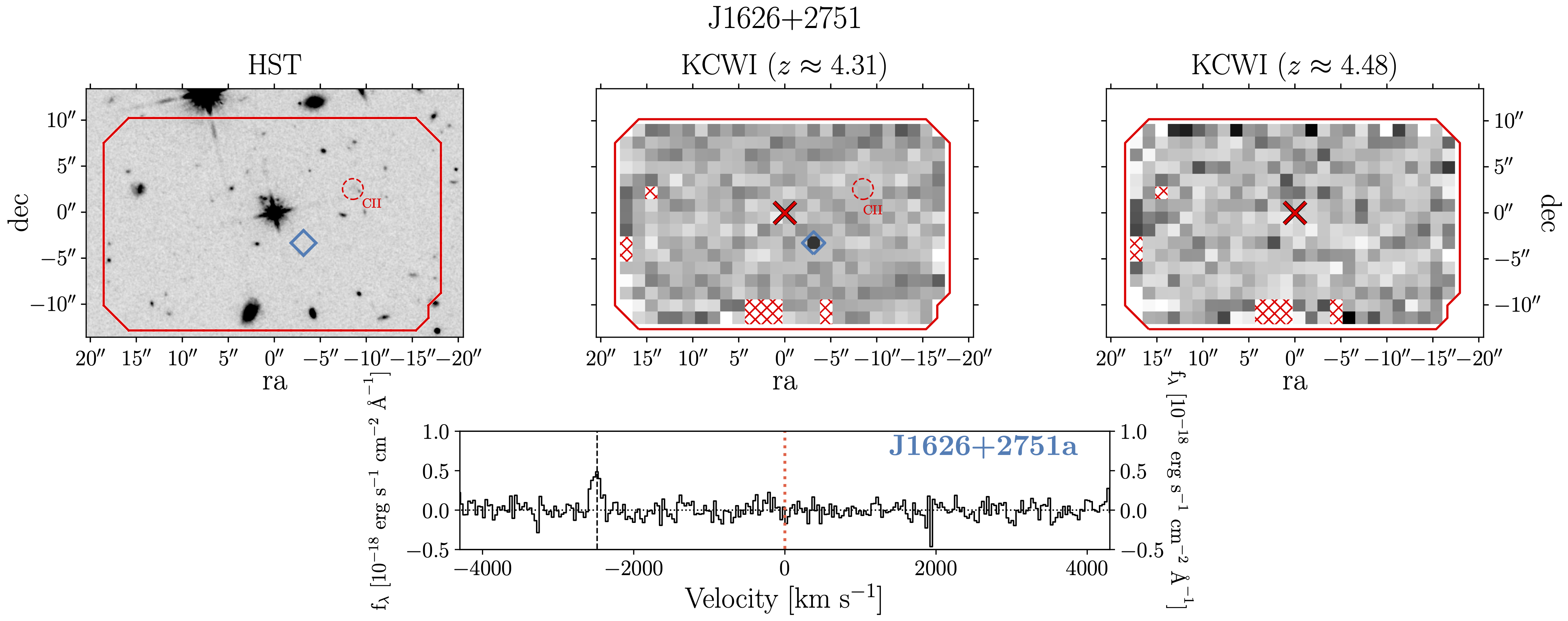}\\
\caption{KAGG data for the QSO sightline J1626+2751. See Appendix \ref{appC} for details.}
\vspace{-0.15in}
\end{figure*}

\newpage

\newpage

\bibliographystyle{aasjournal}
\bibliography{refs.bib}

\end{document}